\RequirePackage{luatex85}
\documentclass{jfm}
\usepackage{amsmath}
\usepackage{amssymb}
\usepackage{graphicx,xcolor}
\usepackage{subcaption}
\usepackage{adjustbox}
\usepackage{multirow}
\usepackage{dcolumn}
\usepackage{bm}
\usepackage{epstopdf, epsfig}

\newcommand{\be}[1]{\begin{equation}\label{#1}}
\newcommand{\ee}{\end{equation}}
\newcommand{\ba}[1]{\begin{eqnarray}\label{#1}}
\newcommand{\ea}{\end{eqnarray}}
\newcommand{\rf}[1]{(\ref{#1})}
\newcommand{\nn}{\nonumber}

\shorttitle{Membrane flutter induced by radiation of surface gravity waves on a uniform flow}
\shortauthor{J. Labarbe and O. N. Kirillov}

\title{Membrane flutter induced by radiation of surface gravity waves on a uniform flow}

\author{Joris Labarbe\aff{1}
 \and Oleg N.  Kirillov\aff{1}  \corresp{\email{oleg.kirillov@northumbria.ac.uk}}}

\affiliation{\aff{1}Northumbria University, Newcastle upon Tyne, NE1 8ST, UK}

\begin{document}

\maketitle

\begin{abstract}
We consider stability of an elastic membrane being on the bottom of a uniform horizontal flow
of an inviscid and incompressible fluid of finite depth with free surface. The membrane is simply supported at the leading and the trailing edges which attach it to the two parts of the horizontal rigid floor. The membrane has an infinite span in the direction perpendicular to the direction of the flow and a finite length in the direction of the flow. For the membrane of infinite length we derive a full dispersion relation that is valid for arbitrary depth of the fluid layer and find conditions for the flutter of the membrane due to emission of surface gravity waves. We describe this radiation-induced instability by means of the perturbation theory of the roots of the dispersion relation and the concept of negative energy waves and discuss its relation to the anomalous Doppler effect.
\end{abstract}


\section{Introduction}

Flutter of membranes is a classical subject for at least seven decades. Membranes submerged in a compressible gas flow occupying a space or a semi-space and their flutter at supersonic speeds have been considered already in the works by \cite{M1947,M1956}, \cite{GL1954}, \cite{TBB1963}, and \cite{B1963}. \cite{B1963}, \cite{S1969}, \cite{DV1970}, and \cite{KD1976} addressed a problem of the so-called membrane flutter paradox on the relation of stability criteria for an elastic plate to that for a membrane. \cite{G1971} demonstrated both theoretically and experimentally that the membrane or elastic plate with the finite chord possesses not only flutter but also a divergence instability. \cite{D1966}, when critically appraising the study by \cite{M1956} of an infinitely long, infinitely wide panel in a compressible flow occupying the upper semi-space, pointed out that the critical wavelength predicted in this study was infinite and the flutter velocity was zero, which was not physically meaningful. This observation has led him to a conclusion that the finite dimension of a membrane or a plate in the flow- or a span direction is critical to the physically meaningful prediction of the instability \citep{D1966}. A similar effect of an elastic foundation was shown both theoretically and experimentally by \cite{DDP1963}. Absolute and convective hydroelastic instabilities of slender elastic structures submerged in a uniform flow were discussed by \cite{T1992}. A comprehensive monograph by \cite{D2015} is a standard reference in the field.

Recent works on membrane flutter are motivated by such diverse applications as stability of membrane roofs in civil engineering \citep{S2007}, flutter of traveling paper webs \citep{B2010,B2019}, aerodynamics of sails and membrane wings of natural flyers \citep{N1991,TR2017}, as well as design of piezoaeroelastic systems for energy harvesting \citep{MA2019}.

Surface gravity waves on a motionless fluid of finite depth is a classical subject as well, going back to the seminal studies of Russell and Kelvin \citep{CR2013}. Numerous generalizations are known taking into account, for instance, a uniform or a shear flow and surface tension \citep{MRS2016}, submerged solids \citep{S1972, AK2016} and hydrofoils \citep{FS2008}, flexible bottom \citep{MS2011} or a flexible plate resting on a free surface \citep{DSM2018a,DSM2018b,DKSM2018c,BLM2016,S1987}. The latter setting has a straightforward motivation in dynamics of sea ice and a less obvious application in analogue gravity experiments \citep{CR2013, AG2011, HRAG2011}.
Recent work \citep{RR2018} discusses effects of viscous dissipation of surface gravity waves to the analogue gravity.

Remarkably,
another phenomenon that is analysed from the analogue gravity perspective is super-radiance \citep{CR2013,AG2011,BCP2015} and its particular form, discovered by Ginzburg and Frank \citep{GF1947,G1996}, known as the anomalous Doppler effect (ADE) \citep{BS1998,Nezlin76,NE1987}. In electrodynamics, the ADE manifests itself when an electrically neutral overall particle, endowed with an internal structure, becomes excited and emits a photon during its uniform but superluminal motion through a medium, even if it started the motion in its ground state; the energy source is the bulk motion of the particle \citep{BS1998}.

The anomalous Doppler effect in hydrodynamics was demonstrated for a mechanical oscillator with one degree of freedom, moving parallel to the border between two incompressible fluids of different densities \citep{GG1983}. It was shown that the oscillator becomes excited due to radiation of internal gravity waves if it moves sufficiently fast. In \cite{AMN1986} the ADE for such an oscillator was demonstrated due to radiation of surface gravity waves in a layer of an incompressible fluid.

\cite{N1986} was the first who considered flutter of an elastic membrane resting at the bottom of a uniform horizontal flow
of an inviscid and incompressible fluid as an anomalous Doppler effect due to emission of long surface gravity waves.
In the shallow water approximation, he investigated both the case of a membrane that spreads infinitely far in both horizontal directions and the case when the length of the membrane in the direction of the flow (or the chord length) is finite whereas the span in the perpendicular direction is infinite.
Nevertheless, the case of flow of arbitrary depth has not been studied in \cite{N1986}, and no numerical computation supporting the asymptotical results has been performed. Another issue that has not been addressed in \cite{N1986} is the relation of stability domains for the membrane of finite length to that for the membrane of infinite length.

\cite{V2004} studied flutter of an elastic plate of finite and infinite length at the bottom of a uniform horizontal flow of a compressible gas occupying the upper semi-space. He performed analysis of the relation of stability conditions for the finite plate with that for the infinite plate using the method of global stability analysis of Kulikovskii \citep{DL2006,V2016}. A single-mode high frequency flutter due to a negative aerodynamic damping and a binary flutter have been identified in \cite{V2016}. However, no connection has been made to the ADE and the concept of negative energy waves.

In the present work we reconsider the setting of Nemtsov in order to address the finite depth of the fluid layer, find flutter domains in the parameter space, analyze them using perturbation of multiple roots of the dispersion relation and investigate the flutter onset for the membrane of infinite chord length. We will explain the radiative instabilities via the interaction of positive and negative energy waves using an explicit expression for the averaged total energy derived rigorously from physical considerations and relate them to the anomalous Doppler effect. We believe that the Nemtsov membrane is as important for understanding the phenomenon of radiation-induced instabilities \citep{HBW2003} as the famous Lamb oscillator coupled to a semi-infinite string  was for understanding the radiative damping \citep{L1900,BC1994}.

\section{Model of a membrane interacting with a free surface}
\subsection{Physical system}

\begin{figure}
\centering
  \includegraphics[scale=.7]{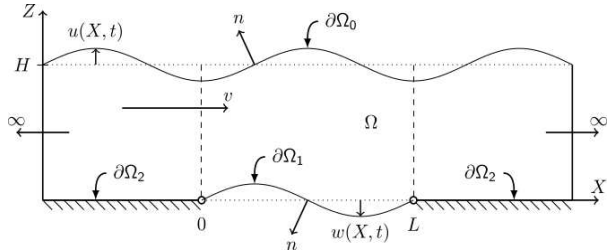}
  \caption{An elastic membrane with chord of length $L$ attached to two rigid walls along its leading $(X=0)$ and trailing $(X=L)$ edges on the bottom of a fluid layer of depth $H$ moving with the velocity $v$. $\Omega$ is the fluid domain and $\partial\Omega_0$, $\partial\Omega_1$, $\partial\Omega_2$ are respectively the free surface, membrane, and rigid wall boundaries.\label{fig:sketch}}
\end{figure}

In a Cartesian coordinate system $OXYZ$, consider an inextensible elastic rectangular membrane strip of constant thickness $h$ and density $\rho_m$, of infinite span in the $Y$-direction, held at $Z=0$ at the leading edge $(X=0)$ and at the trailing edge $(X=L)$ by simple supports, Fig.~\ref{fig:sketch}.

The membrane is initially still and flat, immersed in a layer of inviscid, incompressible fluid of constant density $\rho$, with free surface at the height $Z=H$. The two-dimensional flow in the layer is supposed to
be irrotational and moving steadily with velocity $v$ in the positive $X$-direction. The bottom of the fluid layer at $Z=0$ is supposed to be rigid
and flat for $X\in (-\infty,0] \cup [L,+\infty)$.

\cite{N1986} assumed that vacuum exists below the membrane. In the present study we prefer to consider that a motionless incompressible medium of the same density $\rho$ is present below the membrane with a pressure that is the same as the unperturbed pressure of the fluid \citep{V2004,V2016}.

Assuming small vertical displacement of the membrane $w(X,t)$, where $t$ is time, a constant tension $T$ along the
membrane profile, and neglecting viscous forces, we write the dimensional membrane dynamic equation as \citep{TR2017}
\be{memeq}
\rho_m h \partial^2_{t}w=T\partial^2_{X}w-\Delta P,\quad X\in\left[0,L\right], \, Z=0,
\ee
where $\Delta P(X,t)$ is the pressure difference across the interface $Z=0$. The simply-supported boundary conditions for the membrane are
\be{ssbc}
w(0)=w(L)=0 \quad \text{at} \quad Z=0.
\ee

In general, to recover the pressure $P(X,Z,t)$ of the fluid we write the Euler equation for the vorticity-free flow \citep{CR2013,MRS2016}
\be{euler}
\partial_t \bm{v}+\bm\nabla\left(\frac{\bm{v}\cdot \bm{v}}{2}+\frac{P}{\rho}+gZ \right)=0
\ee
with $\bm{v}=v \bm{e}_X+\bm\nabla \varphi$, where $\varphi(X,Z,t)$ is the potential of the fluid, $\bm{e}_X$ is the unit vector in the $X$-direction, and $g$ stands for the gravity acceleration. This yields the integral of Bernoulli
\be{bp}
\frac{P}{\rho} + \left(\partial_t + v\partial_X\right)\varphi + \frac{1}{2}\bm{\nabla}\varphi\cdot\bm{\nabla}\varphi + gZ = \text{constant}.
\ee
The incompressibility condition takes the form
\be{icc}
\bm{\nabla}^2 \varphi=0.
\ee

From equation \rf{bp} it follows that in the case when a motionless medium of density $\rho$ is present below the membrane with its pressure equal to the unperturbed pressure of the fluid above the membrane, the linear in $\varphi$ expression for the pressure difference, $\Delta P(X,t)$, is
\be{vedp}
\Delta P(X,t) = -\rho\left(\partial_t + v\partial_X\right)\varphi(X,0,t).
\ee
For the sake of completeness, we present also the analogous expression for the pressure difference for the case when there is vacuum below the membrane \citep{N1986}
\be{nemp}
\Delta P(X,t) = -\rho\left(\partial_t + v\partial_X\right)\varphi(X,0,t) - \rho g w(X,t).
\ee

Impermeability of the rigid bottom implies  the condition
\be{irb}
\bm{\nabla}\varphi\cdot\bm{n} = 0 \quad \text{at} \quad Z=0, \, X\in (-\infty,0] \cup [L,+\infty).
\ee
The prescription of normal velocity at the boundaries of moving surfaces allows us to express the kinematic condition for the membrane
\be{imp}
\bm{\nabla}\varphi\cdot\bm{n} = -\left(\partial_t + v\partial_X\right)w \quad \text{at} \quad Z=0, \, X\in [0,L]
\ee
and to specify the same condition at the free surface
\be{kc}
\bm\nabla{\varphi} \cdot \bm{n} = (\partial_t + v\partial_X)u,
\ee
where $u(X,t)$ is the free surface elevation and $\bm{n}$ is the vector of the outward normal to a surface. This implies that the projection of the vector $\bm{\nabla}\varphi$ to the normal will coincide with the pozitive $z$-direction for the free surface and have the opposite direction for the membrane, see Fig.~\ref{fig:sketch}.

Using the Bernoulli integral \rf{bp} at the free surface where $P=0$ and retaining only linear in $\varphi$ terms,
we find
\be{dc}
g u = -(\partial_t + v \partial_X)\varphi.
\ee
Taking $u$ from \rf{dc} and substituting it into \rf{kc} we obtain the boundary condition at the free surface of the liquid that reads
\be{fs}
\bm{\nabla}\varphi\cdot\bm{n} = -\frac{1}{g}\left(\partial_t + v\partial_X\right)^2\varphi \quad \text{at} \quad Z=H.
\ee

\subsection{Dimensionless mathematical model}

Let us choose the height of the fluid layer, $H$, as a length scale, and $\omega_0^{-1}$, where $\omega_0=\sqrt{g/H}$, as a time scale.
Then, we can introduce the dimensionless time and coordinates
\be{dlp1}
\tau=t\omega_0,\quad x=\frac{X}{H}, \quad y=\frac{Y}{H}, \quad z=\frac{Z}{H},
\ee
the dimensionless variables
\be{dlp2}
\xi=\frac{w}{H},\quad \eta=\frac{u}{H},\quad \phi=\frac{\omega_0}{gH}\varphi,
\ee
the dimensionless parameters of the added mass ratio \citep{M1998} and membrane chord length
\be{dlp3}
\alpha=\frac{\rho H}{\rho_m h},\quad \Gamma=\frac{L}{H},
\ee
and the two Mach numbers \citep{V2004,V2016}
\be{dlp4}
M_w=\frac{c}{\sqrt{g H}}, \quad M=\frac{v}{\sqrt{g H}},
\ee
where $c^2=T/(\rho_m h)$ is the squared speed of propagation of elastic waves in the membrane
and $\sqrt{g H}$ is the speed of propagation of long surface gravity waves in the shallow water approximation.
The added mass ratio $\alpha$ is the ratio of the fluid to solid mass contained in the volume delimited by the dashed lines in Fig.~\ref{fig:sketch} and in the membrane \citep{M1998}. In Fig.~\ref{fig:sketch}, $\Omega$ denotes the fluid domain and $\partial\Omega_0$, $\partial\Omega_1$, $\partial\Omega_2$ stand, respectively, for the free surface, membrane, and solid wall borders.

The dimensionless wave equation \rf{memeq} is therefore
\be{dlwe}
\partial^2_{\tau}\xi-M_w^2\partial^2_{x}\xi=-\alpha \frac{\Delta P}{\rho g H},\quad x\in\left[0,\Gamma\right], \,\, z=0.
\ee
Supplementing it with the expression \rf{vedp}, which in the dimensionless time and coordinates has the form
\be{dp}
\frac{\Delta P}{\rho}=-\left(\omega_0\partial_{\tau}+\frac{v}{H}\partial_x \right)\varphi,
\ee
we find
\ba{dlwe2}
\partial^2_{\tau}\xi-M_w^2\partial^2_{x}\xi&=&\alpha \left(\partial_{\tau}+\frac{v}{\omega_0 H}\partial_x \right)\frac{\omega_0}{g H}\varphi\nn\\
&=&\alpha \left(\partial_{\tau}+M\partial_x \right)\phi.
\ea

The dimensionless boundary condition \rf{imp} is
\be{dlkc}
\bm{\nabla}\phi\cdot\bm{n} = - \left(\partial_{\tau} + M\partial_x\right)\xi  \quad \text{at} \quad z=0, \, x\in [0,\Gamma],
\ee
whereas the boundary condition \rf{fs} at the free surface in dimensionless form becomes
\be{dlfs}
\bm{\nabla}\phi\cdot\bm{n} = - \left(\partial_{\tau} + M\partial_x\right)^2\phi \quad \text{at} \quad z=1.
\ee

Collecting together equations \rf{dlwe2}, \rf{dlkc}, \rf{dlfs} and the obvious dimensionless versions of equations \rf{icc} and \rf{irb}
and assuming a time dependence in the form of $\phi,\xi \sim e^{-i \omega \tau}$ results in the following dimensionless set of equations and their boundary conditions for the case when a motionless medium is present below the membrane:
\begin{subequations}
\label{eq:eom}
\begin{align}
\bm{\nabla}^2\phi &= 0, & &\text{in} \, \Omega \label{eoma} \\
\bm{\nabla}\phi\cdot\bm{n} &= -\left( -i\omega + M\partial_x \right)^2\phi, & &\text{on} \, \partial\Omega_0 \label{eomb}\\
\bm{\nabla}\phi\cdot\bm{n} &= V(x), & &\text{on} \, \partial\Omega_1 \label{eomc}\\
\bm{\nabla}\phi\cdot\bm{n} &= 0, & &\text{on} \, \partial\Omega_2 \label{eomd}\\
\left[\omega^2 + M_w^2\partial_x^2\right]\xi &= -\alpha\left(-i\omega + M\partial_x \right)\phi, & &\text{on} \, \partial\Omega_1 \label{eome}\\
\xi(0) = \xi(\Gamma) &= 0, & &\text{on} \, \partial\Omega_1 \label{eomf}
\end{align}
\end{subequations}
where $V(x)=( i\omega - M\partial_x )\xi(x), \, x\in[0,\Gamma]$ is the impermeability condition for the membrane. For simplicity, we retain the same notation for the membrane displacement and the fluid potential after the separation of time.

Therefore, due to the irrotational, incompressible and inviscid character of the fluid, our mathematical model \rf{eq:eom} consists of the Laplace equation for the fluid potential \rf{eoma}, supplemented by the kinematic conditions for the free surface \rf{eomb} and the membrane \rf{eomc}. The pressure at the surface of the fluid is also prescribed as a dynamic condition and therefore closes the system of equations for the fluid in this model: the motion of the membrane is described by a nonhomogeneous wave equation \rf{eome} with the pressure of the fluid (recovered through the Bernoulli principle) as a source term. The membrane is supposed to be simply supported at its extremities as in \rf{eomf}.

\section{Methods and Results}

\subsection{Membrane of the infinite chord length}

Our ultimate goal is to understand the fundamentals of the phenomenon of radiation-induced instabilities in the model \rf{eq:eom} that we see as a reasonable analytically treatable substitute for the famous Lamb system \citep{L1900,BC1994,HBW2003}. In this paper, as a first natural step, we analyse the case when the chord of the membrane is \textit{infinite}, i.e. when the membrane extends from $-\infty$ to $+\infty$ in the $x$-direction.

The extension of the Nemtsov model to the case where the fluid layer presents a \textit{finite depth} is our main concern. In the following we will show that even in the limit of infinite chord length the model \rf{eq:eom} demonstrates physically meaningful radiation-induced flutter that sets in at finite values of the dimensionless flow velocity $M>M_w>0$, no matter what are the values of the wavenumber $\kappa$ and the added mass ratio $\alpha$ in contrast to other known models discussed, e.g. in \cite{M1956} and \cite{D1966}.

\subsubsection{Dispersion relation for the fluid layer of arbitrary depth}

Since the motion of the fluid is two-dimensional in the $(x,z)$-plane and the horizontal extension of the fluid layer is infinite in the $x$-direction too, we can represent the potential of the fluid $\phi$ in the physical space by means of the inverse Fourier transform of the potential $\hat \phi$ in the wavenumber space as
\be{ifift}
\phi(x,z,\omega) = \frac{1}{2\pi}\int_{-\infty}^{+\infty} \hat{\phi}(\kappa,z,\omega) e^{i\kappa x} d\kappa,
\ee
where $\kappa$ is the wavenumber and
\be{fift}
\hat\phi(\kappa,z,\omega) = \int_{-\infty}^{+\infty} \phi(x,z,\omega) e^{-i\kappa x} dx,
\ee
under standard assumption that both $\phi(x)$ and $\hat\phi(\kappa)$ are absolutely integrable functions,
implying their vanishing at infinity.


\begin{figure*}
  \begin{subfigure}{0.25\textwidth}
    \includegraphics[width=\textwidth]{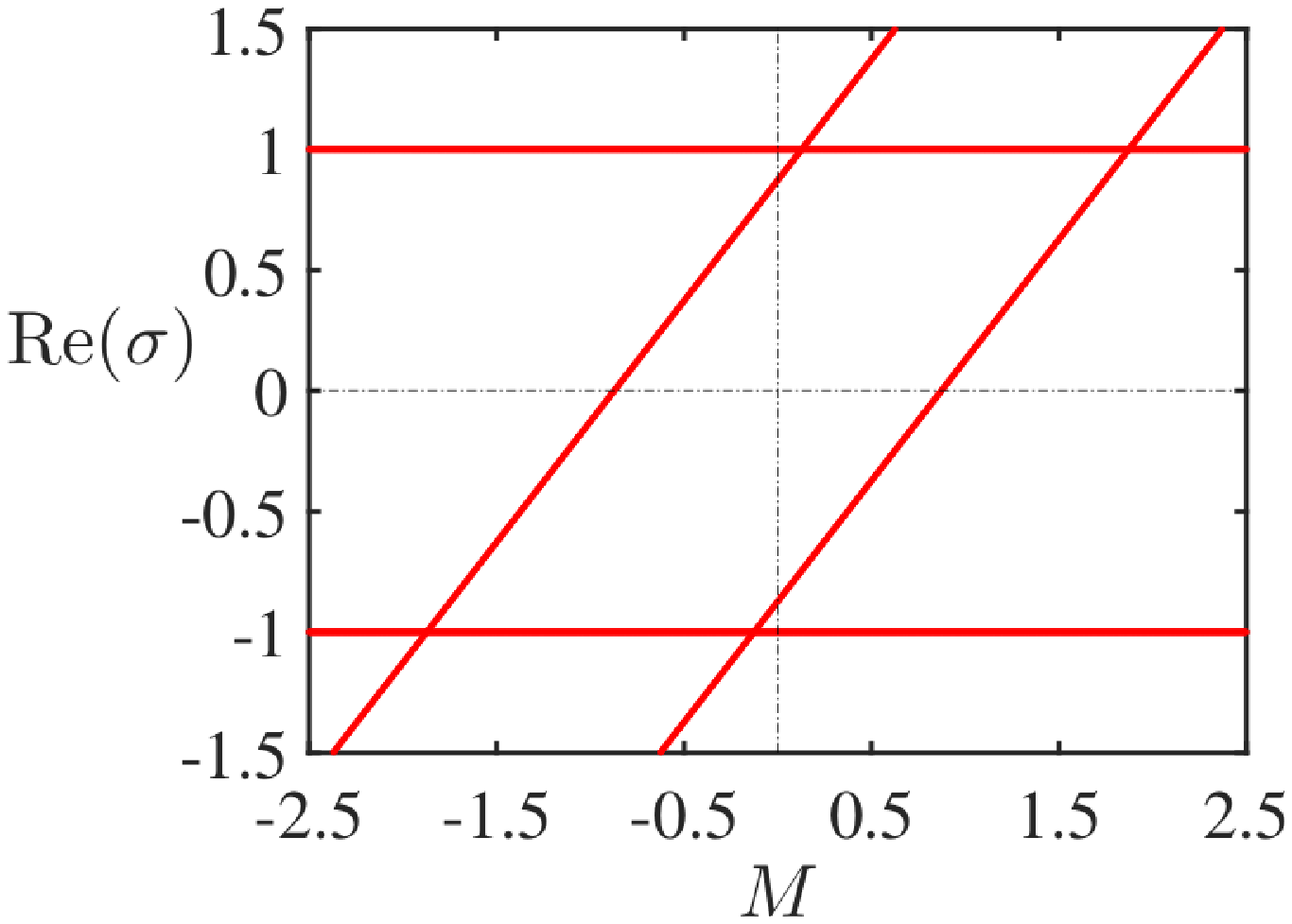}
    \caption{} \label{fig:2a}
  \end{subfigure}%
  \begin{subfigure}{0.25\textwidth}
    \includegraphics[width=\textwidth]{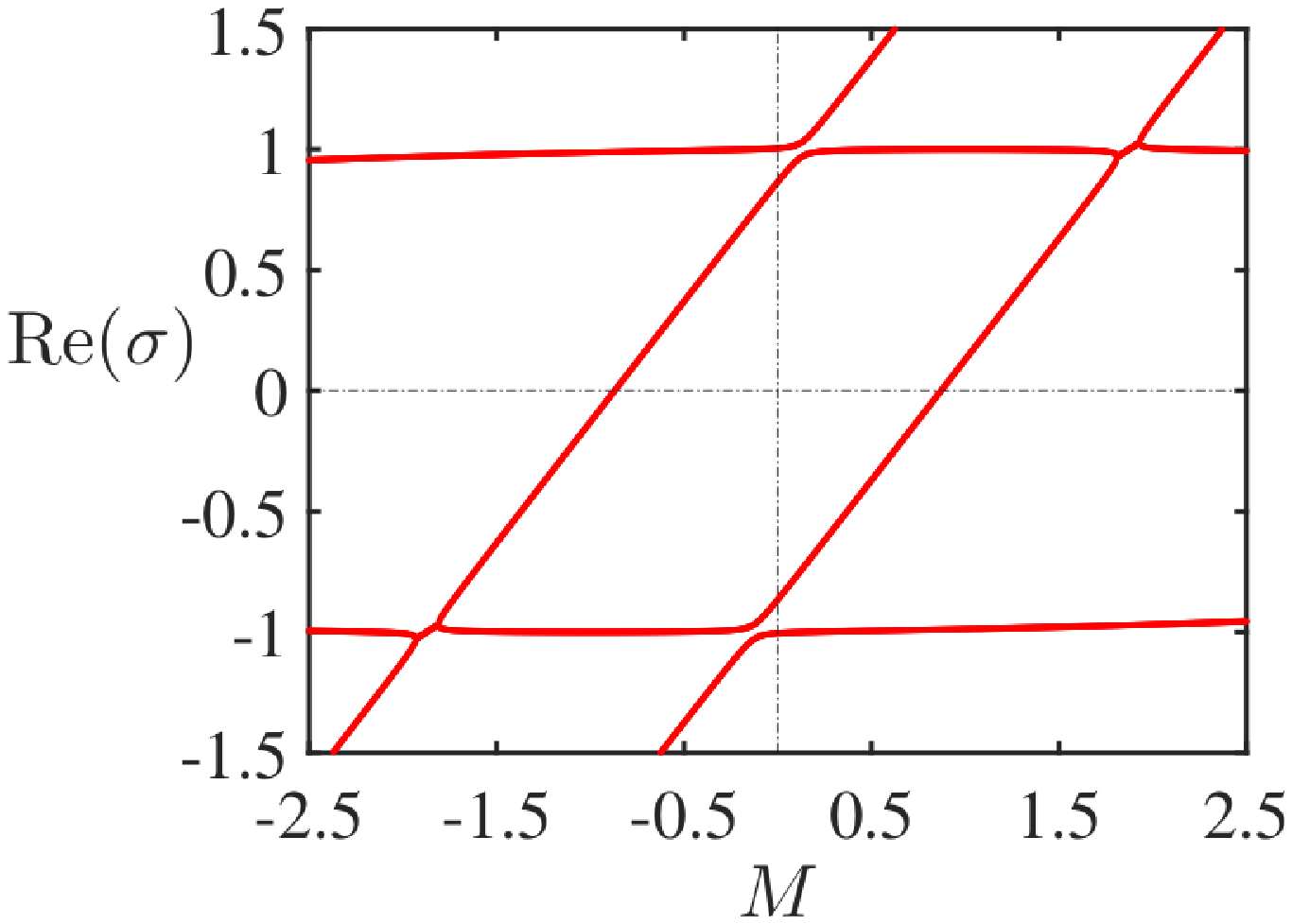}
    \caption{} \label{fig:2b}
  \end{subfigure}%
  \begin{subfigure}{0.25\textwidth}
    \includegraphics[width=\textwidth]{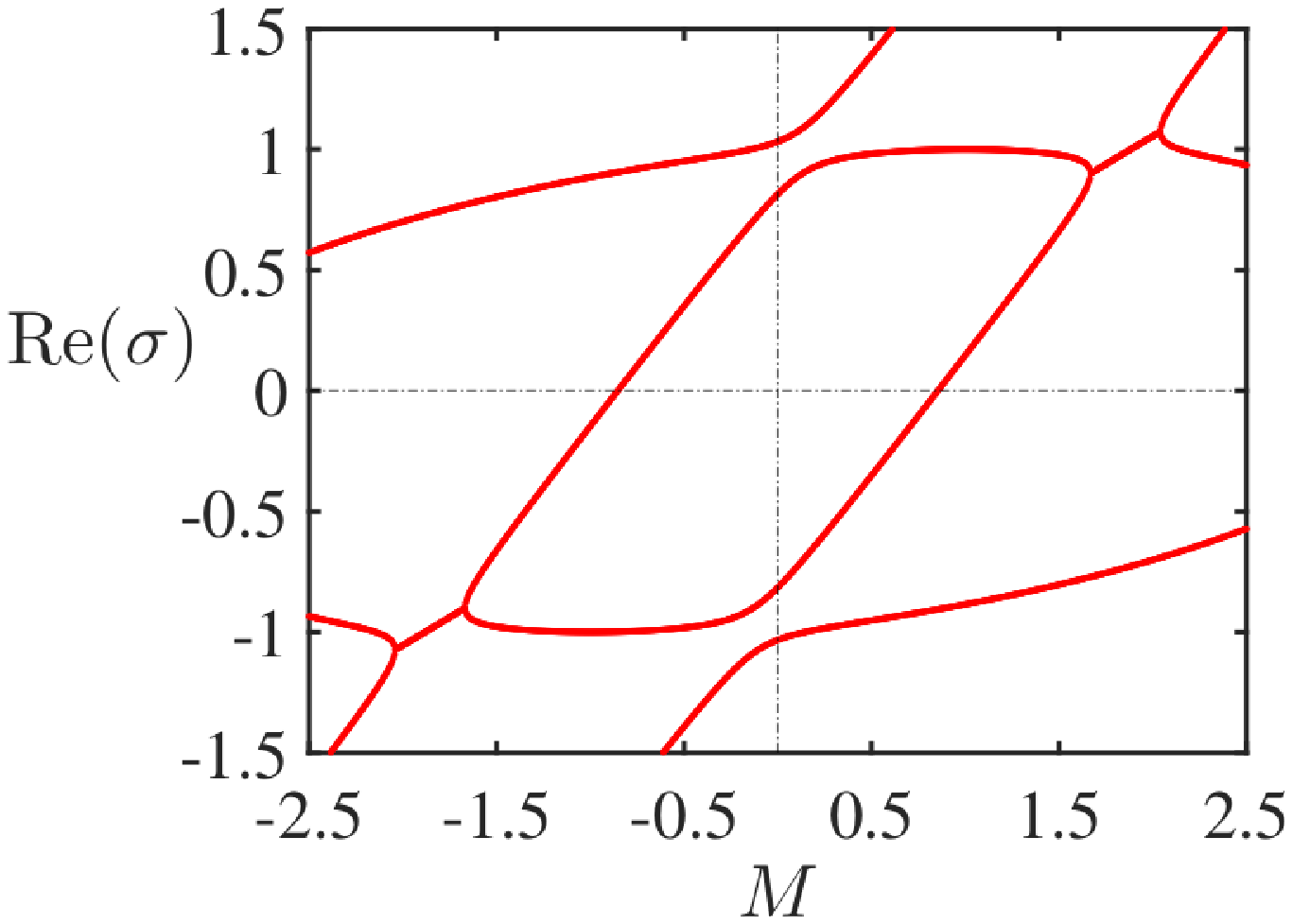}
    \caption{} \label{fig:2c}
  \end{subfigure}%
  \begin{subfigure}{0.25\textwidth}
    \includegraphics[width=\textwidth]{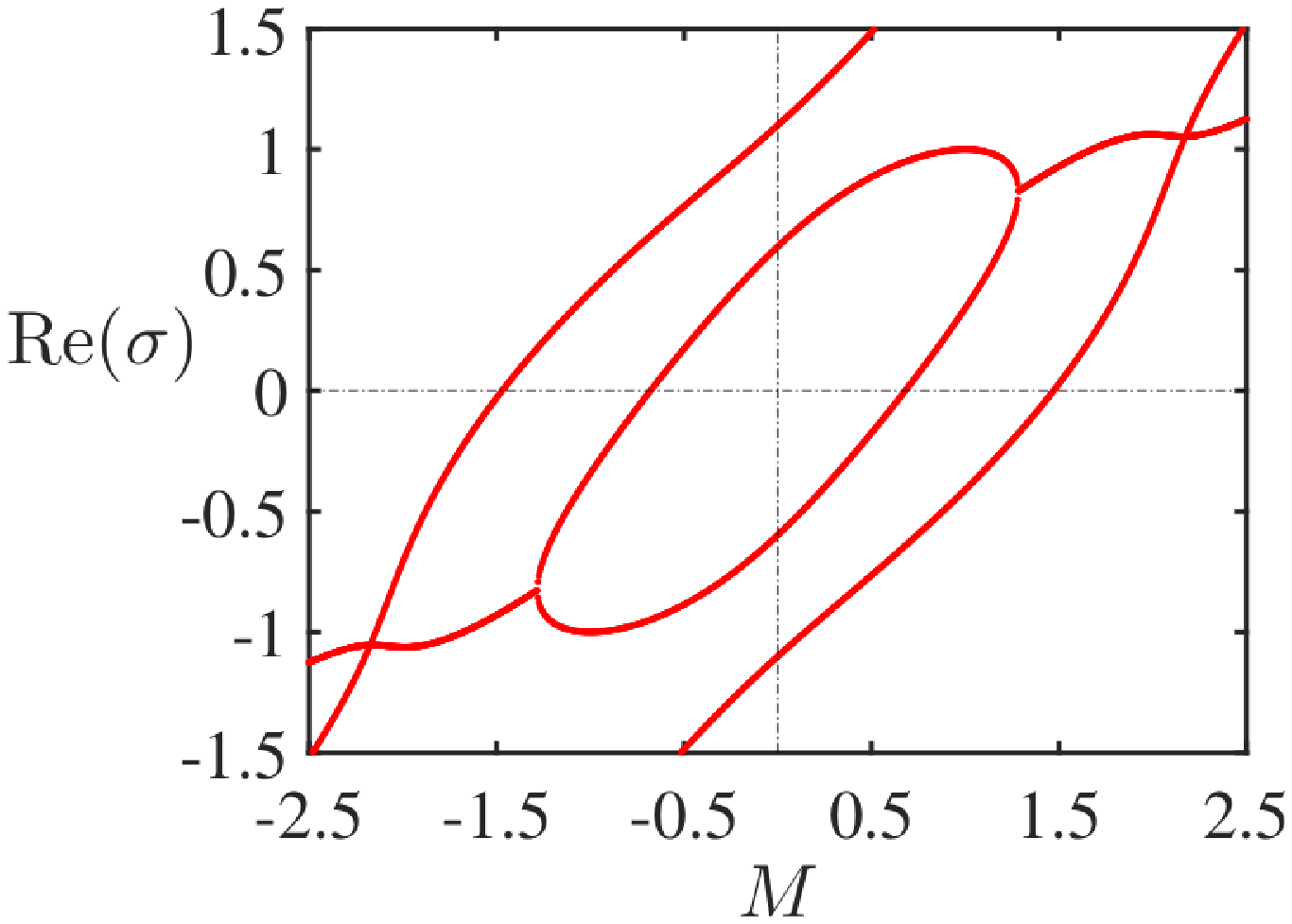}
    \caption{} \label{fig:2d}
  \end{subfigure}\\
  \begin{subfigure}{0.25\textwidth}
    \includegraphics[width=\textwidth]{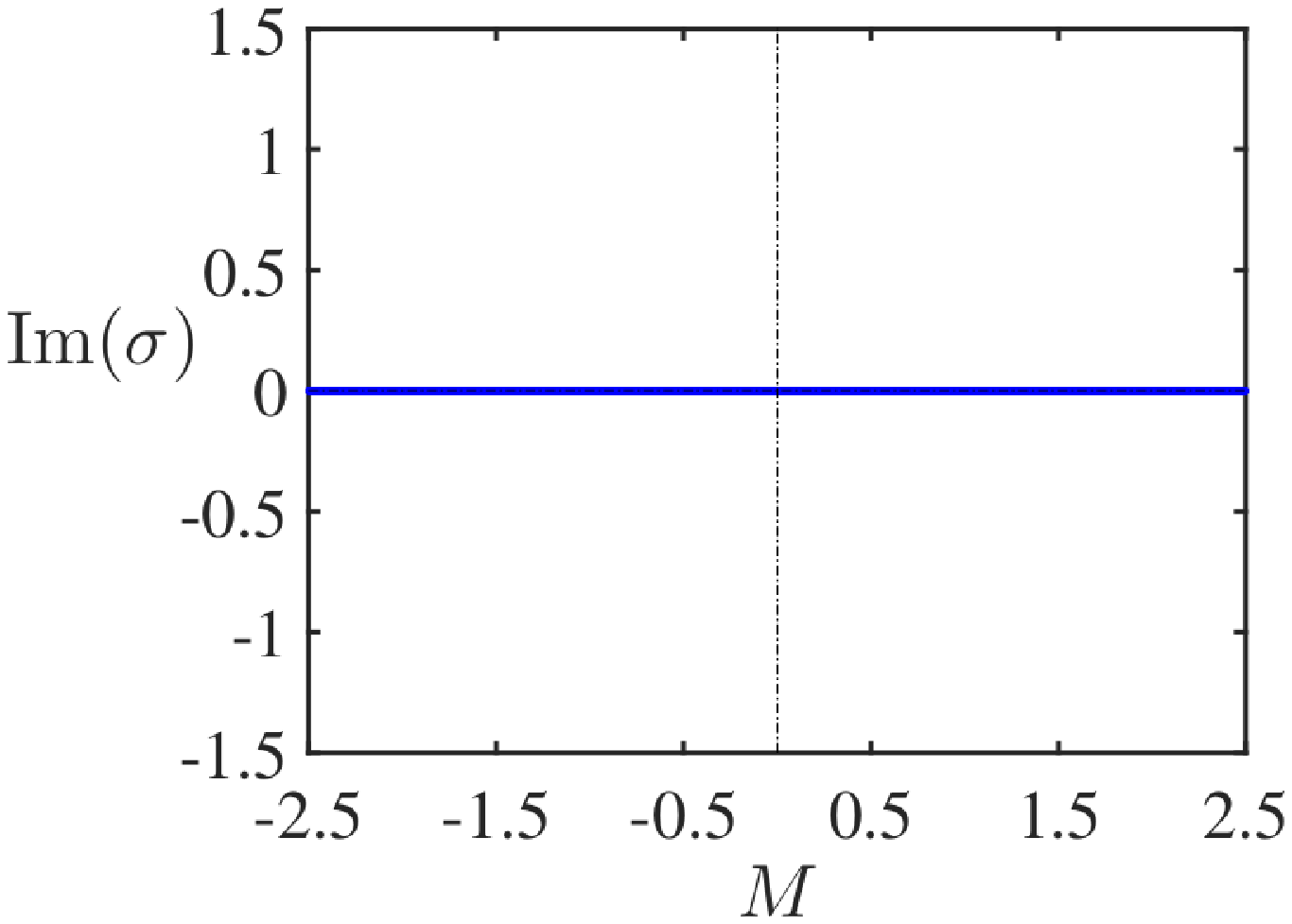}
    \caption{} \label{fig:2e}
  \end{subfigure}%
  \begin{subfigure}{0.25\textwidth}
    \includegraphics[width=\textwidth]{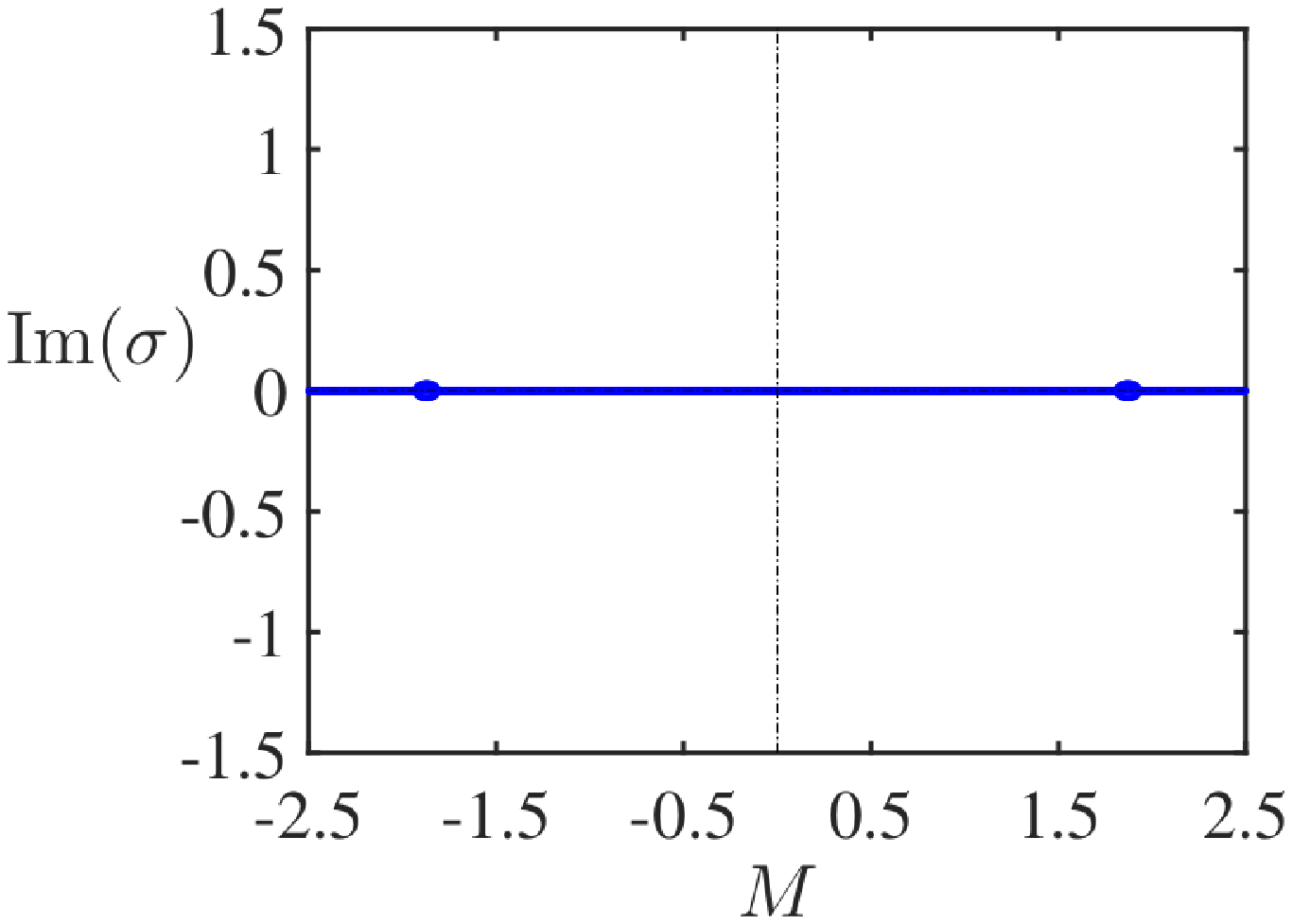}
    \caption{} \label{fig:2f}
  \end{subfigure}%
  \begin{subfigure}{0.25\textwidth}
    \includegraphics[width=\textwidth]{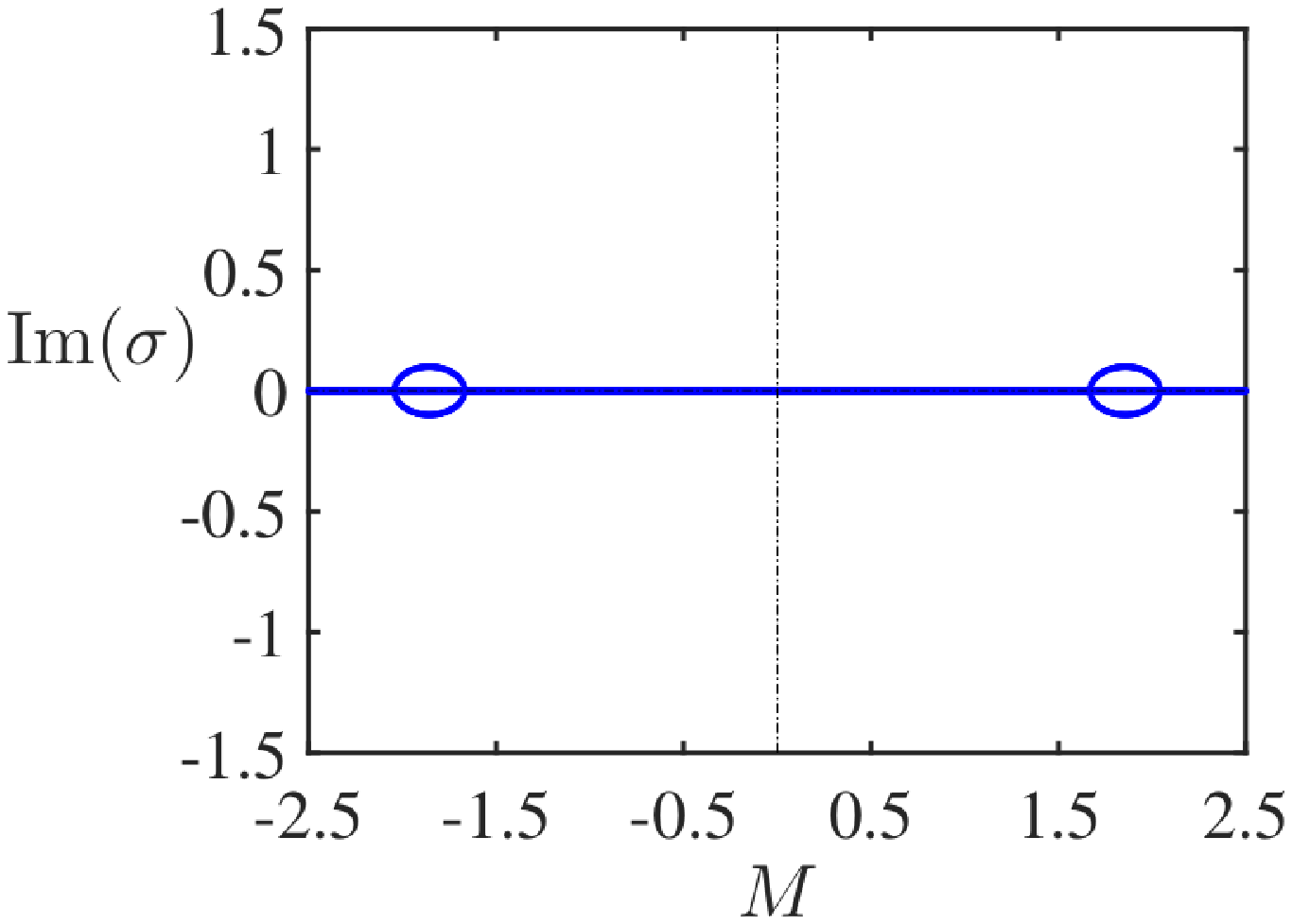}
    \caption{} \label{fig:2g}
  \end{subfigure}%
  \begin{subfigure}{0.25\textwidth}
    \includegraphics[width=\textwidth]{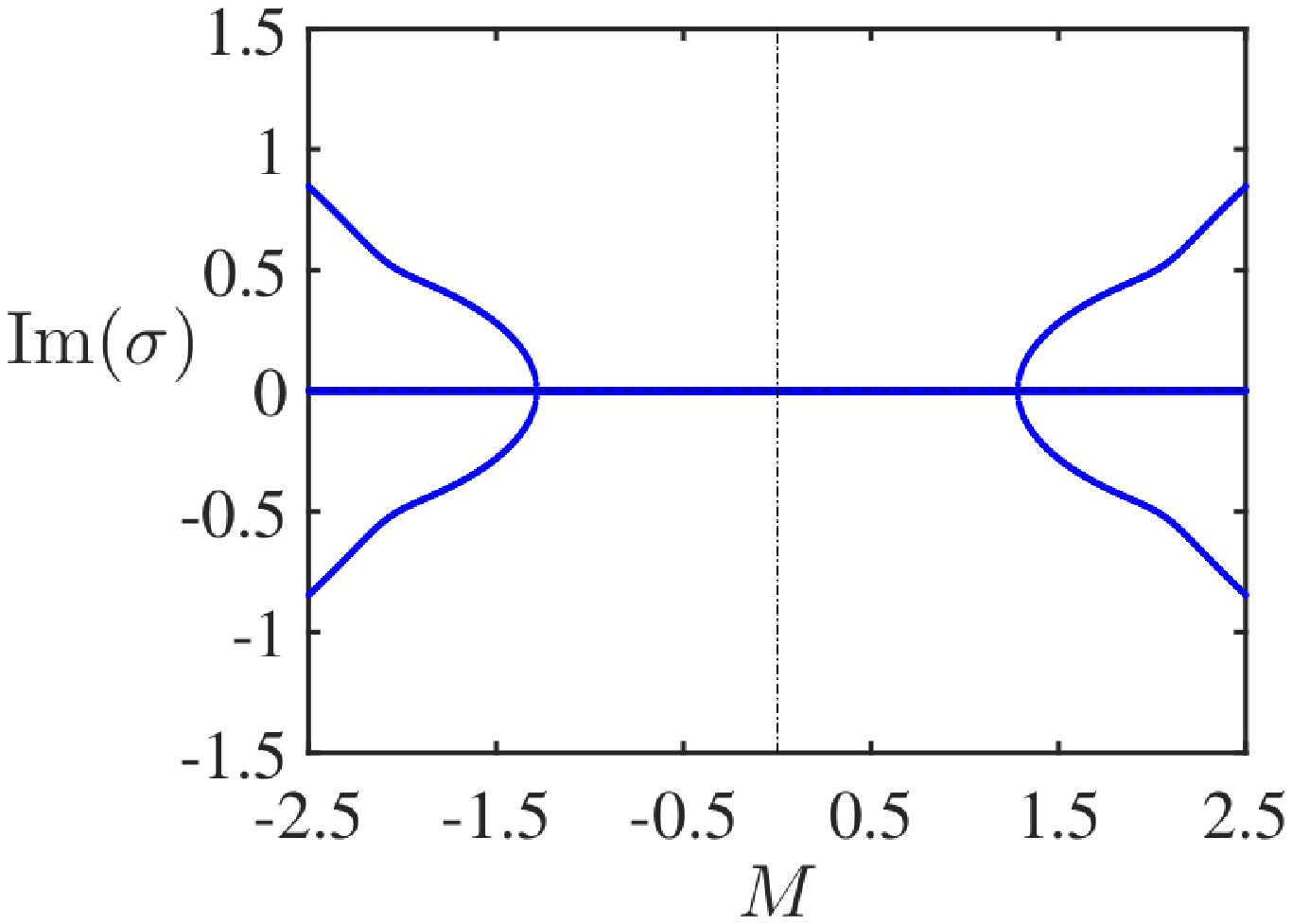}
    \caption{} \label{fig:2h}
  \end{subfigure}%

\caption{Real (red, upper panels) and imaginary (blue, lower panels) parts of the roots of the dispersion relation \rf{eq:DR_g} over the Mach number $M$ for $M_w=1$, $\kappa =1$ and: (a, e) $\beta=0$, (b, f) $\beta=0.01$, (c, g) $\beta=0.1$, and (d, h) $\beta=1$. \label{fig:sigma_k_b}}
\end{figure*}

Assuming that $\partial_x \phi$ is also absolutely integrable, which allows us to use twice the property
$\widehat{\partial_x \phi}=i\kappa \hat\phi$, we find the Fourier transform of the Laplace equation \rf{eoma}
\be{ftle}
\partial_z^2\hat{\phi} - \kappa^2\hat{\phi} = 0.
\ee
The general solution to equation \rf{ftle} is
\be{sftle}
\hat{\phi}(\kappa,z,\omega) = A(\kappa,\omega)e^{\kappa z} + B(\kappa,\omega)e^{-\kappa z},
\ee
where $A(\kappa,\omega)$ and $B(\kappa,\omega)$ are yet to be determined from the Fourier-transformed boundary conditions.

The boundary condition \rf{eomc}, expressing the impermeability of the membrane at $z=0$,
takes the form
\be{eomcm}
-\partial_z \phi=V,
\ee
because the outward direction of the normal vector $\bm{n}$ to the surface of the membrane is opposite to the positive $z$-direction, see Fig.~\ref{fig:sketch}.
The Fourier transform of \rf{eomcm} reads
\be{ftim}
\partial_z\hat{\phi} = -\hat{V}(\kappa,\omega),
\ee
where
\ba{vinf}
\hat{V}(\kappa,\omega) &=& \int_{-\infty}^{+\infty} \left(i\omega\xi(s) - M\partial_{s}\xi(s)\right) e^{-i\kappa s} ds\nn\\
&=&i\left( \omega - \kappa M\right)\hat \xi.
\ea
Substituting \rf{sftle} into \rf{ftim} yields at $z=0$
\be{bc1}
\kappa \left( A - B\right) = -i\left( \omega - \kappa M\right)\hat\xi.
\ee

Similarly transforming the boundary condition \rf{eomb} at the free surface
we find
\be{ftfs}
\partial_z\hat \phi = \left( \omega -\kappa M \right)^2\hat\phi.
\ee
Substituting \rf{sftle} into \rf{ftfs} yields at $z=1$
\be{bc2}
\kappa \left( Ae^{\kappa} - Be^{-\kappa}\right) = \left( \omega - \kappa M\right)^2\left( Ae^{\kappa} + Be^{-\kappa}\right).
\ee
Solving equations \rf{bc1} and \rf{bc2} simultaneously with respect to $A$ and $B$, we obtain
\ba{ab}
A(\kappa,\omega) &=& \frac{-i\hat \xi[\left(\omega - \kappa M\right)^2 + \kappa ]\left(\omega - \kappa M\right)}{\kappa[\left(\omega - \kappa M\right)^2 - \kappa]e^{2\kappa} + \kappa[\left(\omega - \kappa M\right)^2 + \kappa]}, \nn \\
B(\kappa,\omega) &=& \frac{i\hat \xi[\left(\omega - \kappa M\right)^2 - \kappa]\left(\omega - \kappa M\right)}{\kappa[\left(\omega - \kappa M\right)^2 - \kappa] + \kappa [\left(\omega - \kappa M\right)^2 + \kappa]e^{-2\kappa}}.
\ea

The Fourier transform of the non-homogeneous wave equation \rf{eome} for the membrane displacement evaluated at $z=0$ reads
\be{ftnwe}
\left( \omega^2 - \kappa^2 M_w^2 \right)\hat \xi - i\alpha\left(\omega - \kappa M\right)\hat \phi(\kappa,0,\omega) = 0.
\ee

Inserting expression \rf{sftle} for $\hat\phi$ with the coefficients \rf{ab} into \rf{ftnwe}, discarding $\hat \xi$ in the result and introducing new parameters, namely the phase velocity
\be{phasev}
\sigma=\frac{\omega}{\kappa}
\ee
and the coupling parameter
\be{coup}
\beta=\frac{\alpha}{\kappa^2},
\ee
we obtain the following dispersion equation in the case where a medium with constant pressure is present below the membrane
\be{eq:DR_g}
\beta = \frac{(M_w^2 - \sigma^2)\left[\kappa (\sigma -M)^2 - \tanh{\kappa}\right]}{\kappa (\sigma - M)^2\left[\kappa (\sigma - M)^2\tanh{\kappa} - 1\right]}.
\ee

It is instructive to show another way of deriving the dispersion equation \rf{eq:DR_g}. For this, we notice that
\rf{ftim} and \rf{vinf} allow us to express $\hat \xi$ by means of $\partial_z \hat \phi$. Using the result in \rf{ftnwe},
we can obtain a boundary condition for $\hat \phi (z)$ at $z=0$. This new boundary condition together with boundary condition \rf{ftfs}
and equation \rf{ftle} produce a closed-form boundary value problem for the Laplace equation with the Robin boundary conditions:
\ba{rbvp}
\partial_z^2 \hat \phi-\kappa^2 \hat \phi &=&0,\nn\\
\partial_z \hat \phi \left( \omega^2 - \kappa^2 M_w^2 \right) - \alpha \left(\omega - \kappa M\right)^2\hat \phi &=& 0, \quad z=0,\nn\\
\partial_z \hat \phi-(\omega - \kappa M)^2 \hat \phi&=&0, \quad z=1.
\ea
Substituting the general solution \rf{sftle} into the boundary conditions of the problem \rf{rbvp} results in the system of two linear equations with respect to $A$ and $B$,
\ba{abeq}
\kappa(A-B)\left( \omega^2 - \kappa^2 M_w^2 \right) - \alpha \left(\omega - \kappa M\right)^2(A+B) &=& 0,\nn\\
\kappa(Ae^{\kappa} -Be^{-\kappa} )-(\omega - \kappa M)^2 (Ae^{\kappa}+Be^{-\kappa}) &= &0.
\ea
This system can be written in matrix form as
\be{matrixf}
(\omega^2 M_1 + \omega M_2 + M_3)\bm{f} = 0,\quad \bm{f}:=\left(
                                             \begin{array}{c}
                                               A \\
                                               B \\
                                             \end{array}
                                           \right),
\ee
where the $2 \times 2$ matrices involved are
\ba{M123}
M_1 &=&
-\left(
  \begin{array}{cc}
    \alpha-\kappa & \alpha+\kappa \\
    e^{\kappa} & e^{-\kappa} \\
  \end{array}
\right),\nn\\
M_2&=&2\kappa M\left(
  \begin{array}{cc}
    \alpha & \alpha \\
    e^{\kappa} & e^{-\kappa} \\
  \end{array}
\right),\nn\\
M_3&=&-\left(
      \begin{array}{cc}
        \kappa^2(M^2\alpha+M_w^2\kappa) & \kappa^2(M^2\alpha-M_w^2\kappa) \\
        \kappa e^{\kappa}(M^2\kappa-1) & \kappa e^{-\kappa}(M^2\kappa+1) \\
      \end{array}
    \right).
\ea
Computing the determinant of the matrix polynomial we arrive at the dispersion equation
\ba{domba}
D(\omega,\kappa)&=&\det(\omega^2 M_1 + \omega M_2 + M_3)\nn\\
&=&-\alpha(M\kappa-\omega)^2[(M\kappa-\omega)^2\tanh \kappa-\kappa]
+ \kappa (M_w^2\kappa^2-\omega^2) [(M \kappa-\omega)^2 - \kappa \tanh \kappa]\nn\\&=&0,
\ea
which, with the notation $\sigma=\omega/\kappa$ and $\beta=\alpha/\kappa^2$, transforms exactly to \rf{eq:DR_g}.

For the sake of completeness we present also the dispersion relation for the system with vacuum below the membrane
\begin{equation}
\label{eq:DR_v}
\beta = \frac{(M_w^2 - \sigma^2)\left[\kappa (\sigma -M)^2 - \tanh{\kappa}\right]}{\left[\kappa^2 (\sigma - M)^4 - 1\right]\tanh{\kappa}}.
\end{equation}

In the shallow water approximation corresponding to the limit $\kappa \rightarrow 0$, the expression \rf{eq:DR_v} reduces to
\be{nesw}
\beta = (\sigma^2-M_w^2)\left( (\sigma -M)^2 - 1\right),
\ee
which is nothing else but the shallow water dispersion relation derived by \cite{N1986}.

In order to get the dispersion relation \rf{eq:DR_v}, one must take the pressure difference \rf{nemp}, make it nondimensional and use in the expression \rf{dlwe} which then reads as
$$
\partial^2_{\tau}\xi-M_w^2\partial^2_{x}\xi- \alpha \xi=\alpha \left(\partial_{\tau} + M\partial_x\right)\phi.
$$
After separation of time it reduces to the analogue of boundary condition \rf{eome},
$$
\left[-\omega^2 - M_w^2\partial_x^2 - \alpha \right]\xi - \alpha\left(-i\omega + M\partial_x\right)\phi(x,0,t) = 0,
$$
which has the following Fourier transform
\be{ftneme}
\left[\omega^2 -\kappa^2 M_w^2 + \alpha \right]\hat\xi -i \alpha\left(\omega -\kappa M\right)\hat\phi(\kappa,0,\omega) = 0.
\ee
Inserting the expression \rf{sftle} for $\hat\phi$ with the coefficients \rf{ab} into \rf{ftneme} results, after familiar algebraic manipulations, in the dispersion relation \rf{eq:DR_v}.


\begin{figure}
  \begin{subfigure}{0.45\textwidth}
    \includegraphics[width=\textwidth]{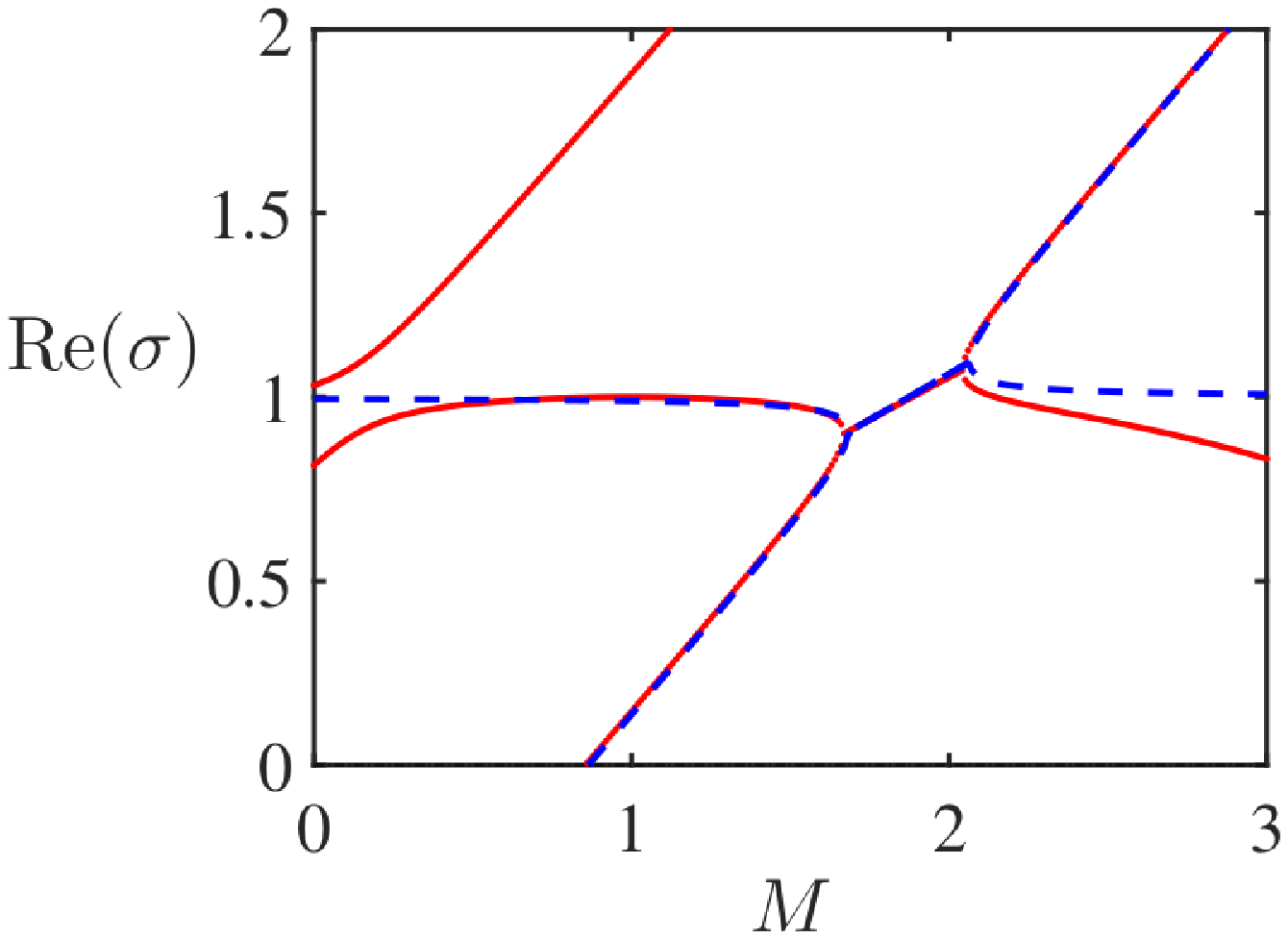}
    \caption{} \label{fig:3a}
  \end{subfigure}%
  \begin{subfigure}{0.45\textwidth}
    \includegraphics[width=\textwidth]{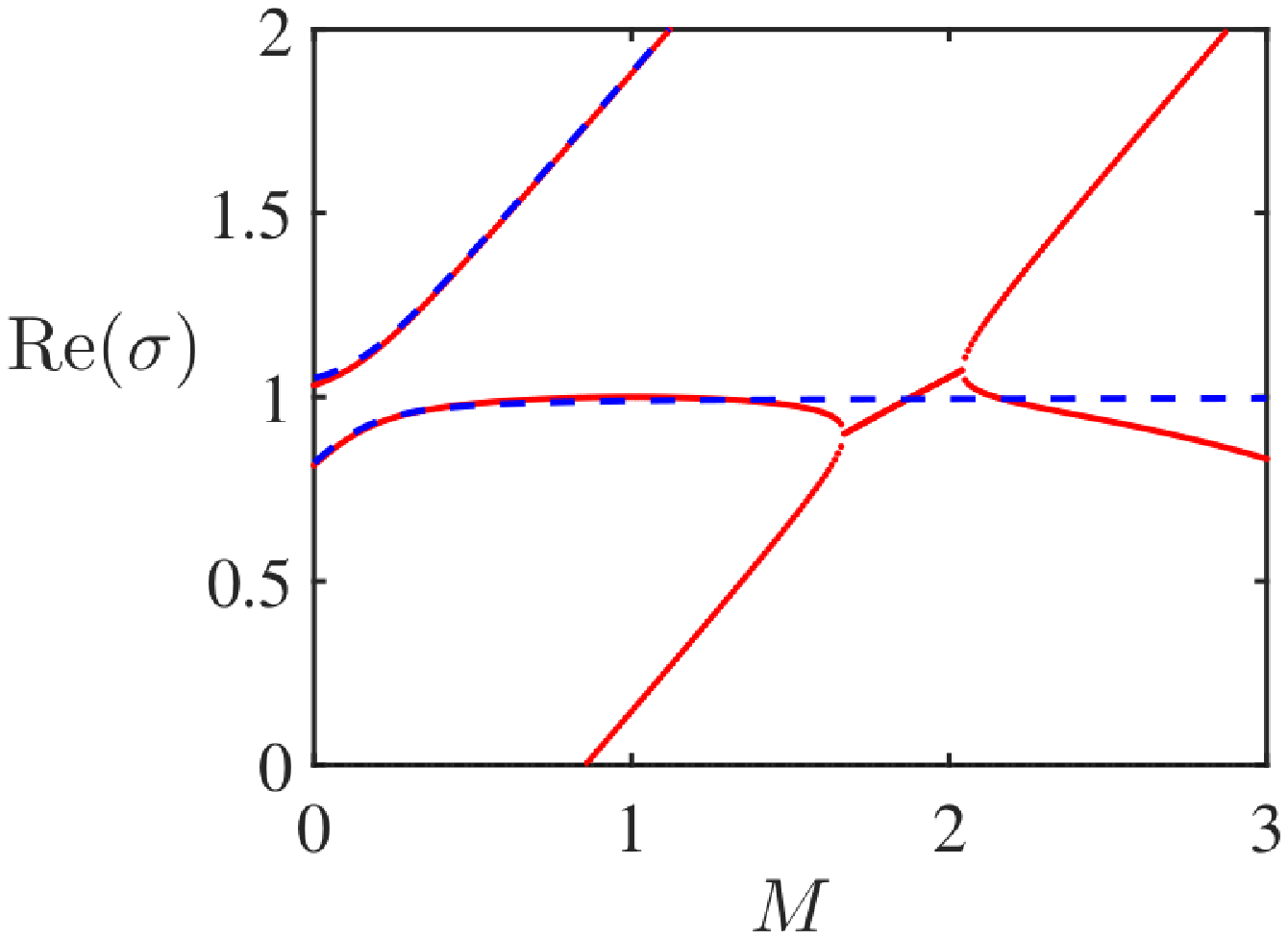}
    \caption{} \label{fig:3b}
  \end{subfigure}\\
  \begin{subfigure}{0.45\textwidth}
    \includegraphics[width=\textwidth]{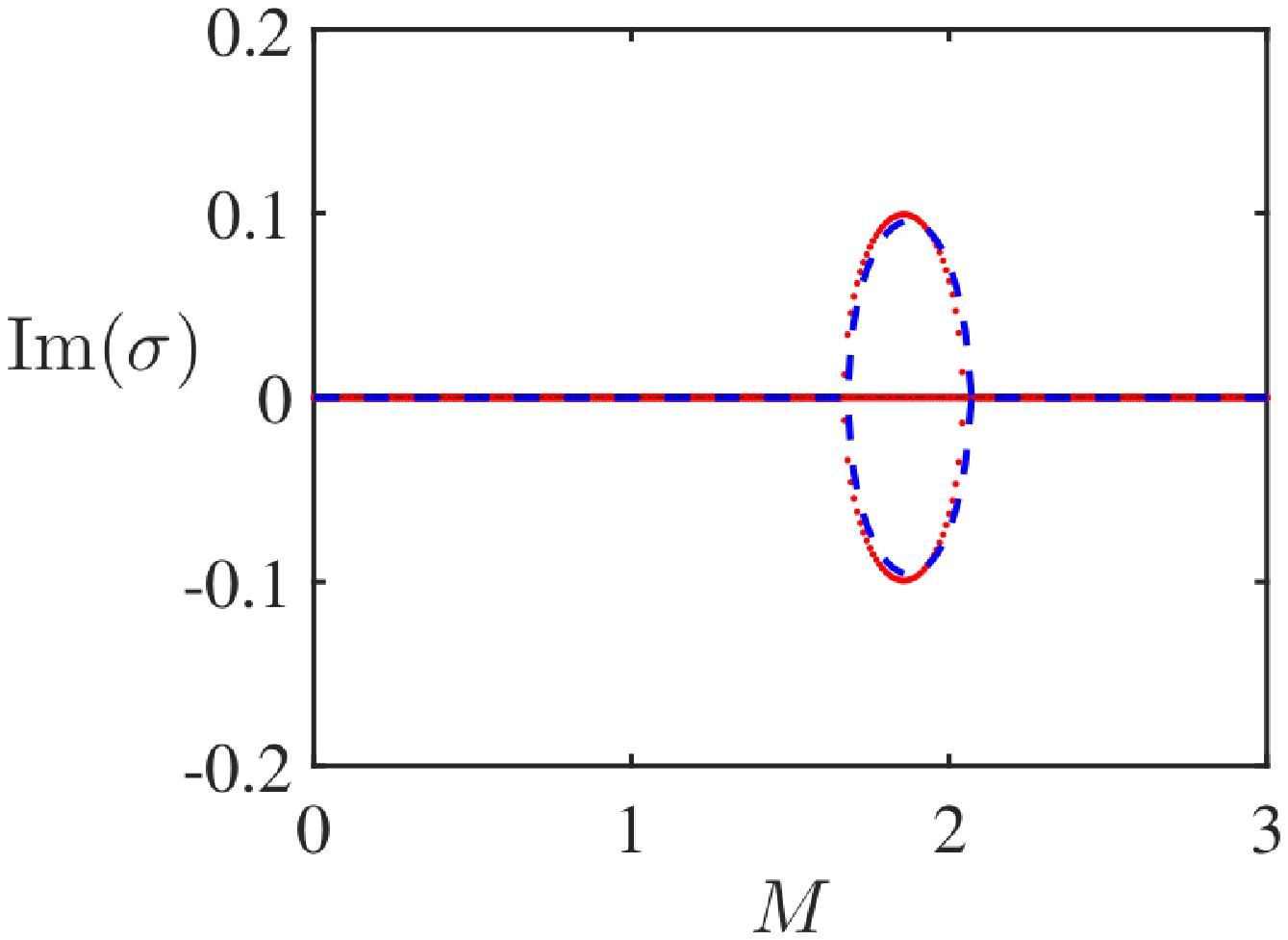}
    \caption{} \label{fig:3c}
  \end{subfigure}%
  \begin{subfigure}{0.45\textwidth}
    \includegraphics[width=\textwidth]{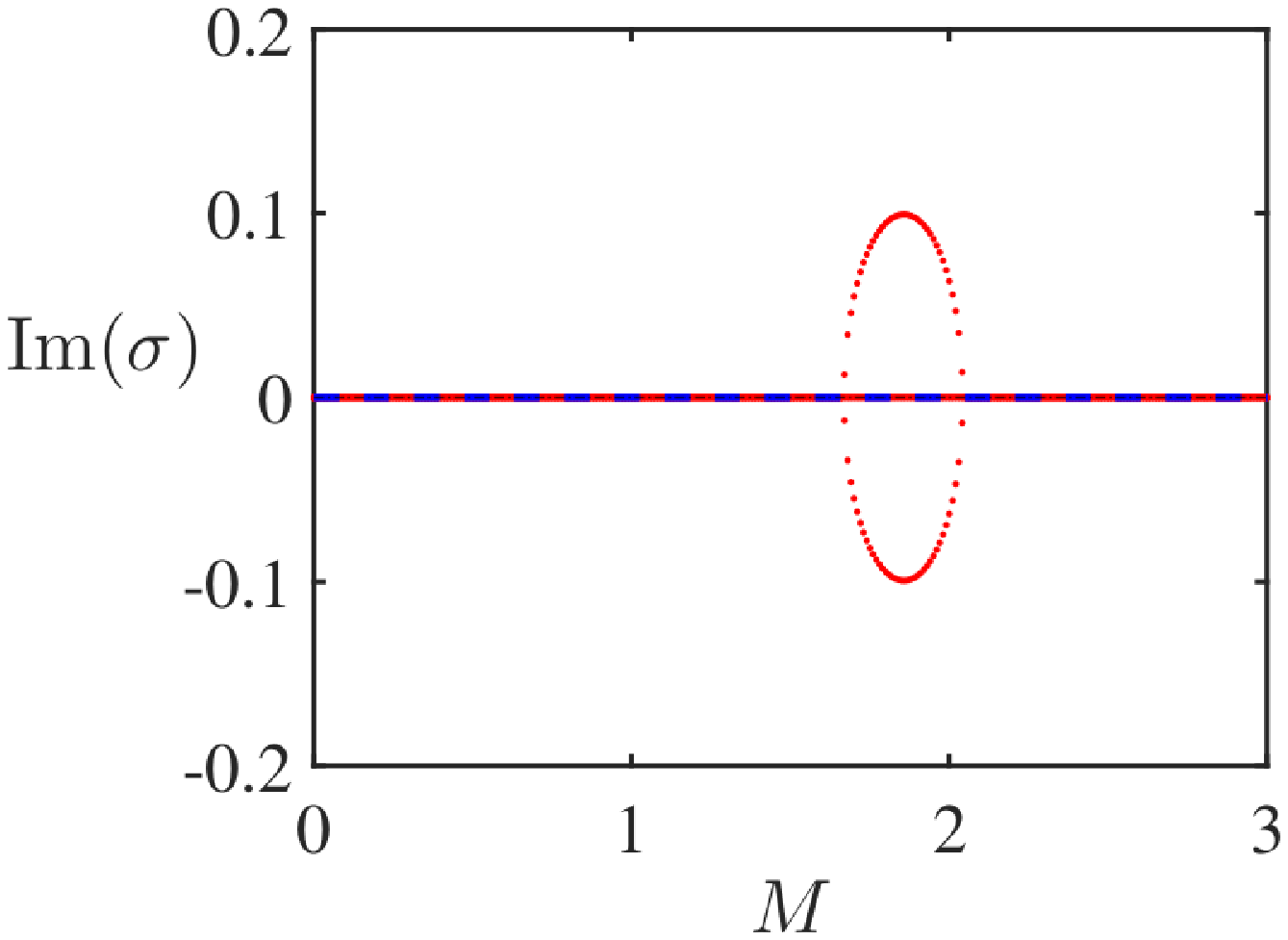}
    \caption{} \label{fig:3d}
  \end{subfigure}%

  \caption{Real and imaginary parts of the roots of the dispersion relation for $M_w=1$, $\kappa=1$, and $\beta=0.1$:  (red) \rf{eq:DR_g} and (blue, dashed) their approximations by  equations \rf{qede3} and \rf{qede3_2} near the crossing points that exist at $\beta=0$, $M=M_0^{\pm}$, $\sigma=\sigma_0$. Notice an avoided crossing above the line ${\rm Re}(\sigma)=M$ and the bubble of instability below this line. \label{fig:approxdr}}
\end{figure}


\begin{figure}
  \begin{subfigure}{0.45\textwidth}
    \includegraphics[width=\textwidth]{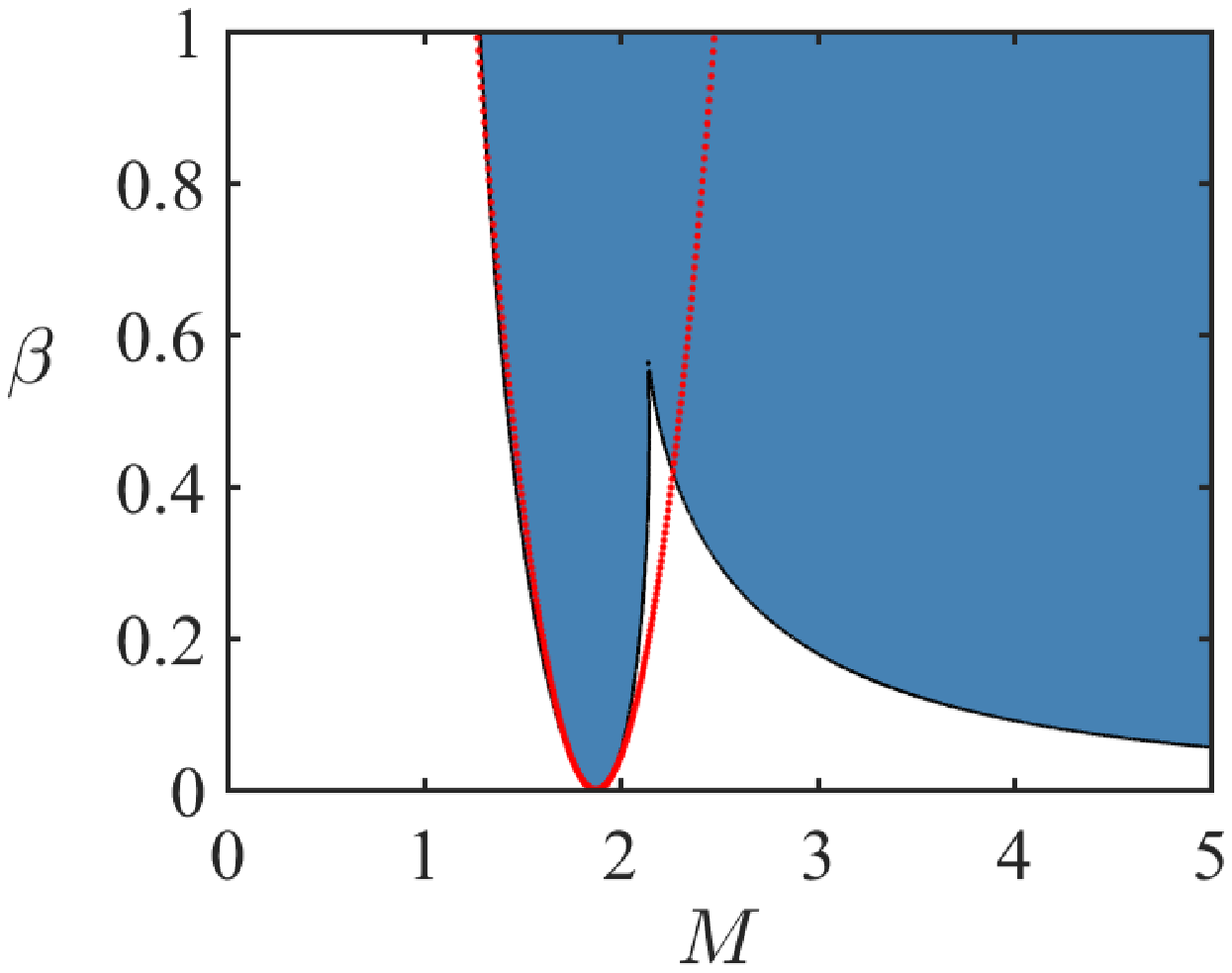}
    \caption{} \label{fig:4a}
  \end{subfigure}%
  \begin{subfigure}{0.45\textwidth}
    \includegraphics[width=\textwidth]{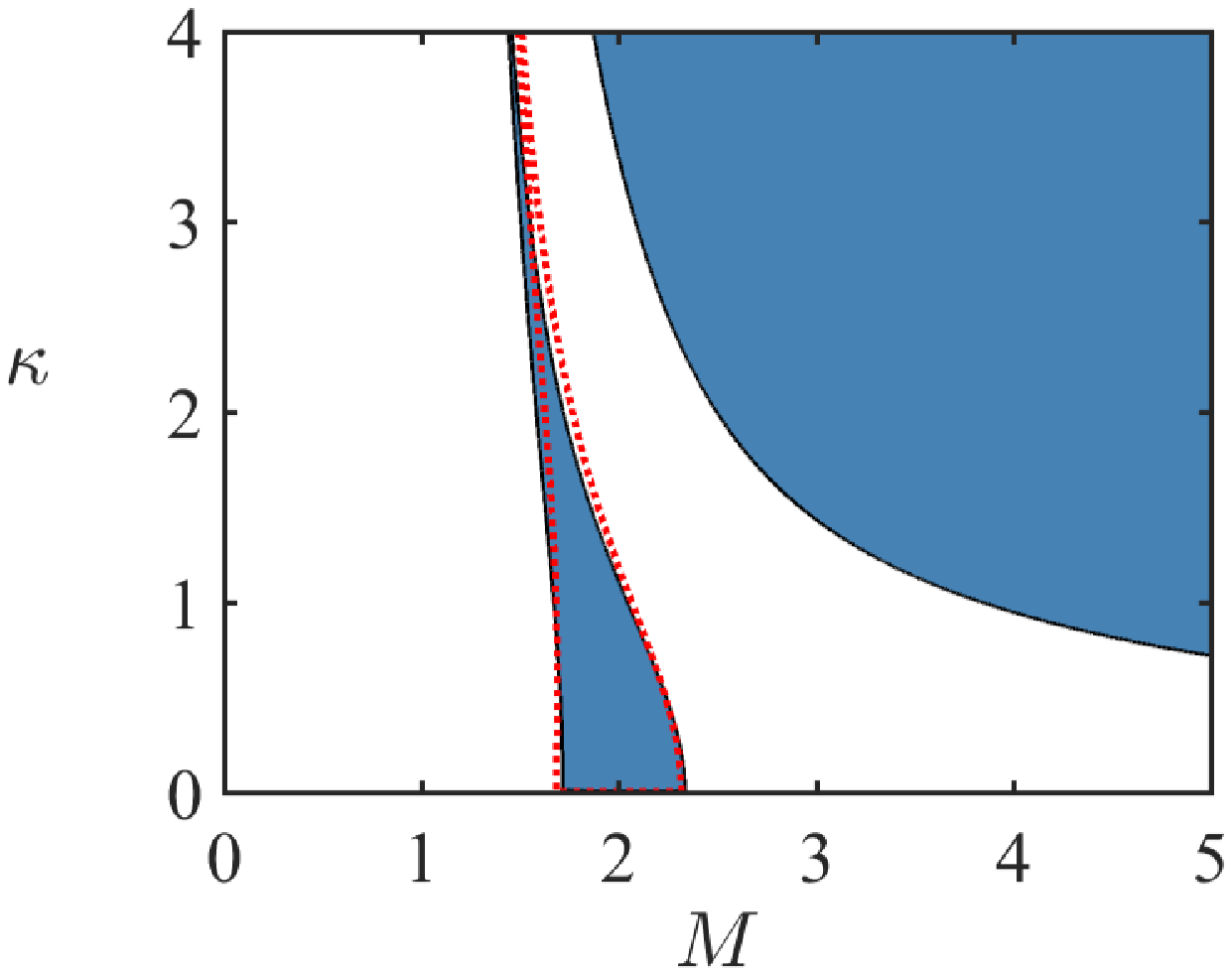}
    \caption{} \label{fig:4b}
  \end{subfigure}\\
  \begin{subfigure}{0.45\textwidth}
    \includegraphics[width=\textwidth]{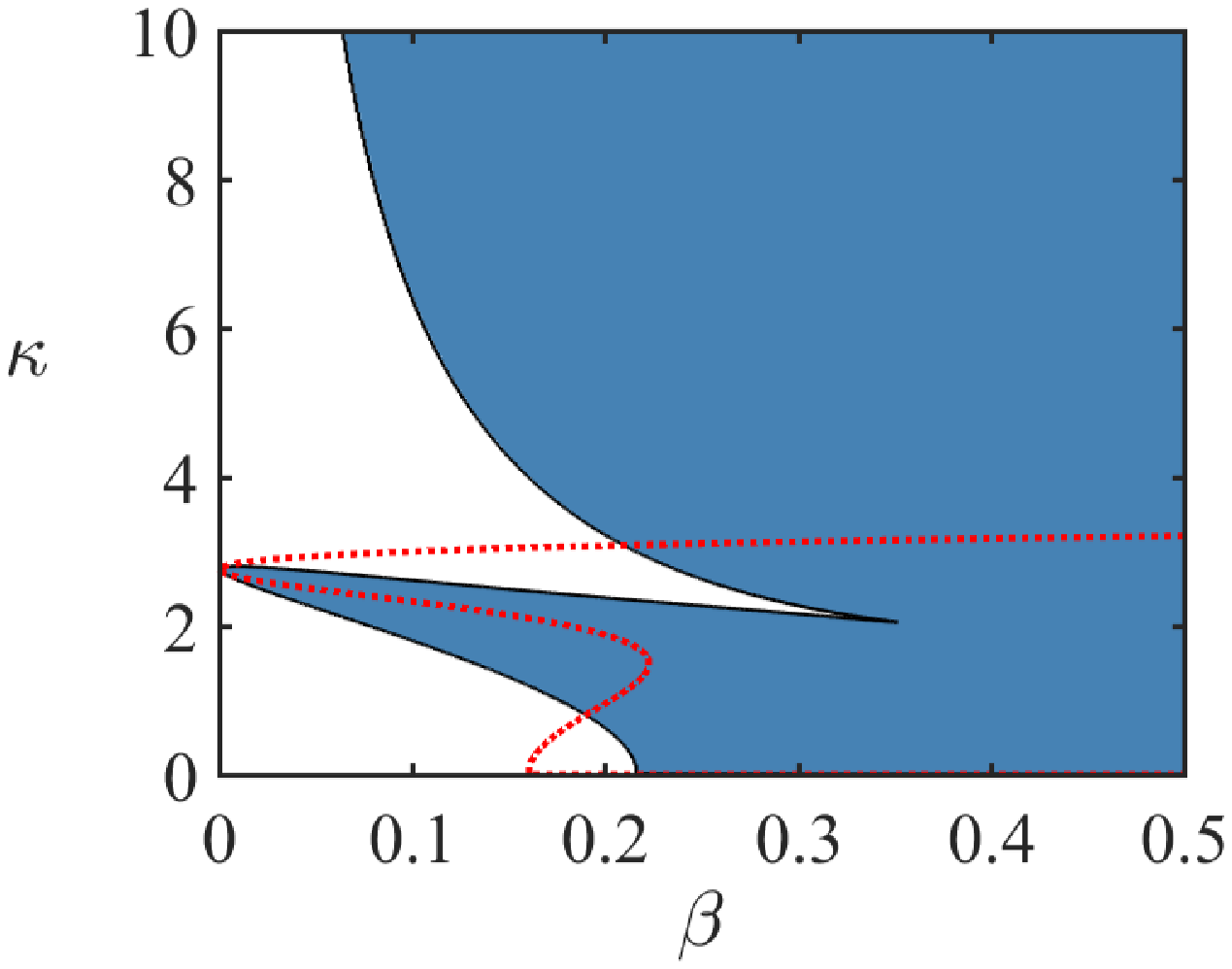}
    \caption{} \label{fig:4c}
  \end{subfigure}%
  \begin{subfigure}{0.45\textwidth}
    \includegraphics[width=\textwidth]{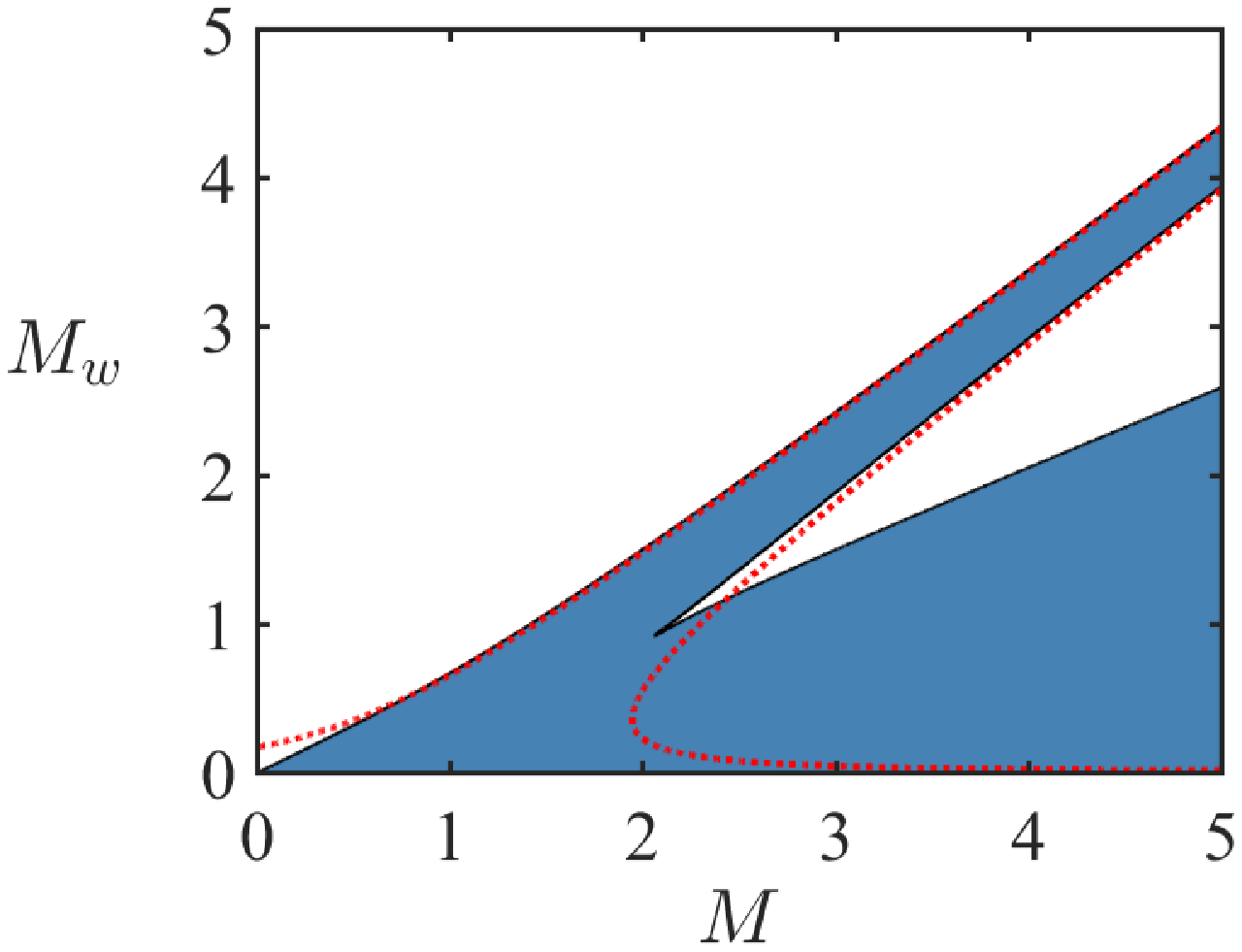}
    \caption{} \label{fig:4d}
  \end{subfigure}%

\caption{Stability maps of the dispersion equation \rf{domb} given by its discriminant for (a) $M_w=1$ and $\kappa=1$, (b) $M_w=1$ and $\beta=0.1$, (c) $M_w=1$ and $M=1.6$, and (d) $\beta=0.5$ and $\kappa=1$.  The regions of real phase speed $\sigma$ are shown in white (stability) and that of the complex $\sigma$ (temporal instability) in blue. The red dotted curves is the approximation \rf{qans}. Notice the absence of instabilities for $M_w>M$ in panel (d). \label{fig:smap}}
\end{figure}

\subsubsection{Analysis of the dispersion equation}

In the  absence of coupling between the free surface and the membrane, i.e. for $\beta=0$, both the dispersion relation \rf{eq:DR_g} and
the dispersion relation \rf{eq:DR_v} reduce to
\be{b0}
(\sigma^2-M_w^2)\left[\kappa (\sigma -M)^2 - \tanh{\kappa}\right]=0,
\ee
which yields the dispersion relation of the elastic waves in the free membrane $\sigma^2=M_w^2$ and that of the surface gravity
waves on a uniform flow:
$
\kappa (\sigma -M)^2 = \tanh{\kappa}
$.
The latter acquires a more familiar traditional form \citep{MRS2016} $$(\omega-\kappa Fr)^2=\kappa \tanh{\kappa}$$
after taking into account that $\sigma=\omega/\kappa$ and that $M$, as defined in \rf{dlp4}, can also be interpreted as the Froude number, $Fr$.

The roots of the decoupled dispersion equation \rf{b0} are real:
\be{rb0}
\sigma_1^{\pm}=\pm M_w, \quad
\sigma_2^{\pm}=M\pm \sqrt{\frac{\tanh{\kappa}}{\kappa}}.
\ee

If we consider the roots \rf{rb0} as functions of the fluid Mach number, $M$,
we find that $\sigma_{1}^{\pm}$ are two horizontal straight lines and $\sigma_{2}^{\pm}$
are two straight lines with the slope equal to $1$, see panel (a) in Fig.~\ref{fig:sigma_k_b}.
One can see that at $\beta=0$ the root branches intersect at four points
forming the double roots $\sigma_0=M_w$ at
\be{phasy}
M_0^{\pm} = M_w\pm\sqrt{\frac{\tanh \kappa}{\kappa}}
\ee
and the double roots $-\sigma_0$ at $-M_0^{\pm}$. The relation $M_w=M_0^+-\sqrt{(\tanh \kappa)/\kappa}=\sigma_2^-=\sigma_1^+=\sigma_0$ following from \rf{phasy} and \rf{rb0} is the condition of `phase synchronism' for the case of arbitrary height of the fluid layer that extends the corresponding result obtained in \cite{N1986} in the shallow water limit, $\kappa \rightarrow 0$.

With the increase in $\beta$ the roots $\pm \sigma_0$ situated at $M=\pm M_0^-$ split into simple real ones
and this splitting is accompanied by unfolding the crossings into avoided crossings, Fig.~\ref{fig:sigma_k_b}.

Quite in contrast, the roots $\pm \sigma_0$ situated at $M=\pm M_0^+$ split into complex-conjugate pairs that form
bubbles of instability at moderate values of $\beta$ that open up with the increase in $\beta$ to develop disconnected complex branches, as is seen in the panels (d) and (h) of
Fig.~\ref{fig:sigma_k_b}

Let us re-write the dispersion relation \rf{eq:DR_g} as follows
\be{domb}
D(\sigma,M,\beta):=\beta\kappa (\sigma - M)^2\left[\kappa (\sigma - M)^2\tanh{\kappa} - 1\right] -
{(M_w^2 - \sigma^2)\left[\kappa (\sigma -M)^2 - \tanh{\kappa}\right]}=0.
\ee
Then, we can apply to it the perturbation theory derived in the Appendix~\ref{sadr}.


\begin{figure}
  \begin{subfigure}{0.33\textwidth}
    \includegraphics[width=\textwidth]{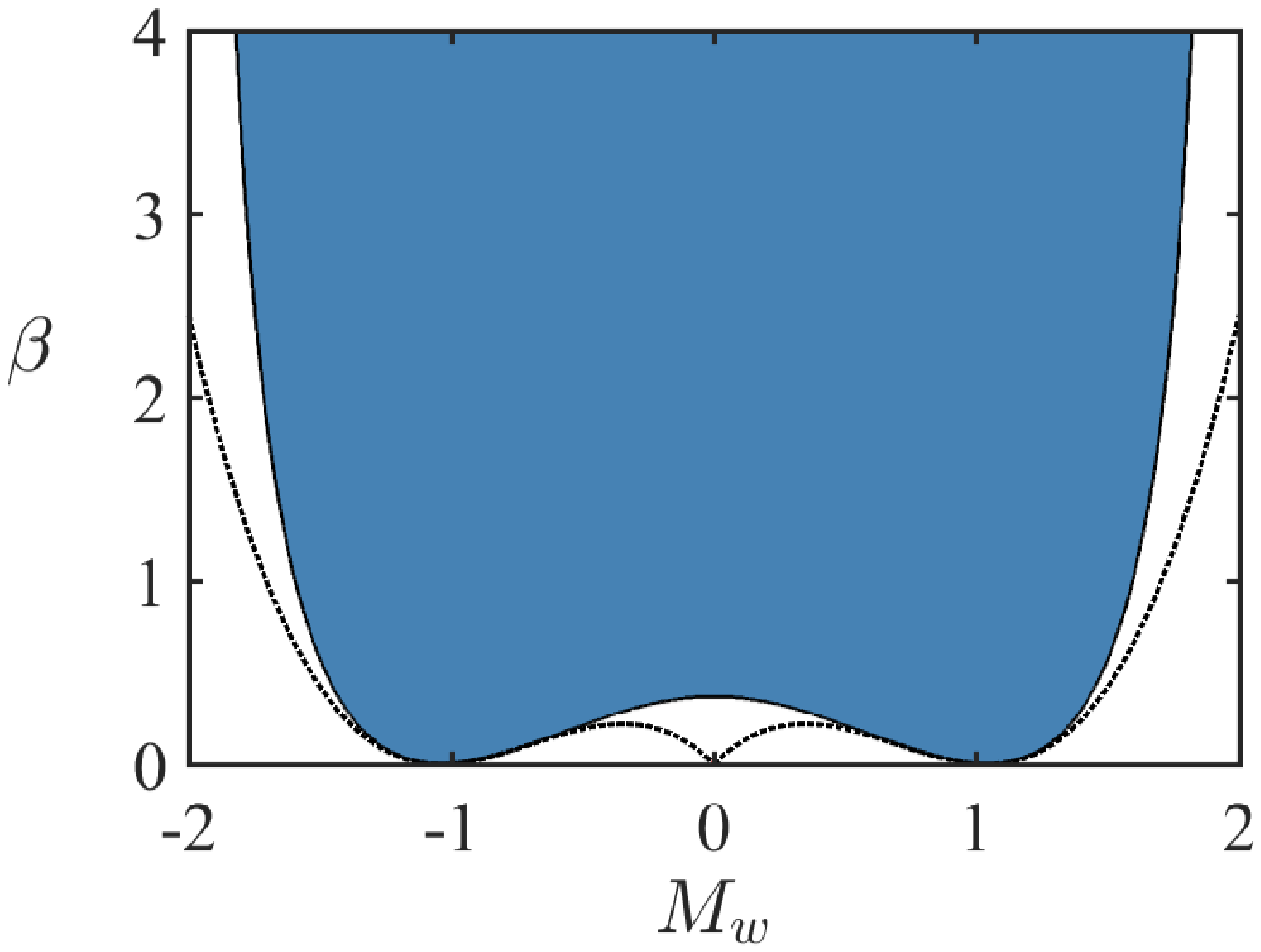}
    \caption{} \label{fig:5a}
  \end{subfigure}%
  \begin{subfigure}{0.33\textwidth}
    \includegraphics[width=\textwidth]{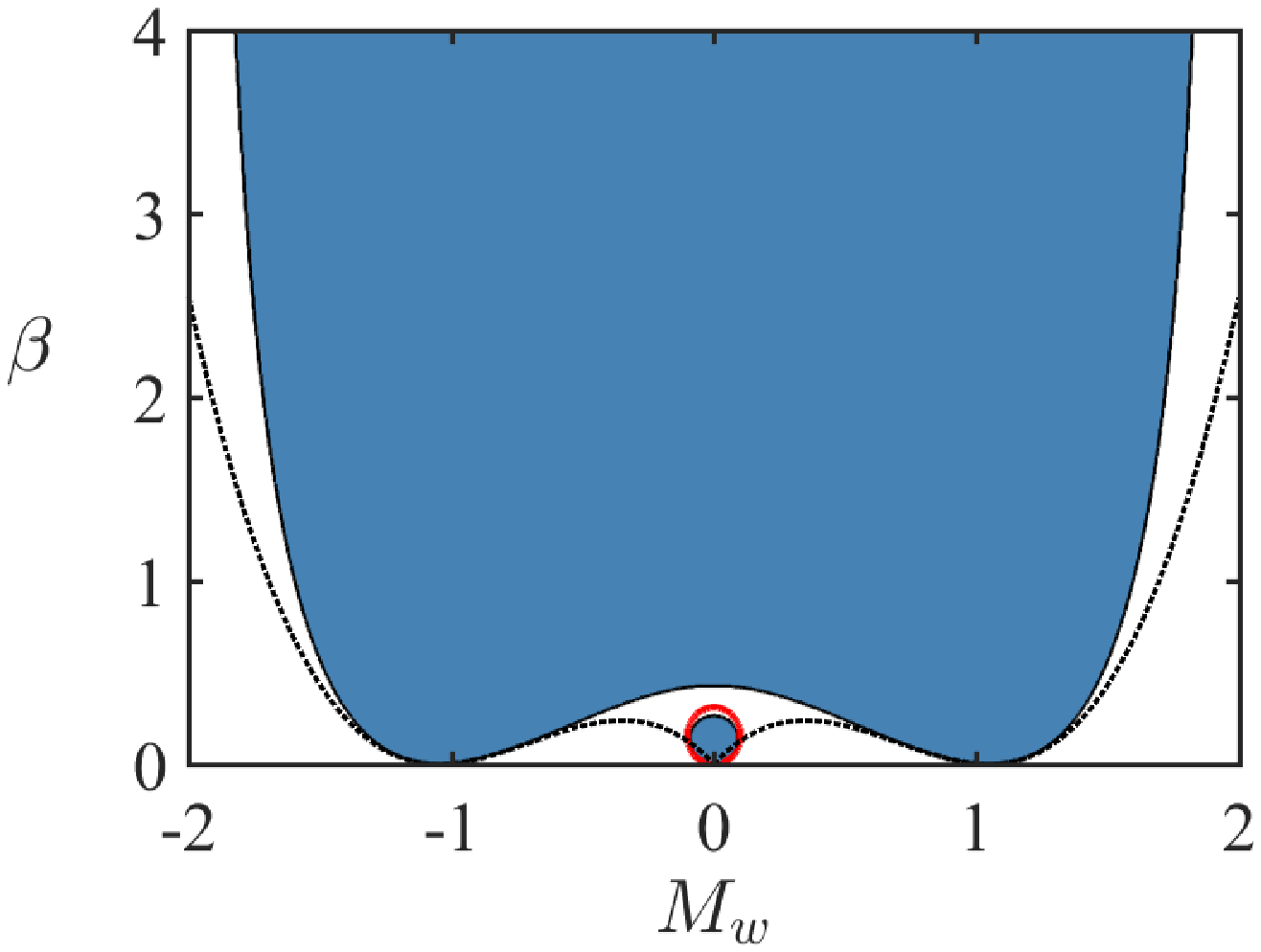}
    \caption{} \label{fig:5b}
  \end{subfigure}%
  \begin{subfigure}{0.33\textwidth}
    \includegraphics[width=\textwidth]{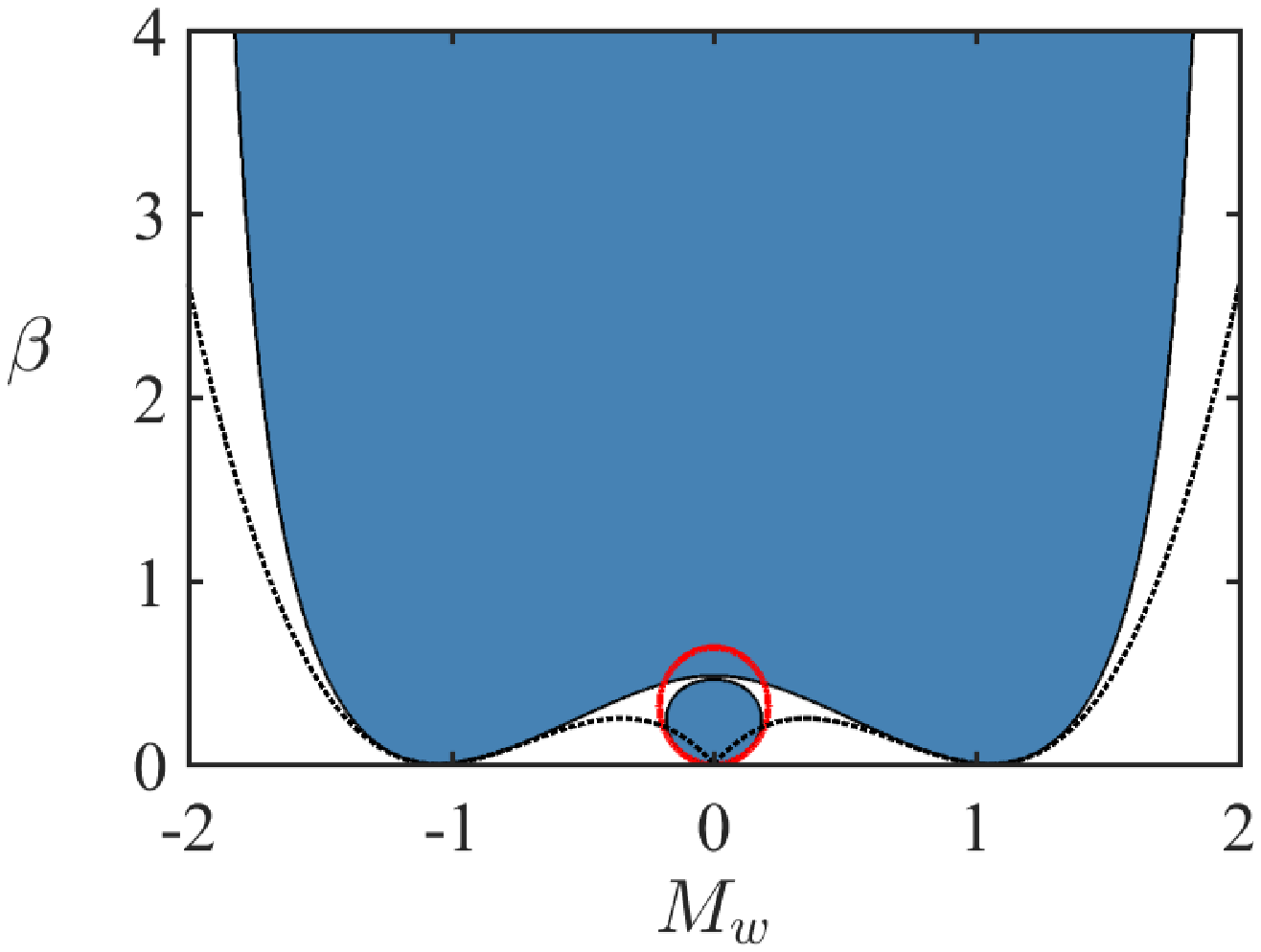}
    \caption{} \label{fig:5c}
  \end{subfigure}\\
  \begin{subfigure}{0.33\textwidth}
    \includegraphics[width=\textwidth]{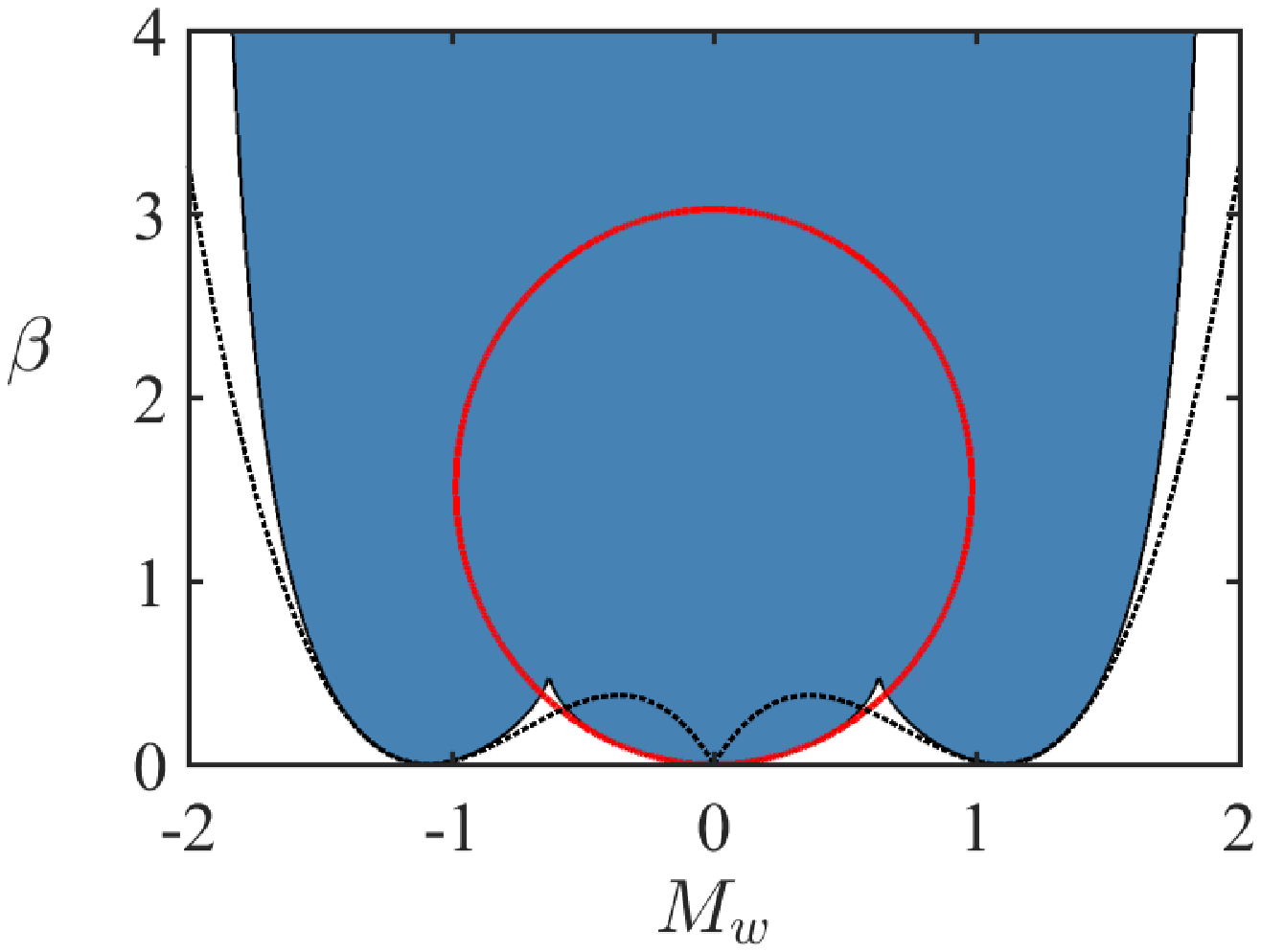}
    \caption{} \label{fig:5d}
  \end{subfigure}%
  \begin{subfigure}{0.33\textwidth}
    \includegraphics[width=\textwidth]{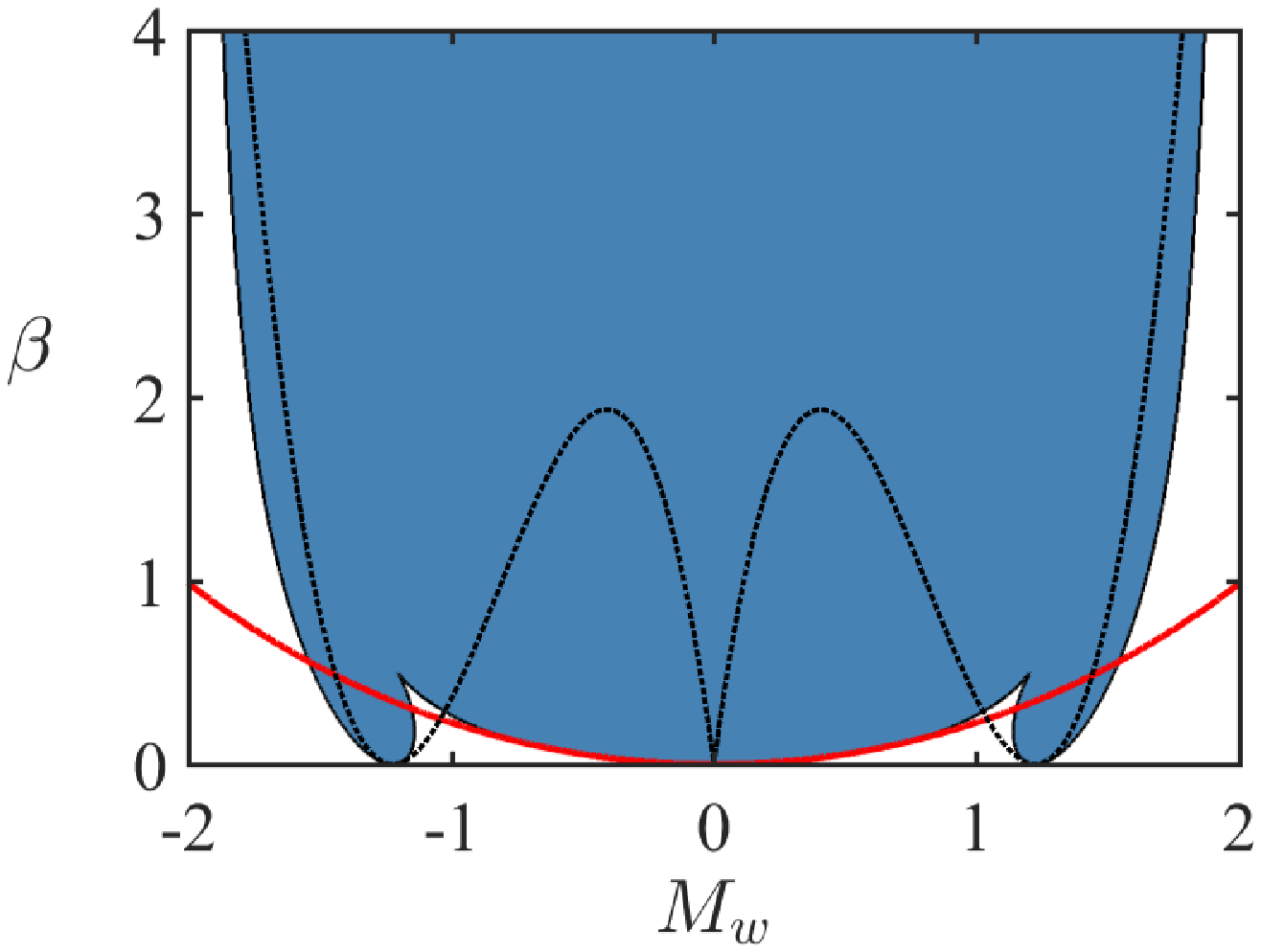}
    \caption{} \label{fig:5e}
  \end{subfigure}%
  \begin{subfigure}{0.33\textwidth}
    \includegraphics[width=\textwidth]{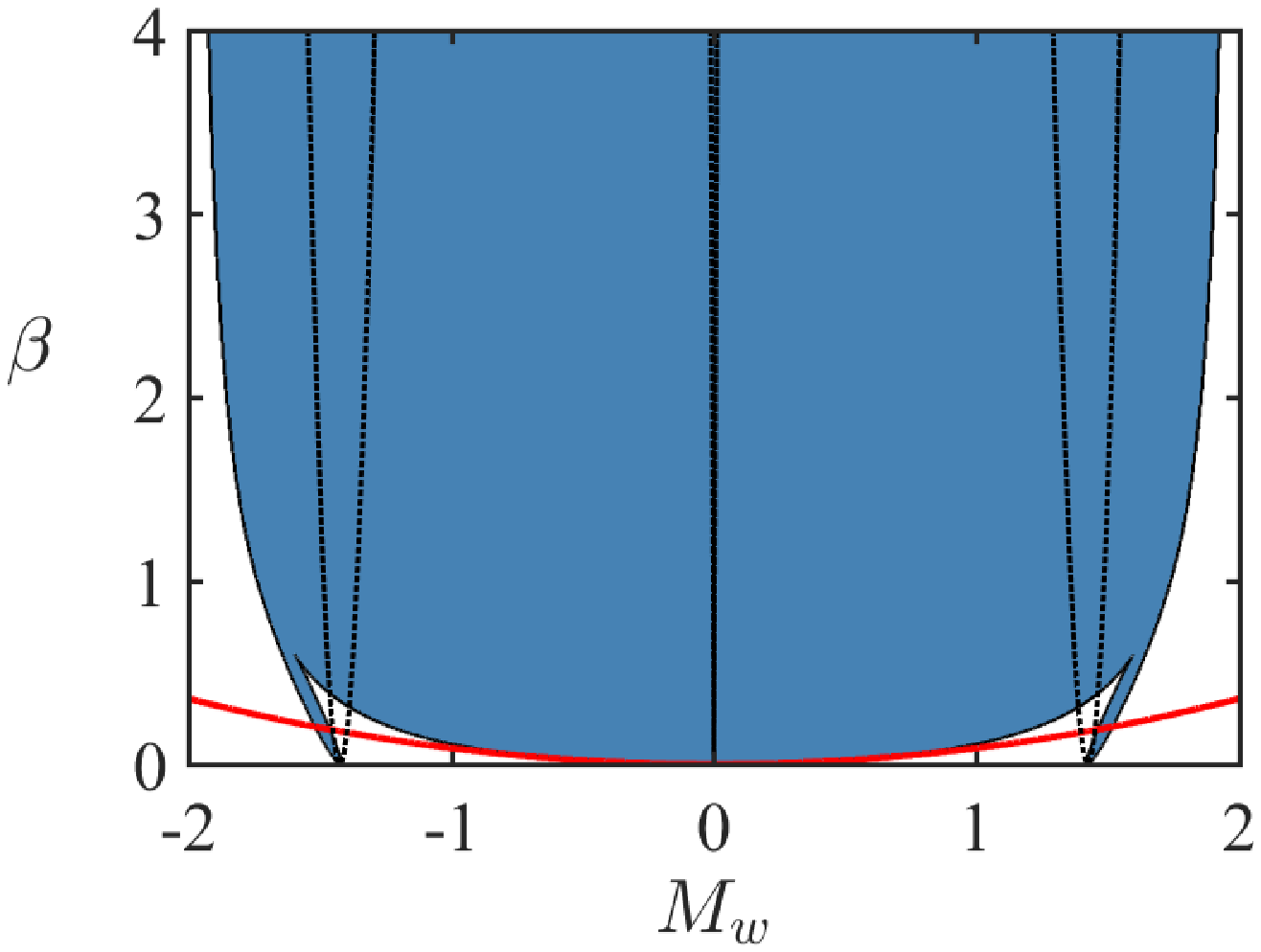}
    \caption{} \label{fig:5f}
  \end{subfigure}%

\caption{Stability maps of the dispersion equation \rf{domb} given by its discriminant for $M=M_0=2$ and: (a) $\kappa=0.5$, (b) $\kappa=0.55$, (c) $\kappa=0.58$, (d) $\kappa=0.8$, (e) $\kappa=1.5$, (f) $\kappa=3$. The regions of real phase speed $\sigma$ are shown in white (stability) and that of the complex $\sigma$ (temporal instability) in blue. The black dotted curves is the approximation \rf{qans} and the solid red ellipse is the conical approximation \rf{conapprox}. When $\kappa \rightarrow \infty$, the central part of the instability domain approximated by \rf{conapprox} dominates over the side parts of the domain. Notice the absence of instabilities for $M_w>M_0$. \label{fig:cone2}}
\end{figure}

Consider the double root $\sigma_0$ at $M= M_0^+$ and $\beta=\beta_0=0$. Adapt the approximate equation \rf{qede}
to our model
\ba{qede2}
&\Delta\sigma(\partial^2_{\sigma M}D \Delta M+\partial^2_{\sigma \beta}D \Delta \beta) + \frac{1}{2}\left[ \partial_M^2 D (\Delta M)^2+2\partial^2_{M\beta} D \Delta M \Delta \beta+ \partial_{\beta}^2 D (\Delta \beta)^2\right]  \nn \\
&+ \frac{1}{2}\partial_{\sigma}^2 D(\Delta\sigma)^2 + \partial_M D \Delta M+\partial_{\beta} D \Delta \beta = 0,
\ea
where $\Delta \sigma=\sigma-\sigma_0$, $\Delta M=M-M_0^+$, and $\Delta \beta=\beta$. Calculating the partial derivatives
at $\sigma=\sigma_0$, $M= M_0^+$, and $\beta=\beta_0=0$, we find
\ba{derqede2}
&\partial_{\sigma}^2 D= -8M_w \kappa\sqrt{\frac{\tanh \kappa}{\kappa}}, \quad \partial_{\sigma M}^2 D= 4M_w\kappa\sqrt{\frac{\tanh \kappa}{\kappa}},&\nn\\
&\partial_{M}^2 D= 0, \quad \partial_{M\beta}^2 D=-\partial_{\sigma\beta}^2 D= 2\kappa(2(\tanh{\kappa})^2-1)\sqrt{\frac{\tanh \kappa}{\kappa}}, \quad \partial_{\beta}^2 D= 0,& \nn\\
&\partial_{M} D= 0, \quad \partial_{\beta} D= (\tanh \kappa)^3-\tanh \kappa.&
\ea
With the derivatives \rf{derqede2} the approximation \rf{qede2} to the dispersion equation \rf{domb} near the crossing takes the form
\be{qede3}
(\sigma-M_w)\left[\sigma-M+\sqrt{\frac{\tanh \kappa}{\kappa}}\right]=\beta\sqrt{\frac{\tanh \kappa}{\kappa}}\frac{(\tanh \kappa)^2-1}{4M_w}.
\ee
For any $\beta>0$ the crossing of the real roots $\sigma$ at $M= M_0^+$ unfolds into two hyperbolic branches of the real roots
\ba{}
\beta\sqrt{\frac{\tanh \kappa}{\kappa}}\frac{[1-(\tanh \kappa)^2]}{4M_w} &=& \frac{1}{4}\left(M-M_w-\sqrt{\frac{\tanh \kappa}{\kappa}}\right)^2 \nn \\
&-& \left( {\rm Re}\, \sigma-\frac{M_w+M}{2}+\frac{1}{2}\sqrt{\frac{\tanh \kappa}{\kappa}}\right)^2, \quad {\rm Im} \,\sigma = 0
\ea
that are connected to the ``bubble'' of complex eigenvalues with the real parts
${\rm Re}\, \sigma = \frac{1}{2}\left(M+M_w-\sqrt{(\tanh{\kappa})/\kappa}\right)$
and with the imaginary parts that form an ellipse in $(M, {\rm Im}\, \sigma)$-plane
\be{bubble}
({\rm Im}\, \sigma)^2+\frac{1}{4}\left(M-M_w-\sqrt{\frac{\tanh \kappa}{\kappa}}\right)^2=\beta\sqrt{\frac{\tanh \kappa}{\kappa}}\frac{[1-(\tanh \kappa)^2]}{4M_w },
\ee
see Fig.~\ref{fig:approxdr}. Equating to zero the discriminant of the quadratic in $\sigma$ equation \rf{qede3}, we arrive at the following quadratic approximation to the neutral stability curve at the crossing point $M= M_0^+$
\be{qans}
\beta = M_w\frac{\left(M-M_w-\sqrt{(\tanh \kappa)/\kappa}\right)^2}{(1-(\tanh \kappa)^2)\sqrt{(\tanh \kappa)/\kappa}}.
\ee
The bubble of instability \rf{bubble} corresponds to the inner points of the instability domain bounded by \rf{qans}.


\begin{figure}
  \begin{subfigure}{0.33\textwidth}
    \includegraphics[width=\textwidth]{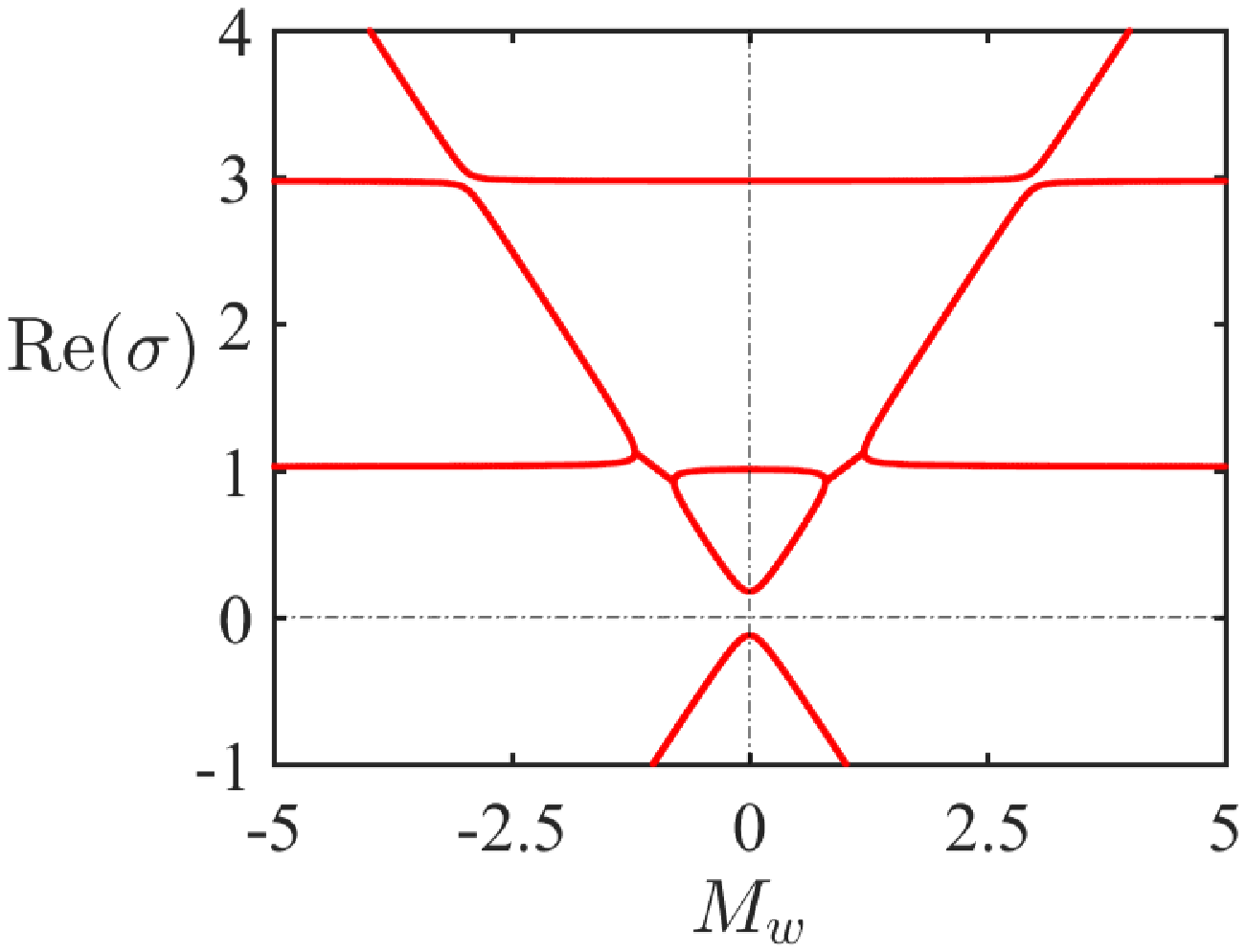}
    \caption{} \label{fig:6a}
  \end{subfigure}%
  \begin{subfigure}{0.33\textwidth}
    \includegraphics[width=0.87\textwidth]{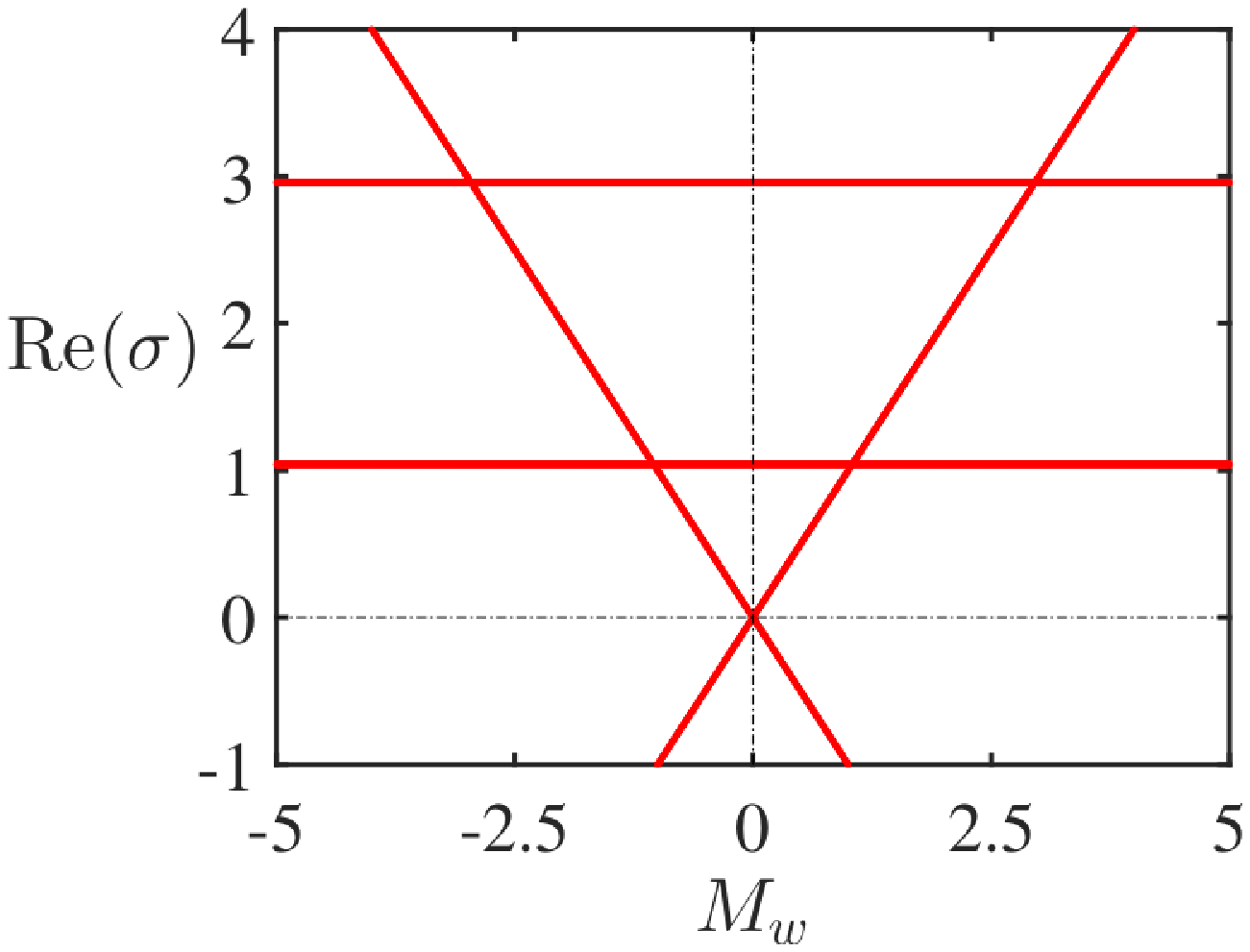}
    \caption{} \label{fig:6b}
  \end{subfigure}%
  \begin{subfigure}{0.33\textwidth}
    \includegraphics[width=\textwidth]{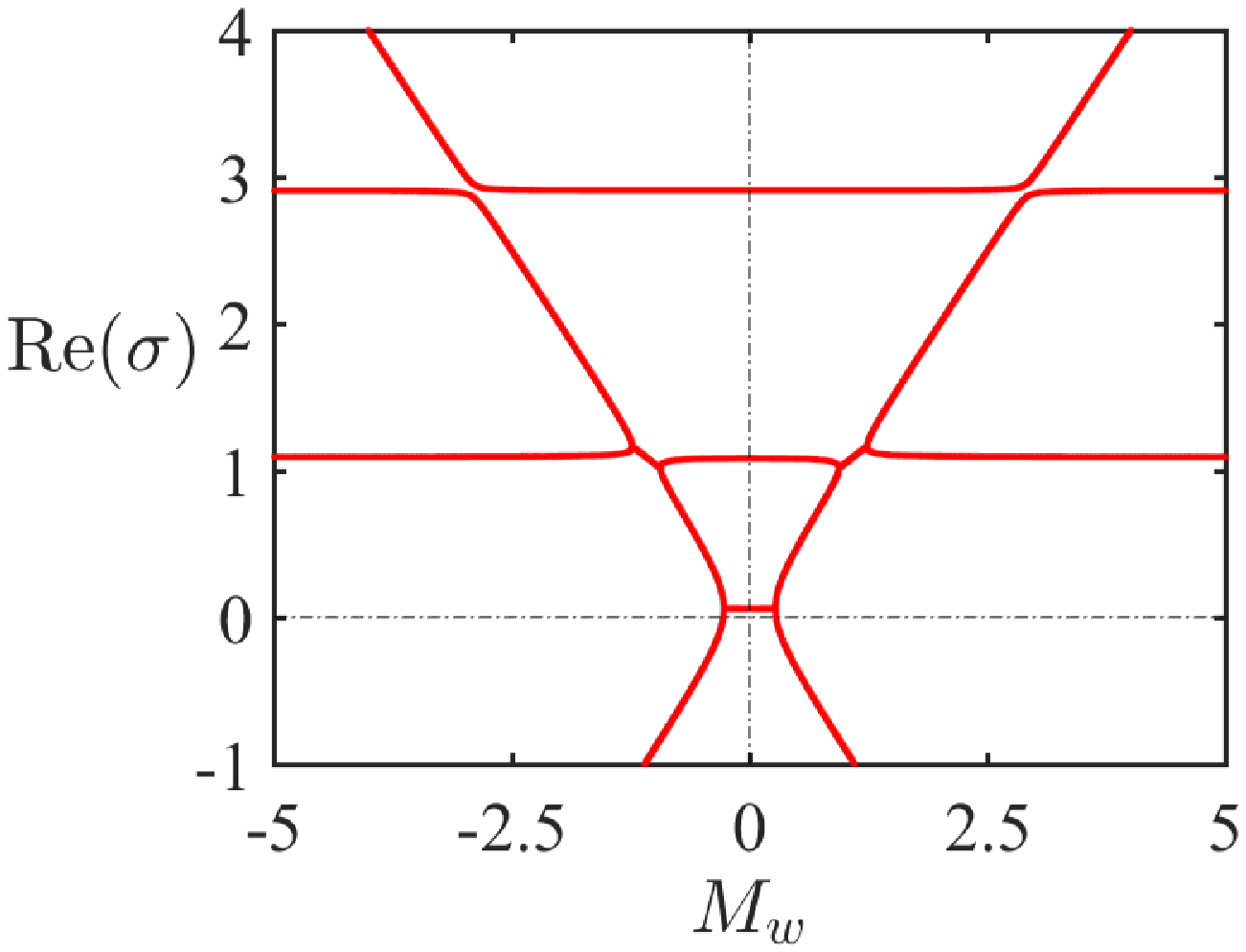}
    \caption{} \label{fig:6c}
  \end{subfigure}\\
  \begin{subfigure}{0.33\textwidth}
    \includegraphics[width=\textwidth]{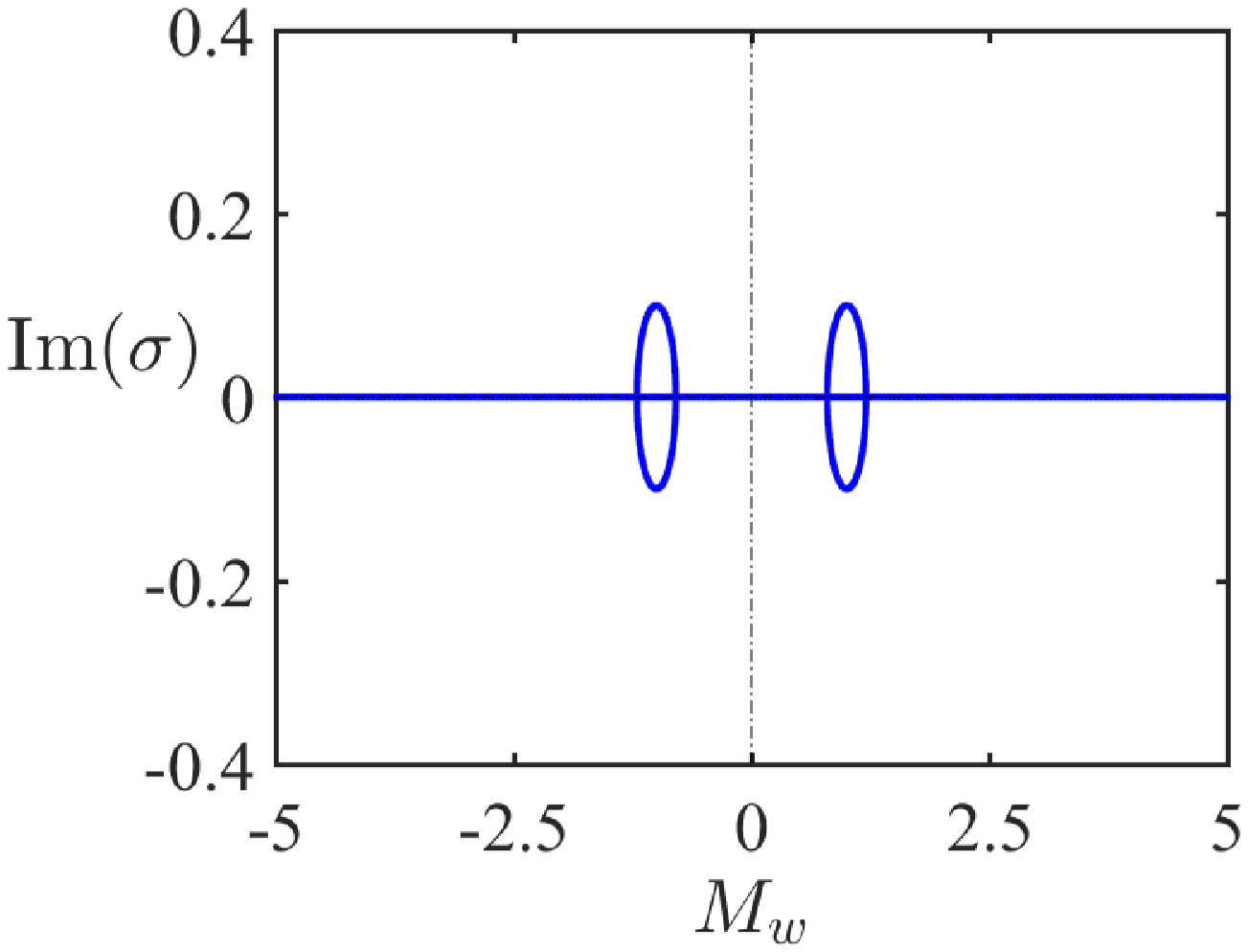}
    \caption{} \label{fig:6d}
  \end{subfigure}%
  \begin{subfigure}{0.33\textwidth}
    \includegraphics[width=0.87\textwidth]{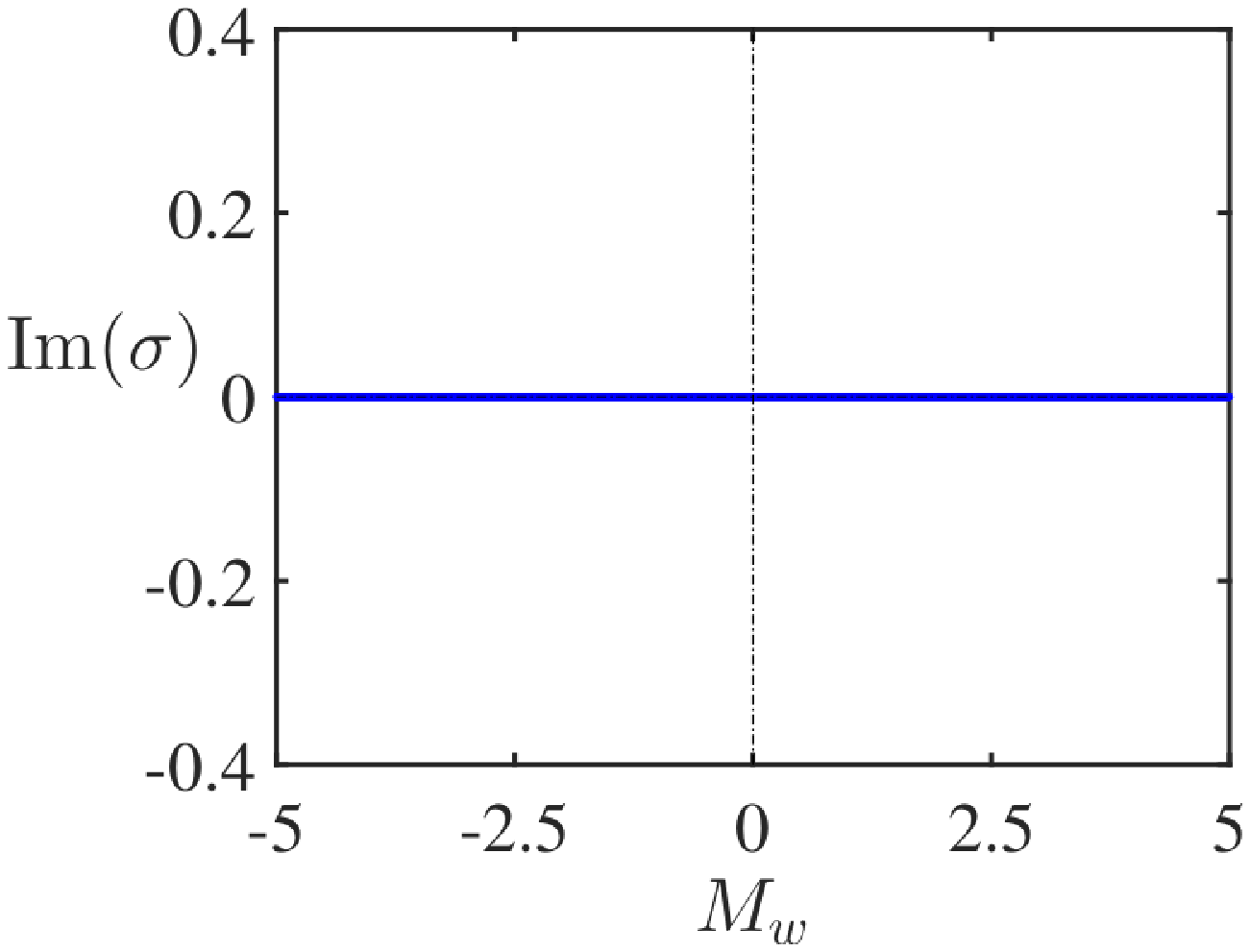}
    \caption{} \label{fig:6e}
  \end{subfigure}%
  \begin{subfigure}{0.33\textwidth}
    \includegraphics[width=\textwidth]{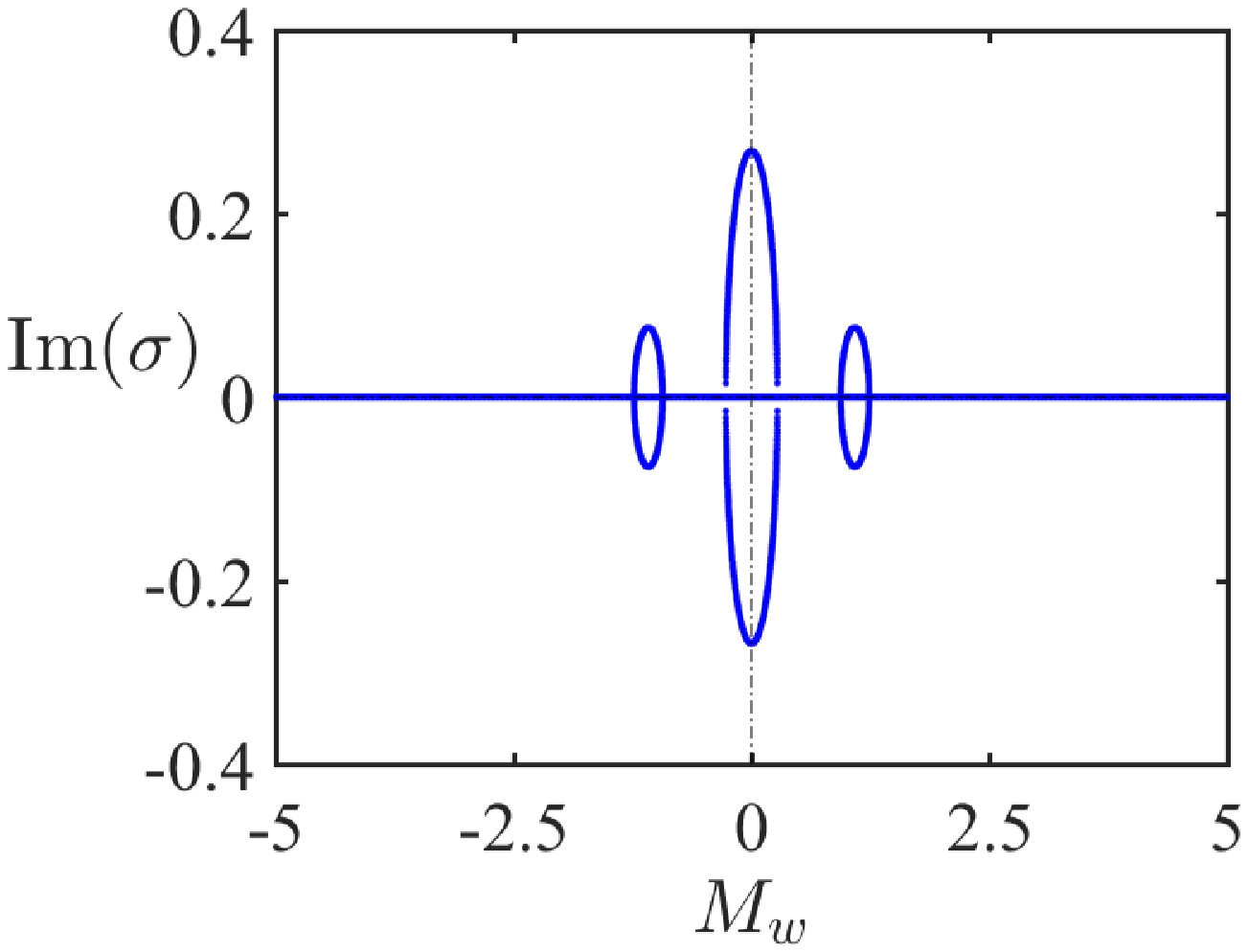}
    \caption{} \label{fig:6f}
  \end{subfigure}%

\caption{Real (upper panels) and imaginary (lower panels) parts of the roots of the dispersion equation \rf{domb} for $M=M_0=2$ and: (a, d) $\beta=0.05$ and $\kappa=\kappa_0-0.1$, (b, e) $\beta=0$ and $\kappa=\kappa_0\approx 0.5218134478$, (c, f) $\beta=0.05$ and $\kappa=\kappa_0+0.3$. Notice that the bubbles of instability develop only for ${\rm Re}(\sigma)<M_0=2$. \label{fig6}}
\end{figure}

Using the same methodology to approximate the avoided crossing close to $M=M_0^-$, $\sigma = \sigma_0$ and $\beta=\beta_0$ by equation \rf{qede2}, we obtain
\be{qede3_2}
(\sigma-M_w)\left[\sigma-M-\sqrt{\frac{\tanh \kappa}{\kappa}}\right]=-\beta\sqrt{\frac{\tanh \kappa}{\kappa}}\frac{(\tanh \kappa)^2-1}{4M_w}.
\ee
Separating real and imaginary parts of $\sigma$ in \rf{qede3_2} similarly to how it has been done in the previous case, one can see that the bubble of instability does not originate for $\beta>0$ in the unfolding of the crossing at $M=M_0^-$, see Fig.~\ref{fig:approxdr}.

In Fig.~\ref{fig:smap} we show that the exact neutral stability boundaries obtained from equating the discriminant of the fourth-order polynomial \rf{domb} in $\sigma$ to zero and their approximation \rf{qans} calculated at the crossing point at $M=M_0^+$ are in a very good agreement.

It is instructive to change the point of view and to look at the critical values of parameters as functions of the Mach number $M_w$ of elastic waves in the membrane. In Fig.~\ref{fig:cone2} we present stability maps of the dispersion equation \rf{domb} given by its discriminant
in the $(M_w,\beta)$-plane for the fixed value of $M=M_0=2$ and increasing values of $\kappa$. We see that for all $\kappa$ the instability is possible only in the interval $|M_w|<M_0=2$, which agrees with Fig.~\ref{fig:smap}. For $\beta=0$, the instability domain touches the $M_w$-axis  at the points  $M_w=M_0-\sqrt{(\tanh \kappa)/\kappa}$ and $M_w=-M_0+\sqrt{(\tanh \kappa)/\kappa}$. In the limit $\kappa\rightarrow 0$, the touching occurs at $M_w=M_0-1=1$ and $M_w=-M_0+1=-1$.

A qualitative change happens when $\kappa\ge \kappa_0$ where
$\kappa_0>0$ is uniquely determined by $M_0>0$ from the equation
\be{M0kappa0}
\kappa_0 \tanh \kappa_0=\frac{1}{M_0^2}.
\ee
For instance, $M_0=2$ yields $\kappa_0\approx 0.5218134478$. At $\kappa=\kappa_0$ a new, isolated, domain of instability originates that touches the $M_w$-axis at $\beta=0$ and grows when $\kappa$ is further increased, Fig.~\ref{fig:cone2}. At some value of $\kappa$ the two domains touch each other and then form a unified domain. At $\kappa \rightarrow \infty$ the central part of the unified domain dominates over its side parts corresponding to the instability found by Nemtsov in the shallow-water approximation when $\kappa \rightarrow 0$ and the coupling $\beta$ is weak, Fig.~\ref{fig:cone2}.

To understand the origin of the new instability, we plot the real and imaginary values of $\sigma$ as functions of $M_w$ in Fig.~\ref{fig6} for a given $M=M_0=2$. The central panel of Fig.~\ref{fig6} corresponding to $\beta=0$ and $\kappa=\kappa_0$ shows four straight lines intersecting at five points, including the origin. The upper horizontal line corresponds to the fast surface gravity wave with $\sigma=M_0+\sqrt{(\tanh \kappa_0)/\kappa_0}\approx 3$, whereas the lower horizontal line to the slow surface gravity wave \citep{N1986} with $\sigma=M_0-\sqrt{(\tanh \kappa_0)/\kappa_0}\approx 1$. The two inclined lines correspond to the forward and backward elastic waves in the membrane with $\sigma=\pm M_w$. When $\beta$ and $\kappa$ depart respectively from zero and $\kappa_0$, all the five crossings unfold either into avoided crossings (as elastic and fast surface gravity waves) or into bubbles of instability (as elastic and slow surface gravity waves) resulting in the high-frequency flutter due to radiation of long surface gravity waves. For $\beta>0$ the crossing at the origin transforms into an avoided crossing at $\kappa<\kappa_0$ or into the bubble of instability at $\kappa>\kappa_0$, which yields low-frequency flutter at short wavelengths $\kappa$.


\begin{figure}
  \begin{subfigure}{0.5\textwidth}
    \includegraphics[width=\textwidth]{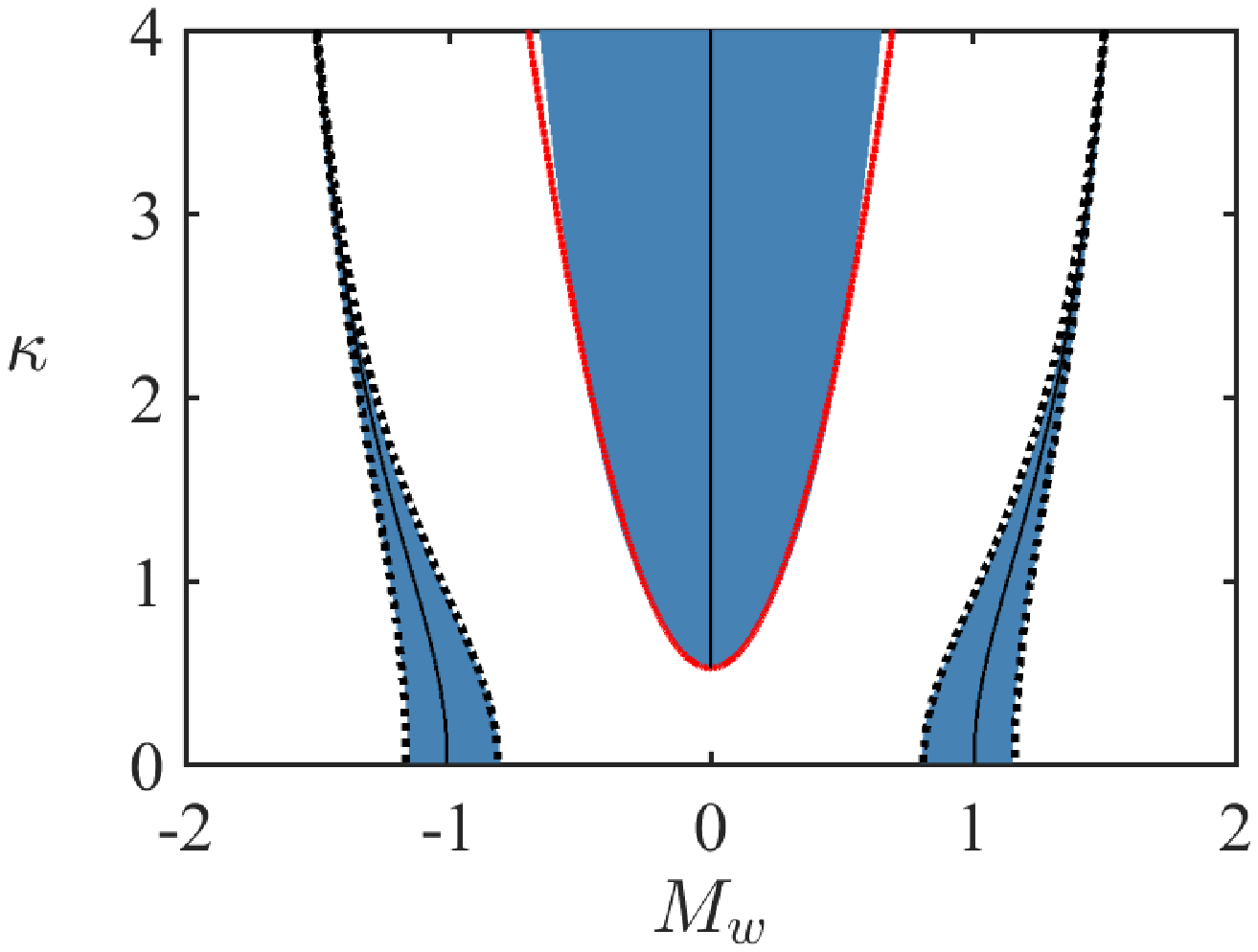}
    \caption{} \label{fig:7a}
  \end{subfigure}%
  \begin{subfigure}{0.5\textwidth}
    \includegraphics[width=\textwidth]{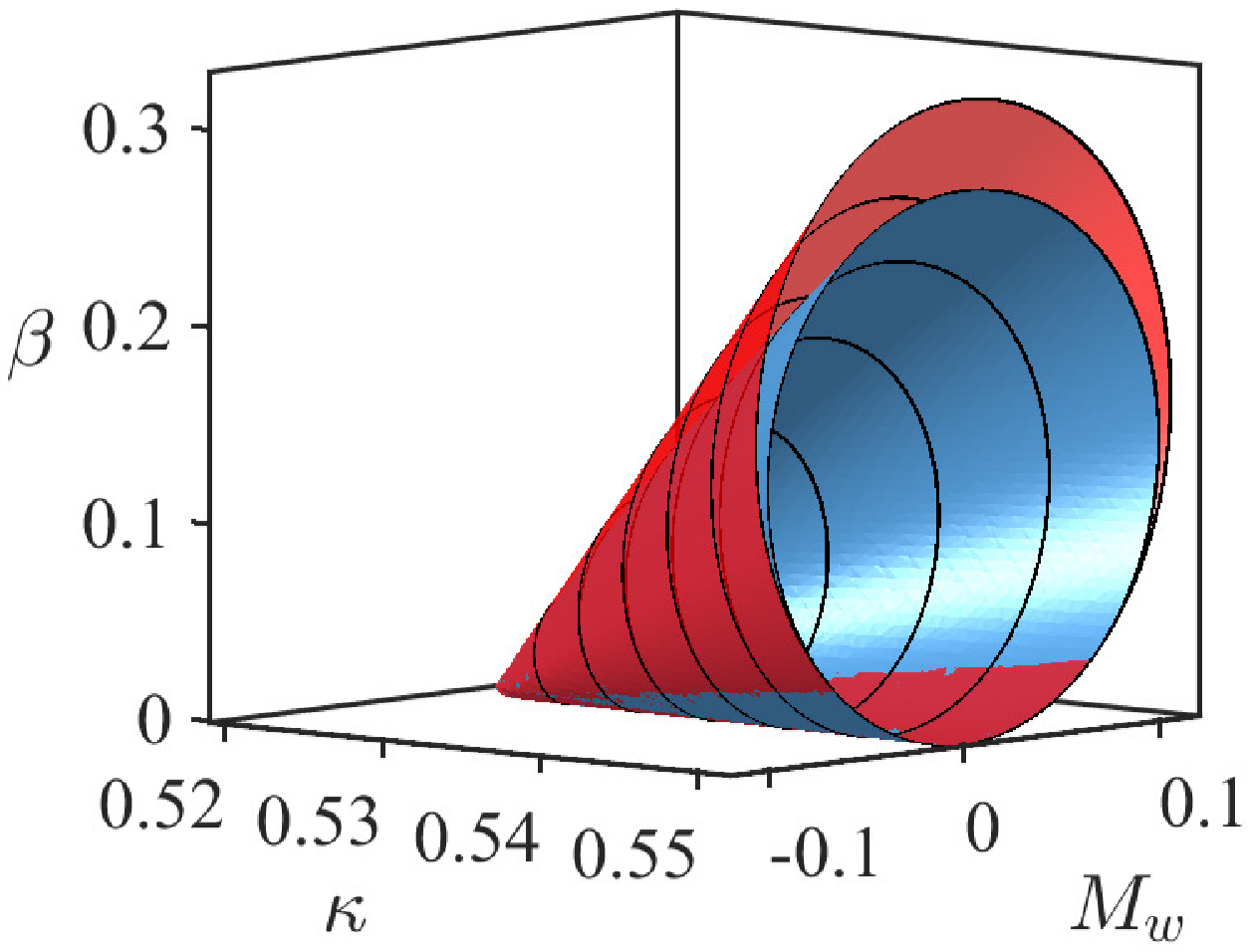}
    \caption{} \label{fig:7b}
  \end{subfigure}%

  \caption{(a) Stability map of the dispersion equation \rf{domb} given by its discriminant for $M=M_0=2$ and $\beta=0.03$. The regions of real phase speed $\sigma$ are shown in white (stability) and that of the complex $\sigma$ (temporal instability) in blue. The black dotted curves correspond to the approximation \rf{qans} and the solid red line is the conical approximation \rf{conapprox}. When $\beta=0$, the blue instability domains degenerate (central) to  the ray $\kappa\ge \kappa_0\approx 0.5218134478$ and (sides) to the curves \rf{beta0} shown as solid black lines. (b) Stability boundary with the conical singularity at $\kappa=\kappa_0$, $\beta=0$ and $M_w=0$, according to (blue, internal surface) the discriminant of the dispersion equation \rf{domb} and (red, external surface) to the approximation of the cone \rf{conapprox}. \label{fig:cone1}}
\end{figure}

The panel (a) in Fig.~\ref{fig:cone1} allows us to track the evolution of the flutter domains as $\kappa$ varies from zero to infinity at $M_0=2$ and $\beta=0.03$. Nemtsov's radiation-induced flutter domain is the widest in the shallow water limit and evolves along the curves (shown as black solid lines in Fig.~\ref{fig:cone1}(b))
\be{beta0}
(M_0\pm M_w)^2=\frac{\tanh \kappa}{\kappa},
\ee
to which the Nemtsov domains degenerate at $\beta=0$. Note that the Nemtsov flutter domain is perfectly approximated by formula \rf{qans} obtained from the unfolding of the eigenvalue crossing corresponding to the slow surface gravity wave and the elastic wave (dotted lines in Fig.~\ref{fig:cone1}(a)).

To understand the central instability domain shown in Fig.~\ref{fig:cone1}(a) for a given $\beta$ we plot it in the $(M_w,\kappa,\beta)$-space in the  panel (b) of Fig.~\ref{fig:cone1}, given $M=M_0$. One can see that the domain is symmetric with respect to the plane $M_w=0$ and has a pronounced conical singularity at $\kappa=\kappa_0$ determined by the equation \rf{M0kappa0} when $\beta=0$ and $M_w=0$. Equation \rf{M0kappa0} follows from the discriminant of the dispersion equation \rf{domb} at $\beta=0$ and $M_w=0$.
The conical singularity of the stability boundary therefore exactly corresponds to the crossing of the eigenvalue curves at the origin in the panel (b) of Fig.~\ref{fig6}. Usually, the conical singularity of the stability boundary is associated with a double semi-simple eigenvalue with two linearly-independent eigenvectors \citep{KS2004, K2009, K2010, K2013dg, GK2006, KGS2009}.


\begin{figure}
  \begin{subfigure}{0.33\textwidth}
    \includegraphics[width=\textwidth]{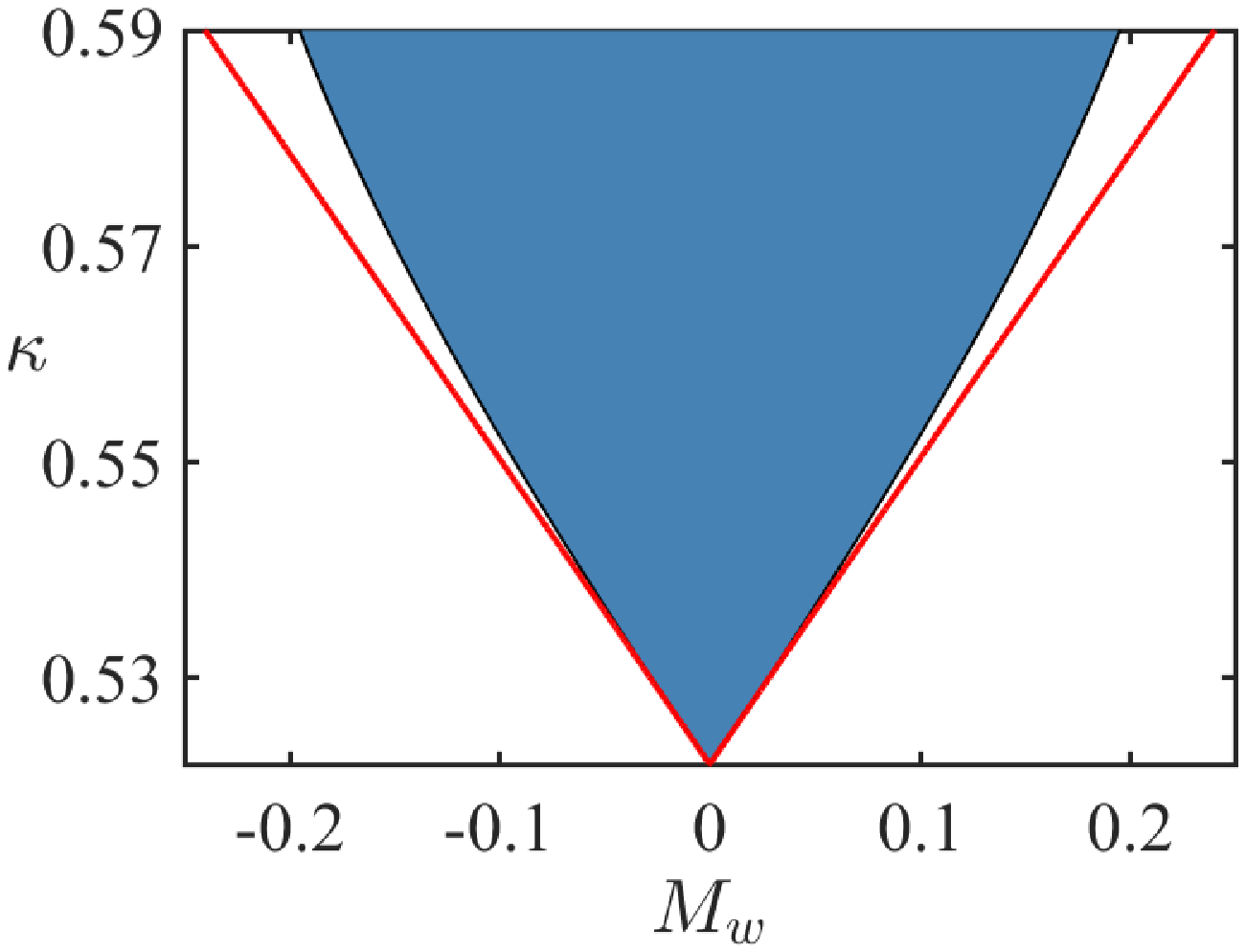}
    \caption{} \label{fig:8a}
  \end{subfigure}%
  \begin{subfigure}{0.33\textwidth}
    \includegraphics[width=\textwidth]{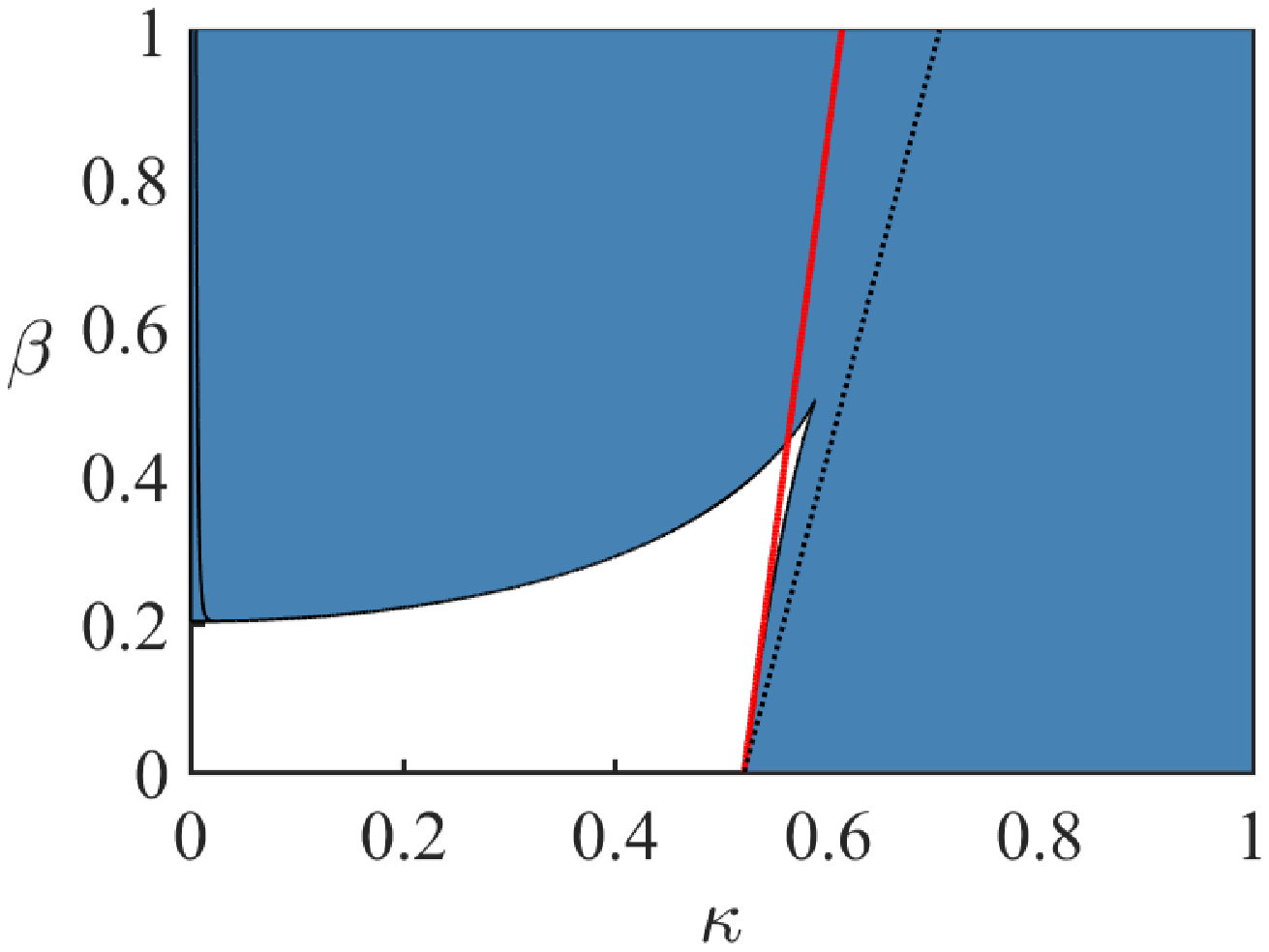}
    \caption{} \label{fig:8b}
  \end{subfigure}%
  \begin{subfigure}{0.33\textwidth}
    \includegraphics[width=\textwidth]{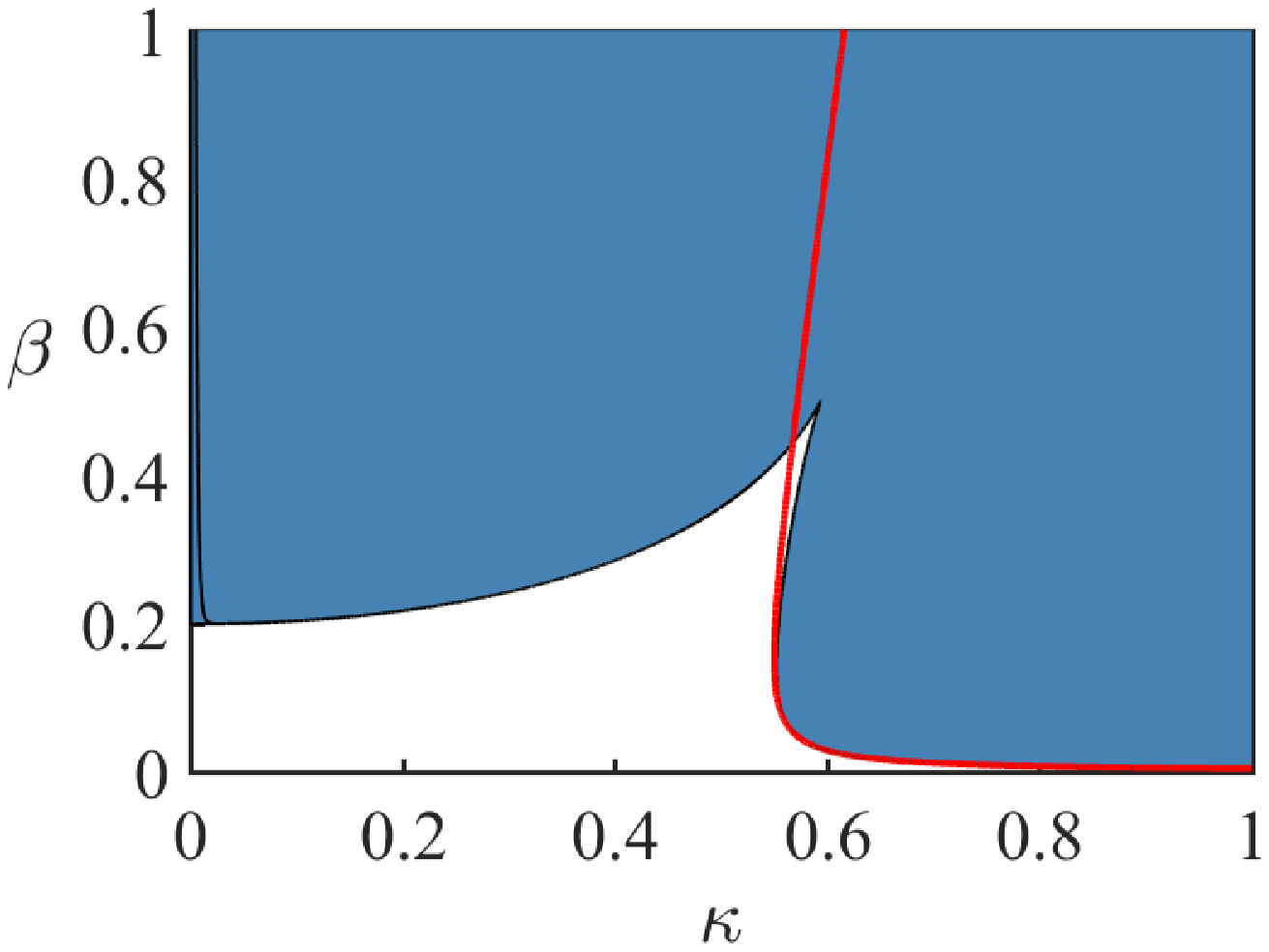}
    \caption{} \label{fig:8b}
  \end{subfigure}%

  \caption{For $M=M_0=2$ (a) cross-section of the instability domain with the conical singularity shown in Fig.~\ref{fig:cone1}(b) by the plane \rf{plane2}. The regions of real phase speed $\sigma$ are shown in white (stability) and that of the complex $\sigma$ (temporal instability) in blue. The red lines crossing at the apex of the cone at $\kappa= \kappa_0\approx 0.5218134478$ is a linear approximation given by Eq.~\rf{section}. (b) Cross-section by the plane $M_w=0$ of the instability domain and (red line) its linear approximation \rf{plane1} at the conical point $\kappa=\kappa_0$. The black dotted line is given by the Eq.~\rf{plane2}. (c) Similar cross-section by the plane $M_w=0.1$ where the red curve is the approximation \rf{conapprox}. \label{fig:cone3}}
\end{figure}

By this reason, we apply the perturbation theory of double eigenvalues presented in Appendix~\ref{sadr} to the double zero eigenvalue $\sigma=\sigma_0=0$ at the crossing shown in the panel (b) of Fig.~\ref{fig6} and corresponding to the values of parameters $\beta=\beta_0=0$, $\kappa=\kappa_0$, $M=M_0$, $M_w=M_{w,0}=0$. A natural extension of the approximation formula \rf{qede} to the case of four parameters $\beta$, $\kappa$, $M_w$, and $M$ yields
\ba{approde}
&(\Delta\sigma)^2\frac{1}{2}\partial_{\sigma}^2 D+\Delta\sigma\left(\partial^2_{\sigma \beta}D \Delta \beta+\partial^2_{\sigma \kappa}D \Delta \kappa+\partial^2_{\sigma M}D \Delta M+\partial^2_{\sigma M_w}D \Delta M_w\right) +\partial^2_{\beta \kappa} D \Delta \beta \Delta \kappa &\nn \\
&+\frac{1}{2}\left[ \partial_{\beta}^2 D (\Delta \beta)^2+ \partial_{\kappa}^2 D (\Delta \kappa)^2+ \partial_{M}^2 D (\Delta M)^2+ \partial_{M_w}^2 D (\Delta M_w)^2\right] +\partial^2_{M \kappa} D \Delta M \Delta \kappa &\nn \\
&+\partial^2_{M_w \kappa} D \Delta M_w \Delta \kappa+\partial^2_{\beta M} D \Delta \beta \Delta M+\partial^2_{\beta M_w} D \Delta \beta \Delta M_w+\partial^2_{\beta M} D \Delta M_w \Delta M &\nn \\
&+\partial_{\beta} D \Delta \beta+\partial_{\kappa} D \Delta {\kappa}+\partial_{M} D \Delta {M}+\partial_{M_w} D \Delta {M_w}=0.&
\ea

Computing the corresponding partial derivatives of the left part of the dispersion equation \rf{domb}, and evaluating them at $\beta=\beta_0=0$, $\kappa=\kappa_0$, $M=M_0$, $M_w=M_{w,0}=0$, where $M_0$ and $\kappa_0$ are related by the equation \rf{M0kappa0},
we find that the only non-zero derivatives are
\ba{nonzede}
&\partial^2_{\sigma} D=-\partial^2_{M_w} D= 2 \kappa_0 M_0^2-\frac{2}{\kappa_0 M_0^2}, \quad \partial^2_{M \beta} D=-\partial^2_{\sigma \beta} D= 2 \kappa_0 M_0,& \nn \\
&\partial^2_{\kappa \beta} D = M_0^4\kappa_0^2+M_0^2-1.&
\ea
Taking this into account in \rf{approde}, we find a simple approximation describing the unfolding of the double zero eigenvalue
\be{approeig}
(M_0^4\kappa_0^2-1)(\sigma^2-M_w^2)-2\kappa_0^2M_0^3\sigma\beta+\kappa_0M_0^2(M_0^4\kappa_0^2+M_0^2-1)(\kappa-\kappa_0)\beta
+2\kappa_0^2M_0^3(M-M_0)\beta=0.
\ee

Let us further assume that $M=M_0$ is fixed. Then the last term in \rf{approeig} vanishes, and the discriminant of the resulting quadratic polynomial in $\sigma$ produces the equation of a cone with the apex at $\kappa=\kappa_0$, $M_w=0$, and $\beta=0$
\be{conapprox}
M_0^6 \kappa_0^4 \beta^2-M_0^2 \kappa_0 (M_0^4\kappa_0^2+M_0^2-1)(M_0^4\kappa_0^2-1)(\kappa-\kappa_0)\beta+M_w^2(M_0^4\kappa_0^2-1)^2=0.
\ee
The cone \rf{conapprox} is shown in red in the panel (b) of Fig.~\ref{fig:cone1}. With $\beta=0.03$, $M_0=2$ and $\kappa_0$ computed by means of the equation \rf{M0kappa0}, the approximation \rf{conapprox} fits the boundary of the exact instability domain with remarkable precision, as is evident in Fig.~\ref{fig:cone1}(a).

It is easy to see that in the plane $M_w=0$ the cone \rf{conapprox} defines the two lines
\be{plane1}
\beta = \frac{(M_0^4\kappa_0^2-1)(M_0^4\kappa_0^2+M_0^2-1)}{M_0^4\kappa_0^3}(\kappa-\kappa_0),\quad \beta=0
\ee
that approximate the instability domain near $\kappa=\kappa_0$, see panel (b) in Fig.~\ref{fig:cone3}.
As soon as $M_w$ deviates from zero, the cone \rf{conapprox} again provides a very good fit to the actual stability boundary, Fig.~\ref{fig:cone3}(c).
In the plane
\be{plane2}
\beta = \frac{(M_0^4\kappa_0^2-1)(M_0^4\kappa_0^2+M_0^2-1)}{2M_0^4\kappa_0^3}(\kappa-\kappa_0)
\ee
the cross-section of the cone \rf{conapprox} is described by the two lines
\be{section}
\kappa = \kappa_0\pm M_w\frac{2\kappa_0M_0}{M_0^4\kappa_0^2+M_0^2-1}
\ee
that constitute a linear approximation to the stability boundary shown in Fig.~\ref{fig:cone3}(a).

\subsubsection{Wave energy of the Nemtsov system for membrane of infinite chord length}
Let us use physical considerations to derive the expression for the averaged over the wave period energy of the Nemtsov system with the membrane of infinite chord length, by combining the approaches of the works by \cite{MRS2016} and \cite{S1987}.

In the linear wave theory, the energy is a function of the squared wave amplitude \citep{MRS2016}. Therefore,
the total energy per surface area of the membrane resulting both from the wave velocity of the structure and the elastic energy due to its tension is
\be{e_m}
\mathcal{E}_m = \mathcal{K}_m + \mathcal{P}_m = \frac{1}{2} \left(\text{Re}[\partial_{\tau}\xi(x,\tau)]\right)^2 + \frac{1}{2} M_w^2 \left(\text{Re}[\partial_x\xi(x,\tau)]\right)^2,
\ee
where $\text{Re}$ stands for the real part of the vibration amplitude that is complex-valued because of the assumed plane wave solution
\be{ansatz}
[\phi(x,z,\tau),\eta(x,\tau),\xi(x,\tau)] \sim [\hat{\phi}(z),\hat{\eta},\hat{\xi}] e^{i(\kappa x - \omega \tau)}.
\ee
Recall that $\hat{\phi}(z)$ is determined by the expression \rf{sftle} with the coefficients \rf{ab} and $\hat{\eta}$, $\hat{\xi}$ are, respectively, displacement amplitudes of the free surface and the membrane.

The energy of the fluid depends on whether we assume a vacuum below the membrane \citep{N1986} or a quiescent medium of the same density as the fluid above the membrane and with the pressure equal to the pressure of the unperturbed fluid \citep{V2004,V2016}. The gravitational potential energy of the free surface is the only term contributing to the total potential energy of the fluid in the latter context. Therefore,
\be{p_f}
\mathcal{P}_f = \frac{1}{2} \alpha \left(\text{Re}[\eta(x,\tau)]\right)^2.
\ee

The kinetic energy of the flow per unit area is determined by the velocity field $\bm{u} = \bm{\nabla}\phi + M\bm{e}_x$, where $\bm{u}={\bm{v}}/{\sqrt{gH}}$, that needs to be directly integrated within the limits given by the surface of the membrane and the free surface of the fluid,
\ba{k_f}
\mathcal{K}_f = \frac{1}{2} \alpha \int_{\text{Re}\, \xi}^{\text{Re}\, \eta} \vert\vert\text{Re}(\bm{u})\vert\vert^2 dz &=& \frac{1}{2} \alpha \int_{\text{Re}\, \xi}^{\text{Re}\,\eta} \left[ \left(\text{Re}[\bm{\nabla}\phi]\right)^2 + 2M\text{Re}[\partial_x\phi] + M^2 \right] dz \nn \\
&=& \frac{1}{2} \alpha \int_{\text{Re}\, \xi}^{\text{Re}\, \eta} \left[ \left(\text{Re}[\partial_x\phi]\right)^2 + \left(\text{Re}[\partial_z\phi]\right)^2 \right] dz \nn \\
&+& \alpha M \int_{\text{Re}\, \xi}^{\text{Re}\, \eta} \left[ \text{Re}[\partial_x\phi] + \frac{M}{2} \right] dz.
\ea

From assumption \rf{ansatz} and the explicit form of the complex amplitude $\hat{\phi}(z)$ determined by \rf{sftle} with the coefficients \rf{ab}, it follows that
\ba{re_parts}
\text{Re}[\partial_x\phi] = i\kappa\hat{\phi}(z)\cos{(\kappa x - \omega \tau)} ,
&\quad&
\text{Re}[\partial_z\phi] = i\partial_z\hat{\phi}(z)\sin{(\kappa x - \omega \tau)}, \nn \\
\text{Re}\, \xi = \hat{\xi}\cos{(\kappa x - \omega \tau)} ,
&\quad&
\text{Re}\, \eta = \hat{\eta}\cos{(\kappa x - \omega \tau)}.
\ea

Taking into account the expressions \rf{re_parts} in \rf{k_f}, we find
\ba{ibp}
\int_{\text{Re}\, \xi}^{\text{Re}\, \eta} \left(\text{Re}[\partial_x\phi]\right)^2 dz = -\kappa^2 \cos^2{(\kappa x - \omega \tau)} \int_{\text{Re}\, \xi}^{\text{Re}\, \eta} \hat{\phi}(z)^2 dz.
\ea

Similarly, with the help of integration by parts, the Laplace equation \rf{ftle}, and expressions \rf{re_parts}, we obtain
\ba{ibp1}
\int_{\text{Re} \, \xi}^{\text{Re}\, \eta} \left(\text{Re}[\partial_z\phi]\right)^2 dz &=& - \sin^2{(\kappa x - \omega \tau)} \left\lbrace \left[ \hat{\phi}\partial_z\hat{\phi} \right]_{\text{Re}\, \xi}^{\text{Re}\, \eta} - \int_{\text{Re}\, \xi}^{\text{Re}\, \eta} \hat{\phi}(\partial_z^2\hat{\phi}) dz \right\rbrace \nn \\
&=& - \sin^2{(\kappa x - \omega \tau)} \left\lbrace \left[ \hat{\phi}\partial_z\hat{\phi} \right]_{\text{Re}\, \xi}^{\text{Re}\, \eta} - \kappa^2 \int_{\text{Re}\,\xi}^{\text{Re}\,\eta} \hat{\phi}(z)^2 dz \right\rbrace .
\ea

Finally, following \cite{MRS2016}, we evaluate the last integral term in \rf{k_f} with the help of the Lagrange mean value theorem, which is justified by the assumption that $\eta$ and $\xi$ are infinitesimally small perturbations of the surface boundaries $\partial\Omega_0$ and $\partial\Omega_1$. Performing this procedure, and then taking into account expressions \rf{re_parts}, we obtain
\ba{lmvf}
&&\int_{\text{Re}\, \xi}^{\text{Re}\, \eta} \left[ \text{Re}[\partial_x\phi] + \frac{M}{2} \right] dz \nn\\
&=& \int_{\text{Re}\, \xi}^{0} \text{Re}[\partial_x\phi] dz + \int_{0}^{1} \text{Re}[\partial_x\phi] dz + \int_{1}^{\text{Re}\, \eta} \text{Re}[\partial_x\phi] dz + \frac{1}{2} \int_{\text{Re}\,\xi}^{\text{Re}\, \eta} M dz \nn \\
&=& \text{Re}\, \eta\left.\text{Re}[\partial_x\phi]\right|_{z=1} - \text{Re}\, \xi\left.\text{Re}[\partial_x\phi]\right|_{z=0} + \int_{0}^{1} \text{Re}[\partial_x\phi] dz + \frac{M}{2} \text{Re}\left( \eta - \xi \right) \nn \\
&=& i\kappa \left[ \hat{\eta}\hat{\phi}(1) - \hat{\xi}\hat{\phi}(0) \right] \cos^2{(\kappa x - \omega \tau)} + \left[ i \kappa\int_0^1 \hat \phi(z)dz + \frac{M}{2} ( \hat{\eta} - \hat{\xi} ) \right] \cos{(\kappa x - \omega \tau)}.\nn\\
\ea
Notice that the right-hand sides in the expressions \rf{ibp}, \rf{ibp1}, and \rf{lmvf} are $T$-periodic functions of time, where $T=2\pi/\omega$. Averaging these expressions over the wave period $T$ according to the rule
\be{mean}
\left\langle f(\tau) \right\rangle = \frac{1}{T} \int_{0}^{T} f(\tau) d\tau,
\ee
we deduce the mean kinetic energy of the fluid
\be{kf_a}
\left\langle \mathcal{K}_f \right\rangle  = \frac{1}{4} \alpha \left\lbrace -\left[ \hat{\phi}\partial_z\hat{\phi} \right]_{\partial \Omega_1}^{\partial \Omega_0}  + 2i\kappa M \left[ \hat{\eta}\hat{\phi}(1) - \hat{\xi}\hat{\phi}(0) \right] \right\rbrace.
\ee
The term $\hat{\phi}\partial_z\hat{\phi}$ in \rf{kf_a} is evaluated with the help of the Bernoulli principle \rf{dc} and the free surface kinematic condition \rf{kc} at $\partial\Omega_0$, and the wave equation \rf{eome} with the impermeability condition \rf{eomc} at $\partial\Omega_1$. This yields, respectively,
\ba{bound_val}
\left.\hat{\phi}\right|_{\partial\Omega_0} = \frac{\hat{\eta}}{i(\omega - \kappa M)},&\qquad&
 \left.\hat{\phi}\right|_{\partial\Omega_1} = \frac{\omega^2 - \kappa^2 M_w^2}{i\alpha(\omega - \kappa M)}\hat{\xi},\nn\\
\left.\partial_z\hat{\phi}\right|_{\partial\Omega_0} = -i(\omega - \kappa M)\hat{\eta},&\qquad&
 \left.\partial_z\hat{\phi}\right|_{\partial\Omega_1} = -i(\omega - \kappa M)\hat{\xi}.
\ea
Substituting expressions \rf{bound_val} into \rf{kf_a} we obtain the final expression for the mean kinetic energy of the fluid
\be{kf_a2}
\left\langle \mathcal{K}_f \right\rangle  = \frac{1}{4} \left\lbrace \alpha\hat{\eta}^2 - \left(\omega^2 - \kappa^2 M_w^2\right) \hat{\xi}^2 + 2i\alpha\kappa M \left[ \hat{\eta}\hat{\phi}(1) - \hat{\xi}\hat{\phi}(0) \right] \right\rbrace.
\ee
The other energies of the system, after taking into account \rf{re_parts} and time-averaging \rf{mean}, become
\be{av_ene}
\left\langle\mathcal{P}_m\right\rangle = \frac{1}{4}\kappa^2M_w^2\hat{\xi}^2, \quad
\left\langle\mathcal{K}_m\right\rangle = \frac{1}{4}\omega^2\hat{\xi}^2, \quad
\left\langle\mathcal{P}_f\right\rangle = \frac{1}{4}\alpha\hat{\eta}^2.
\ee
Notice that in the absence of the background flow $(M=0)$ the system respects the equipartition of energy
$$
\left\langle\mathcal{P}_f\right\rangle+\left\langle\mathcal{P}_m\right\rangle=
\left\langle\mathcal{K}_f\right\rangle+\left\langle\mathcal{K}_m\right\rangle,
$$
in accordance with the virial theorem \citep{LL1987}, because the flow is irrotational and thus derived from a fluid potential \citep{S1987}.


\begin{figure}
  \begin{subfigure}{0.25\textwidth}
    \includegraphics[width=\textwidth]{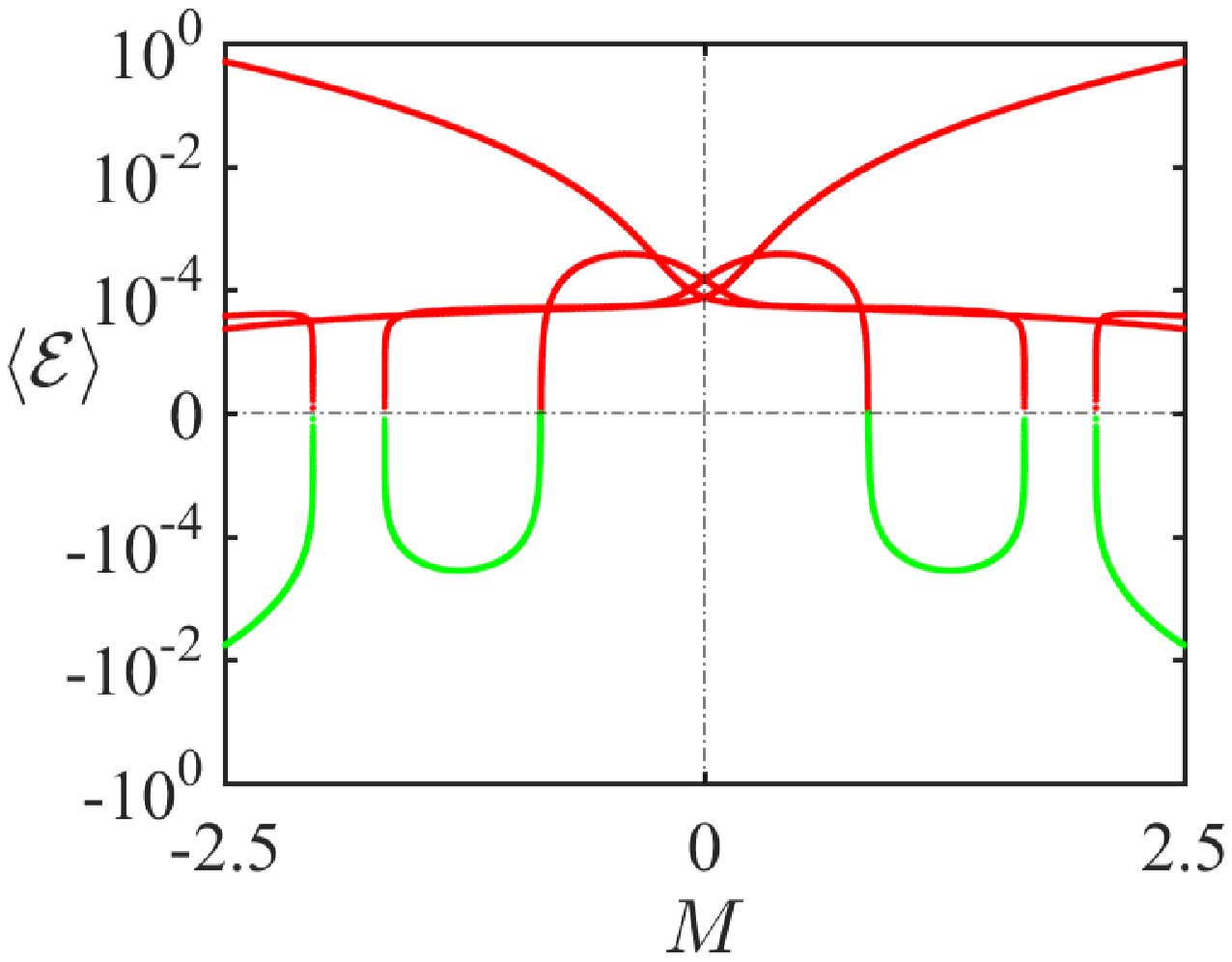}
    \caption{} \label{fig:9a}
  \end{subfigure}%
  \begin{subfigure}{0.25\textwidth}
    \includegraphics[width=\textwidth]{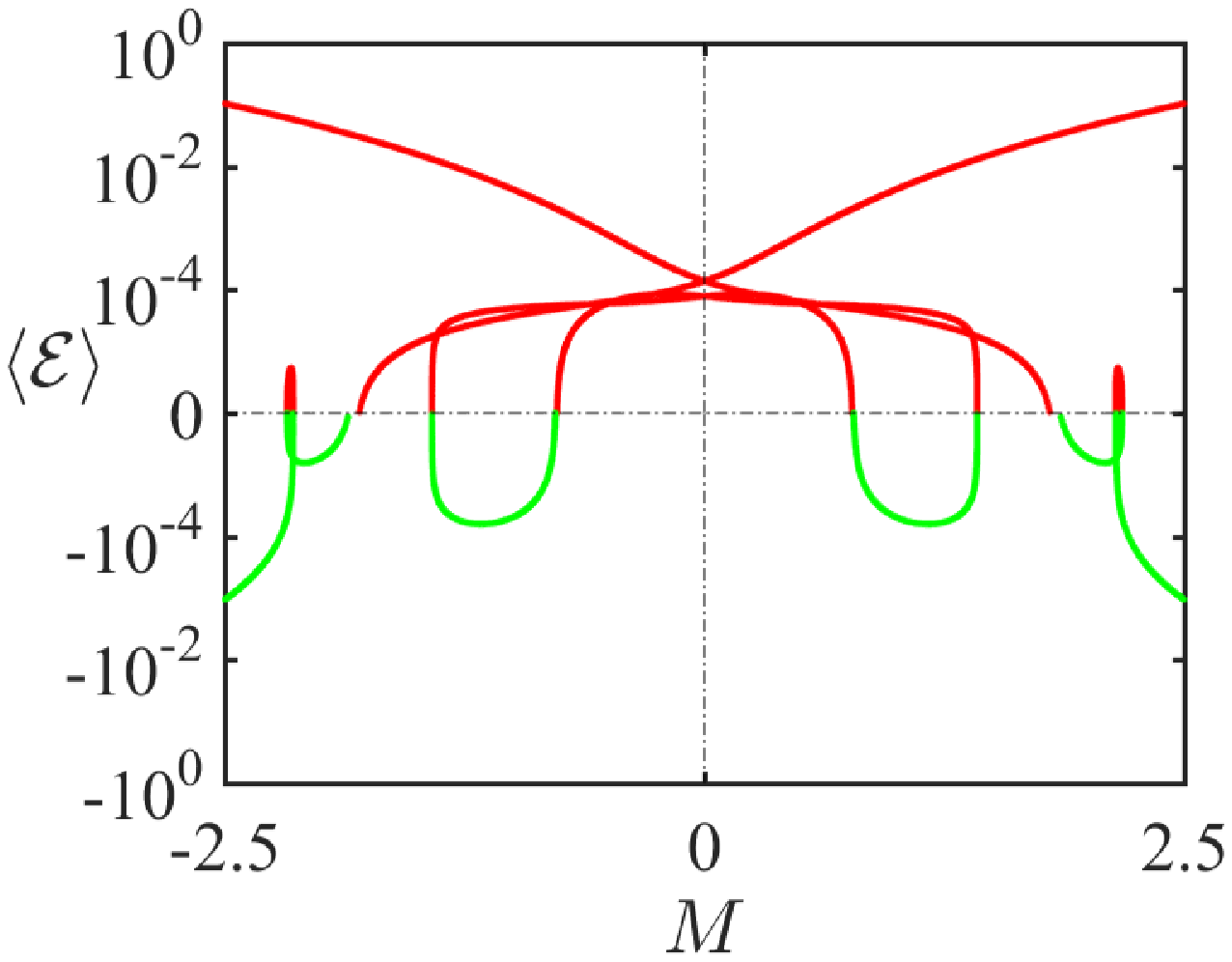}
    \caption{} \label{fig:9b}
  \end{subfigure}%
  \begin{subfigure}{0.25\textwidth}
    \includegraphics[scale=.225]{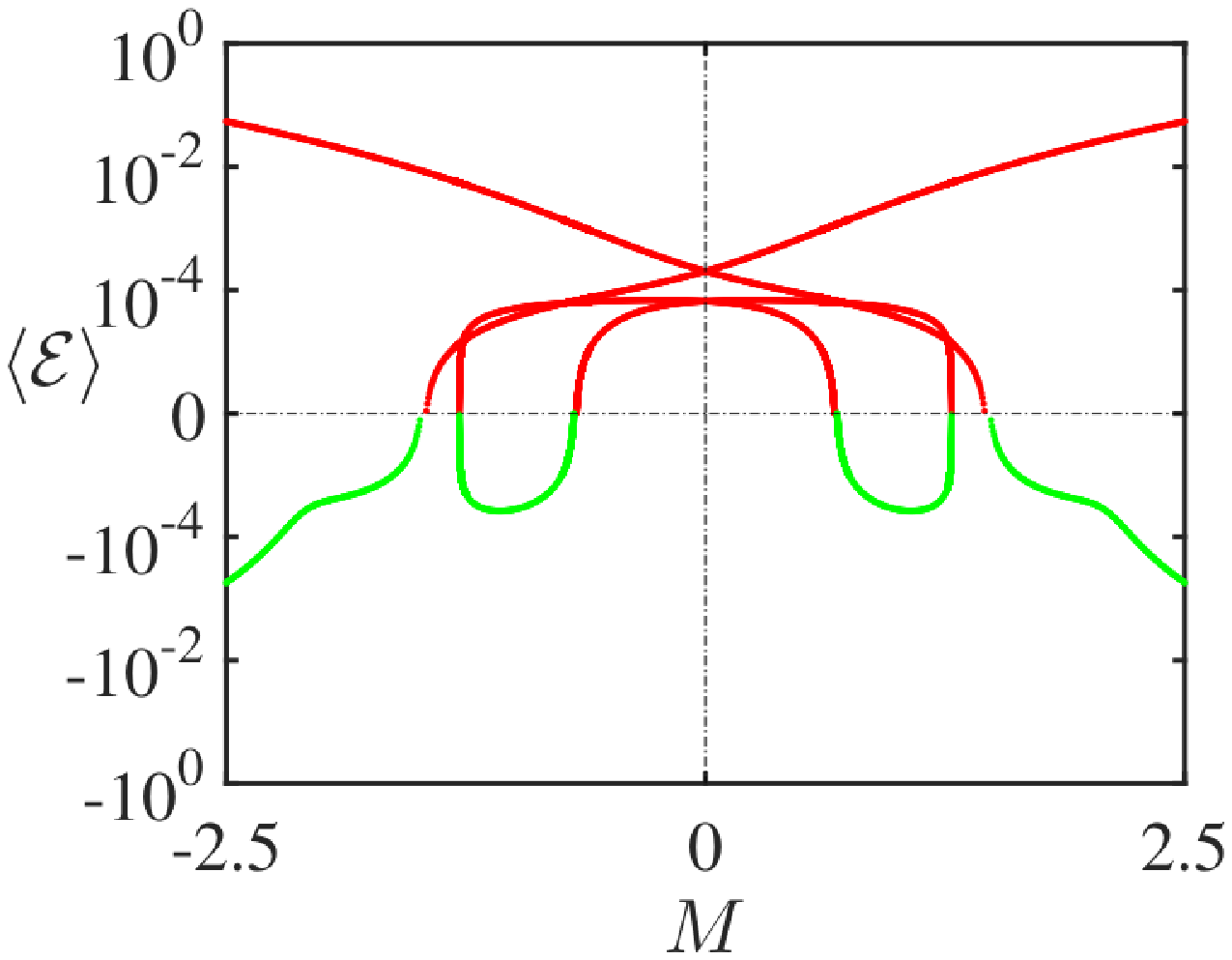}
    \caption{} \label{fig:9c}
  \end{subfigure}%
  \begin{subfigure}{0.25\textwidth}
    \includegraphics[width=\textwidth]{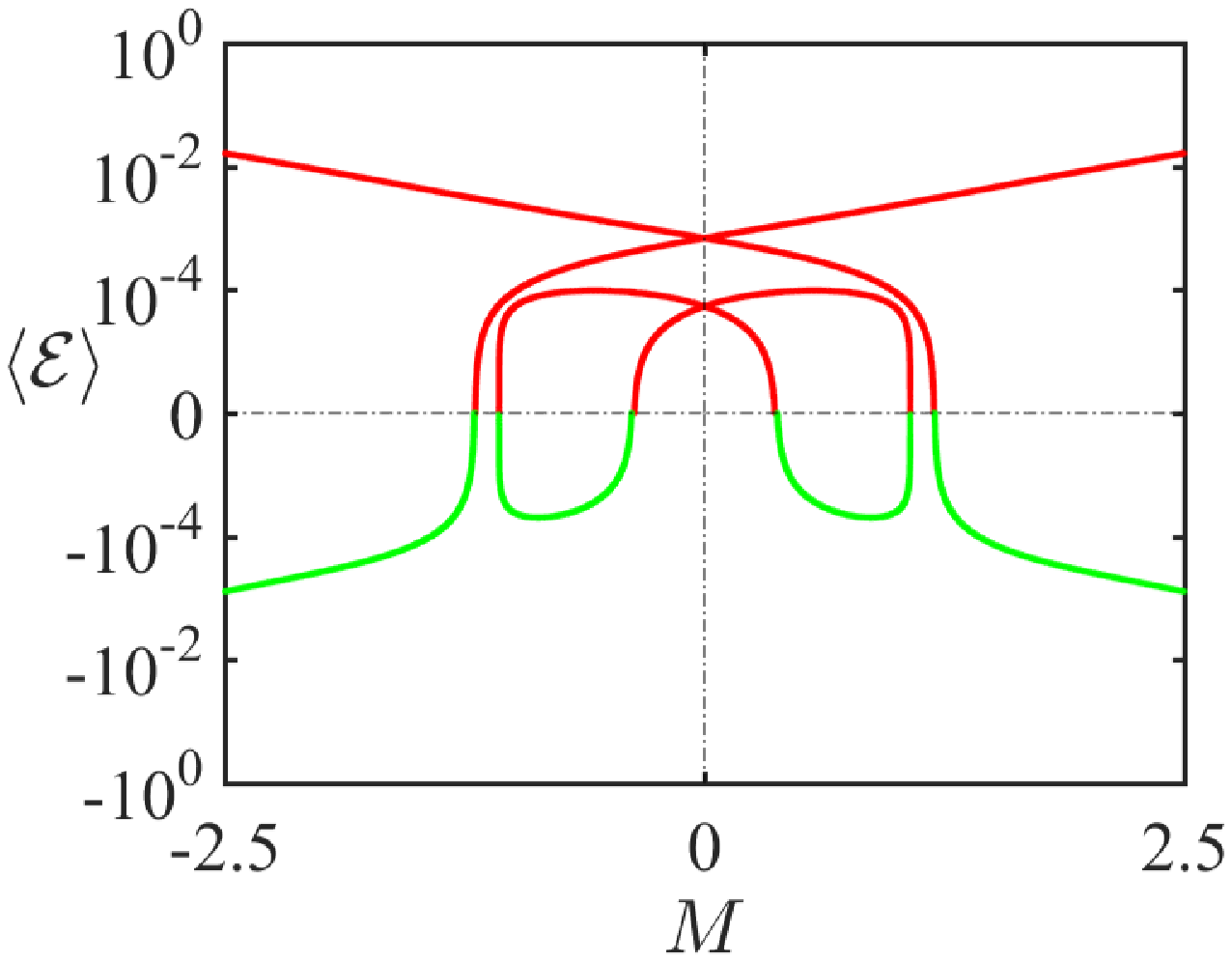}
    \caption{} \label{fig:9d}
  \end{subfigure}\\
  \begin{subfigure}{0.25\textwidth}
    \includegraphics[width=\textwidth]{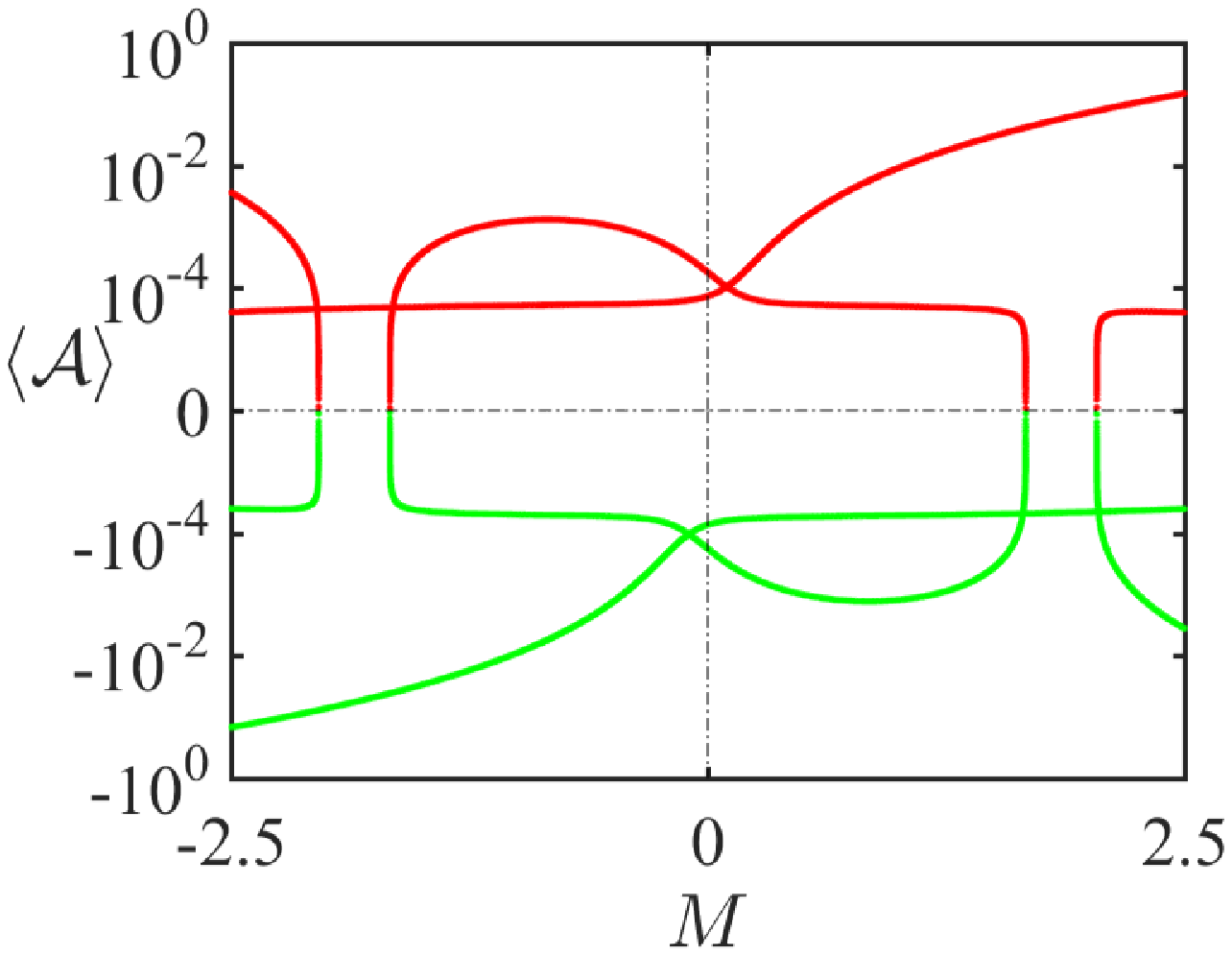}
    \caption{} \label{fig:9e}
  \end{subfigure}%
  \begin{subfigure}{0.25\textwidth}
    \includegraphics[width=\textwidth]{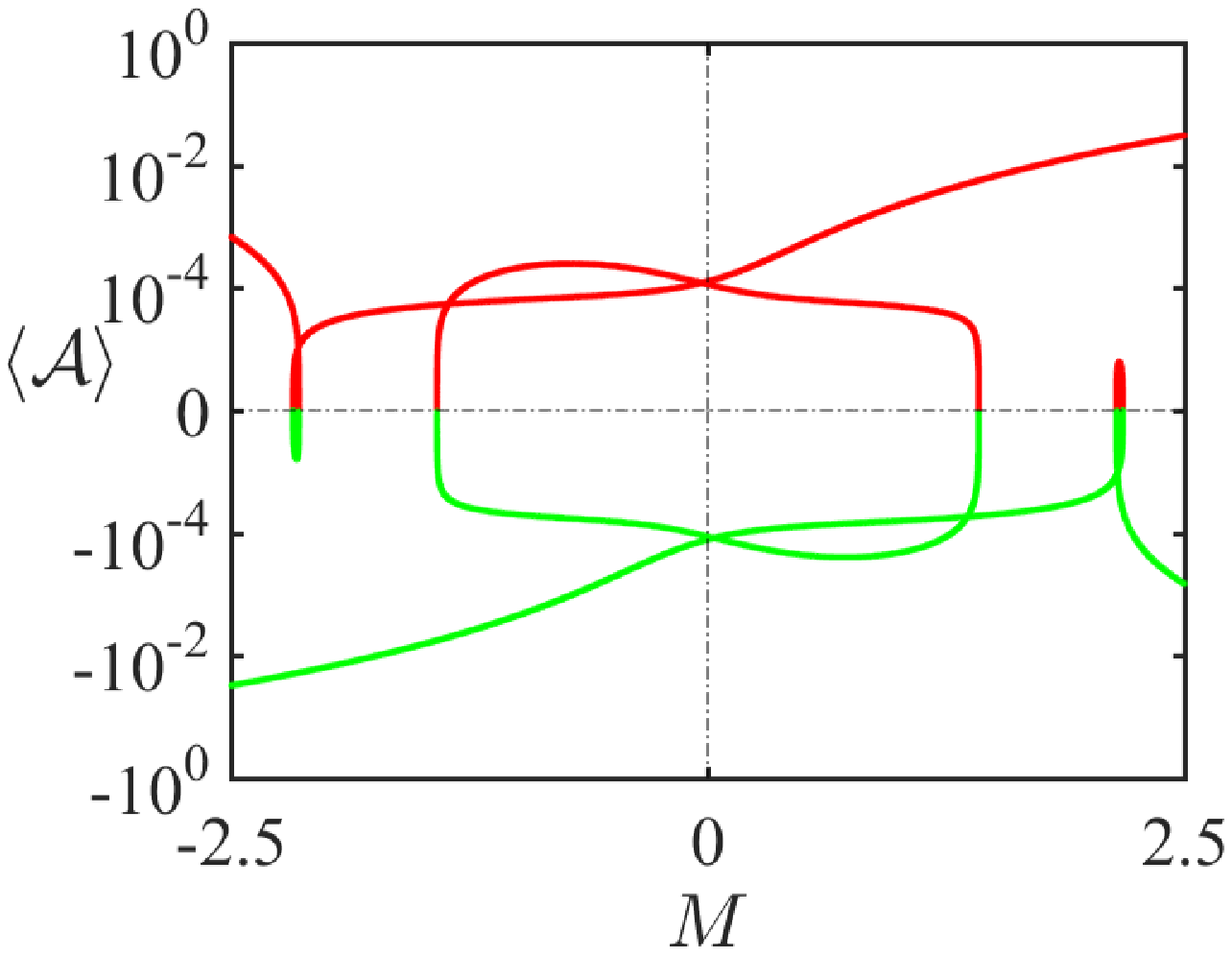}
    \caption{} \label{fig:9f}
  \end{subfigure}%
  \begin{subfigure}{0.25\textwidth}
    \includegraphics[scale=.225]{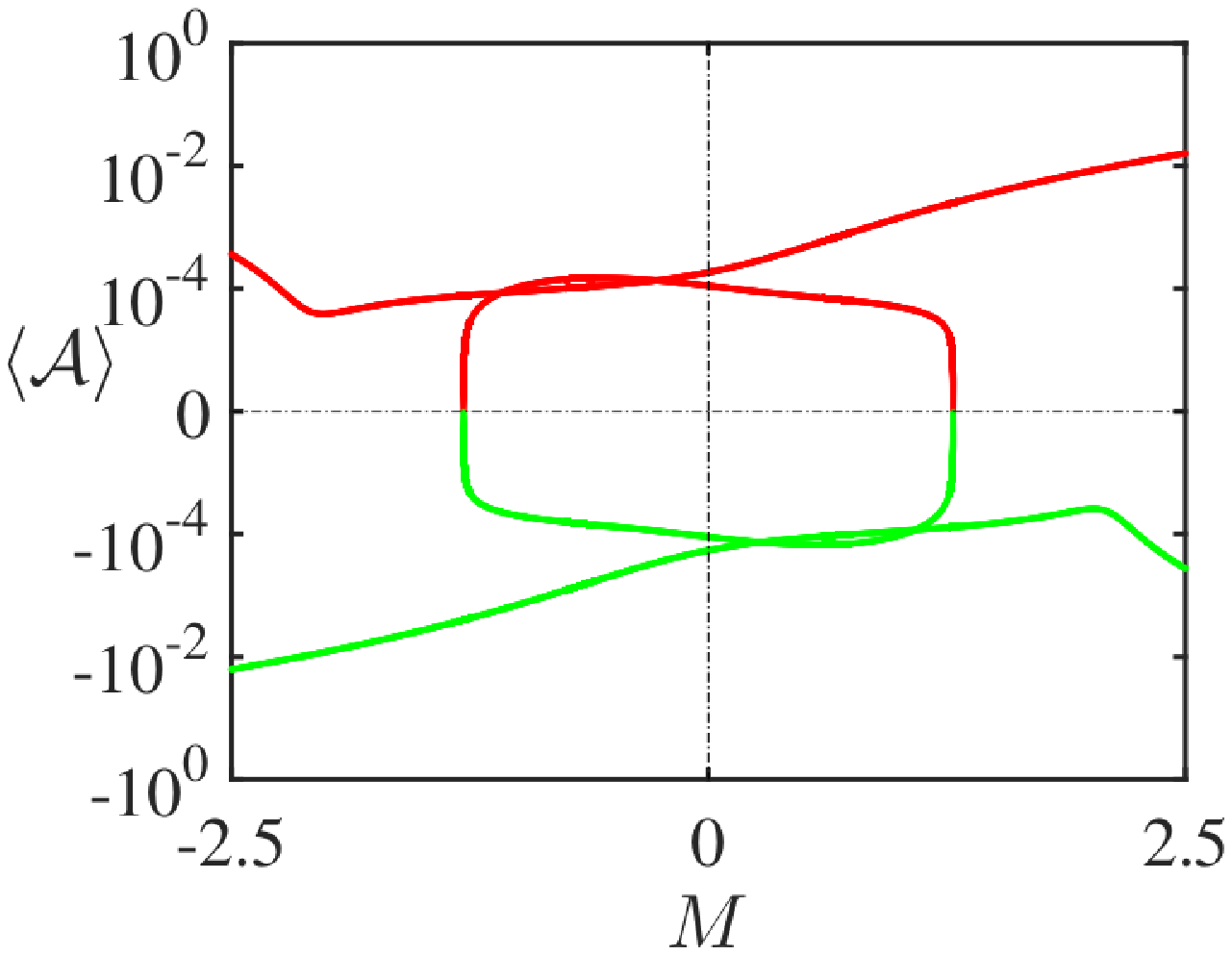}
    \caption{} \label{fig:9g}
  \end{subfigure}%
  \begin{subfigure}{0.25\textwidth}
    \includegraphics[width=\textwidth]{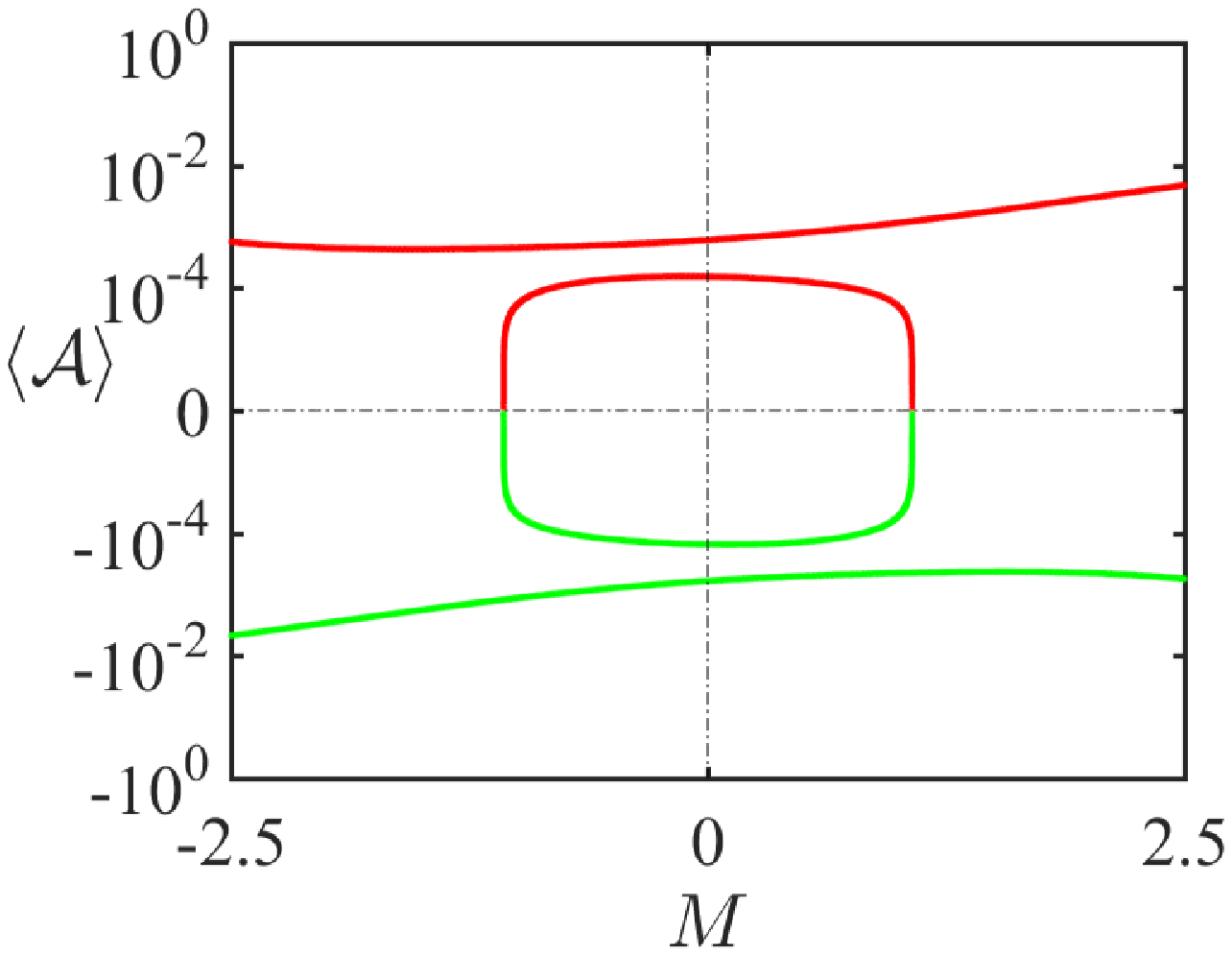}
    \caption{} \label{fig:9h}
  \end{subfigure}%

\caption{The averaged wave energy (upper panels) $\left\langle\mathcal{E}\right\rangle$ given by the expression \rf{e_axi1} and the action (lower panels) $\left\langle\mathcal{A}\right\rangle = \left\langle\mathcal{E}\right\rangle / \omega$ over the Mach number $M$ evaluated for $M_w=1$, $\kappa=1$, $\hat{\xi}=0.01$ and: (a, e) $\alpha=0.1$, (b, f) $\alpha=0.5$, (c, g) $\alpha=1$ and (d, h) $\alpha=5$. Positive (respectively negative) energy/action is represented in red (respectively green). \label{fig:energy}}
\end{figure}

After summing up all the different terms given by equations \rf{kf_a2} and \rf{av_ene} we obtain the total averaged energy
\be{e_a}
\left\langle \mathcal{E} \right\rangle  = \frac{1}{2} \left\lbrace \kappa^2 M_w^2 \hat{\xi}^2 +  \alpha \hat{\eta}^2 + i\alpha\kappa M \left[ \hat{\eta}\hat{\phi}(1) - \hat{\xi}\hat{\phi}(0) \right] \right\rbrace,
\ee
thus providing an extension to the case when the velocity field contains a background flow ($M\ne 0$).

A more suitable expression for the mean total energy can be obtained by expressing the different amplitudes of the system in \rf{e_a} in terms of a unique one, for instance, $\hat{\xi}$. From the kinematic condition \rf{kc} on the free surface with the plane wave solution \rf{ansatz} and the coefficients \rf{ab}, it is straightforward to express the surface amplitude $\hat{\eta}$ as
\be{eta}
\hat{\eta} = \frac{i\kappa}{(\omega - \kappa M)}\left[ Ae^{\kappa} - Be^{-\kappa} \right] = \frac{(\omega - \kappa M)^2\hat{\xi}}{(\omega-\kappa M)^2\cosh{\kappa} - \kappa\sinh{\kappa}}.
\ee

Substituting \rf{eta} into \rf{e_a} and using the complex amplitude $\hat{\phi}(z)$ recovered from the boundary value problem \rf{rbvp}, we find
\ba{e_axi}
\left\langle \mathcal{E} \right\rangle  &=& \frac{\hat{\xi}^2}{2}\left\lbrace \kappa^2M_w^2 + \alpha\frac{(\omega - \kappa M)^4(1-(\tanh{\kappa})^2)}{\left[(\omega-\kappa M)^2 - \kappa\tanh{\kappa})\right]^2} \right. \nn \\
&&\left. + \alpha M (\omega - \kappa M) \frac{\left[(\omega -\kappa M)^4 + \kappa^2\right]\tanh{\kappa} - 2\kappa(\omega -\kappa M)^2(\tanh{\kappa})^2}{\left[(\omega-\kappa M)^2 - \kappa\tanh{\kappa})\right]^2} \right\rbrace .
\ea

Next, expressing the term $\kappa^2M_w^2$ by means of the dispersion relation \rf{domba} and substituting the result into \rf{e_axi} yields a more compact formula for the total energy:
\be{e_axi1}
\left\langle \mathcal{E} \right\rangle  =\frac{1}{4}\omega\left\lbrace 2\omega+\frac{\alpha}{\kappa}\frac{2(\omega-\kappa M)\tanh\kappa\left[(\omega-\kappa M)^4+\kappa^2-2\kappa(\omega-\kappa M)^2\tanh\kappa\right]}{[(\omega-\kappa M)^2-\kappa\tanh\kappa]^2}\right\rbrace\hat{\xi}^2.
\ee

Notice that the term in the braces in Eq.~\rf{e_axi1} is nothing else but the partial derivative ${\partial\mathcal{D}}/{\partial\omega}$ of the
dispersion relation \rf{domba} written in the following equivalent form
\be{dr_cairns}
\mathcal{D}(\omega,\kappa) :=  \mathcal{D}_m(\omega,\kappa) + \frac{\alpha}{\kappa} \frac{(\omega - \kappa M)^2[(\omega - \kappa M)^2\tanh{\kappa} - \kappa]}{\mathcal{D}_f(\omega,\kappa)}=0,
\ee
where $\mathcal{D}_m = \omega^2-\kappa^2M_w^2 $ and $\mathcal{D}_f = [(\omega - \kappa M)^2 - \kappa\tanh{\kappa}]$ stand for the dispersion relation of, respectively, the free membrane and the free surface flow with a rigid boundary at the bottom. This proves that our total energy per unit area, averaged over the wave period, possesses the following simple representation in terms of the dispersion relation
\be{e_cairns}
\left\langle \mathcal{E} \right\rangle = \frac{1}{4} \omega \frac{\partial\mathcal{D}}{\partial\omega} \hat{\xi}^2.
\ee


\begin{figure}
  \begin{subfigure}{0.33\textwidth}
    \includegraphics[width=\textwidth]{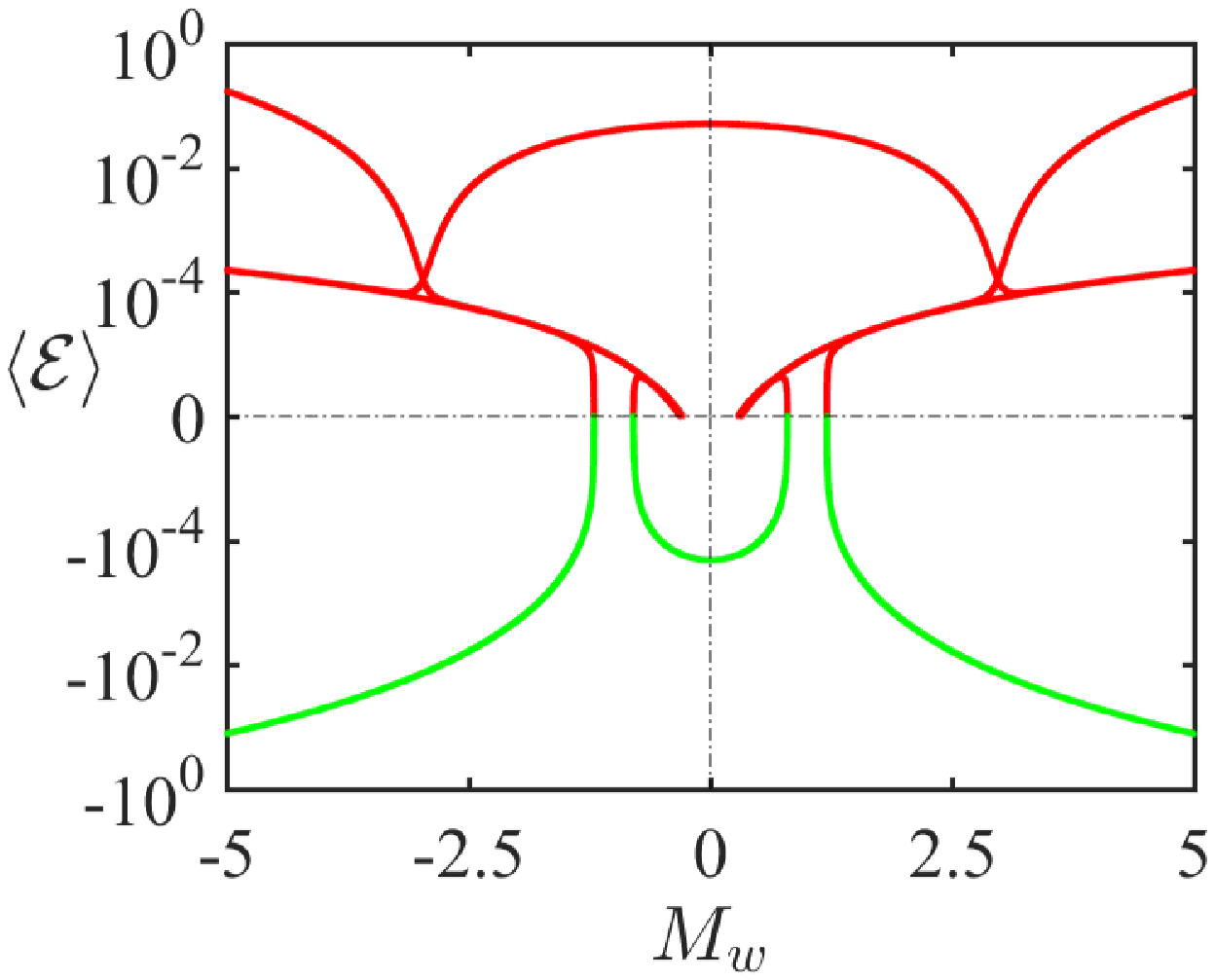}
    \caption{} \label{fig:10a}
  \end{subfigure}%
  \begin{subfigure}{0.33\textwidth}
    \includegraphics[width=\textwidth]{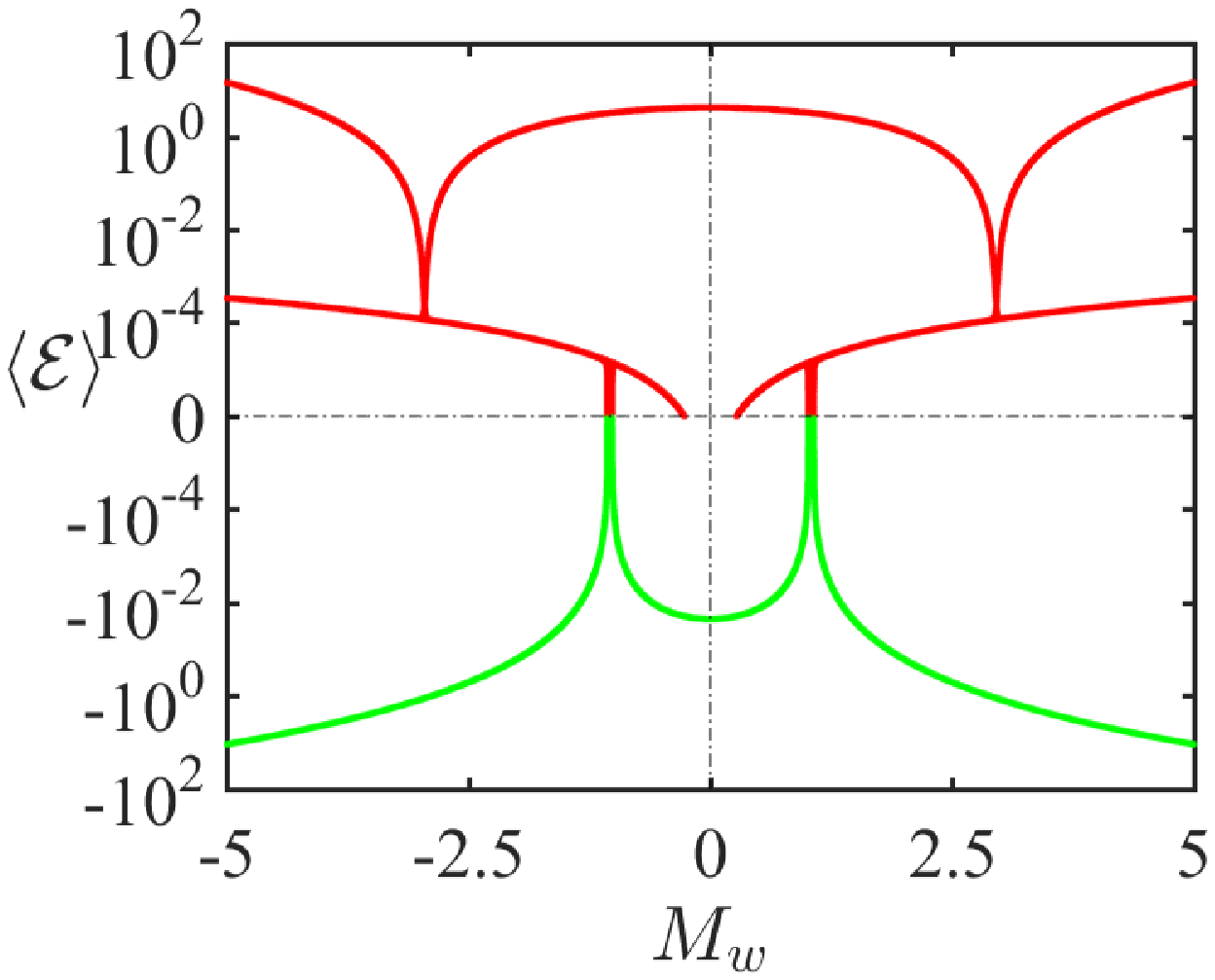}
    \caption{} \label{fig:10b}
  \end{subfigure}%
  \begin{subfigure}{0.33\textwidth}
    \includegraphics[width=\textwidth]{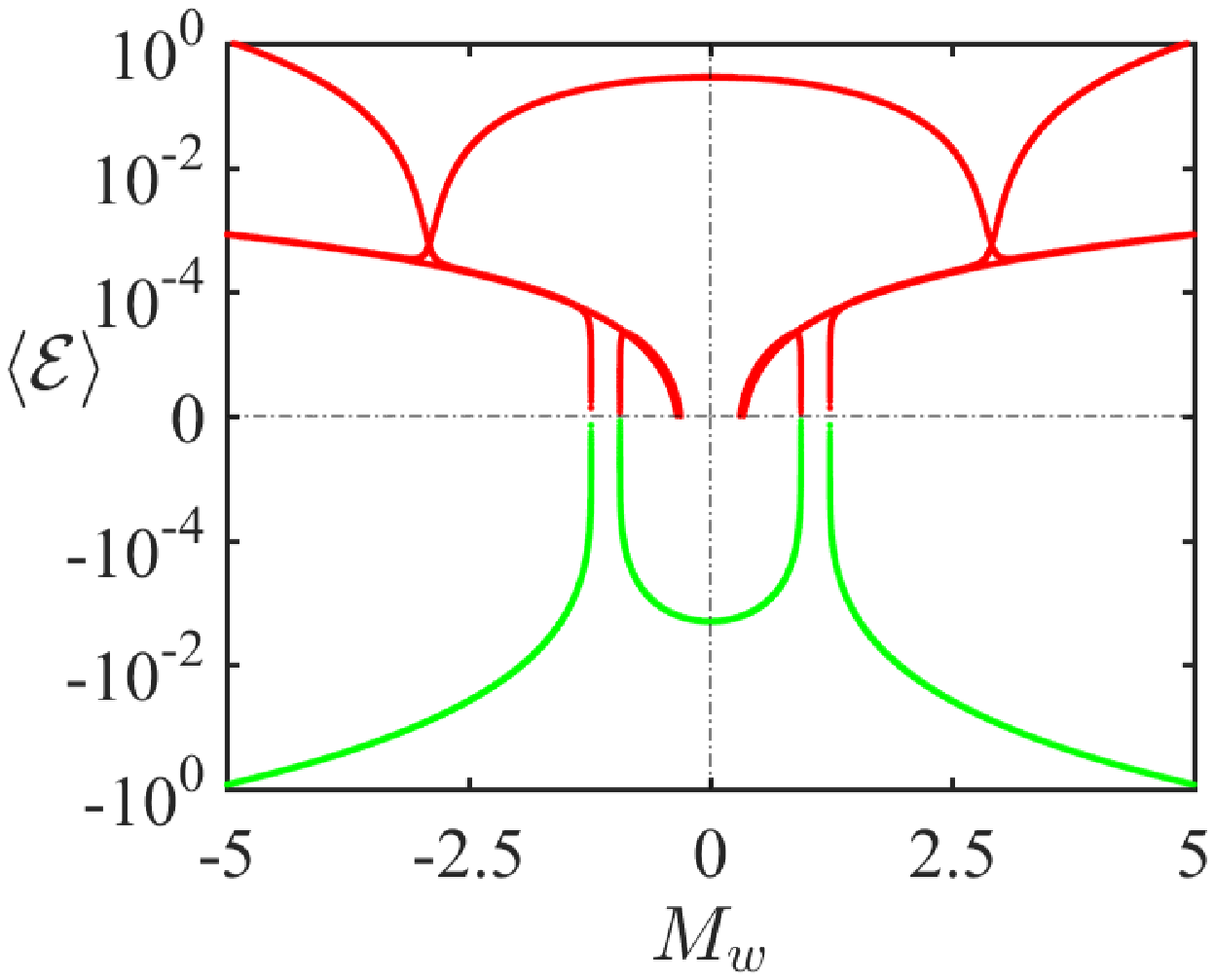}
    \caption{} \label{fig:10c}
  \end{subfigure}\\
  \begin{subfigure}{0.33\textwidth}
    \includegraphics[width=\textwidth]{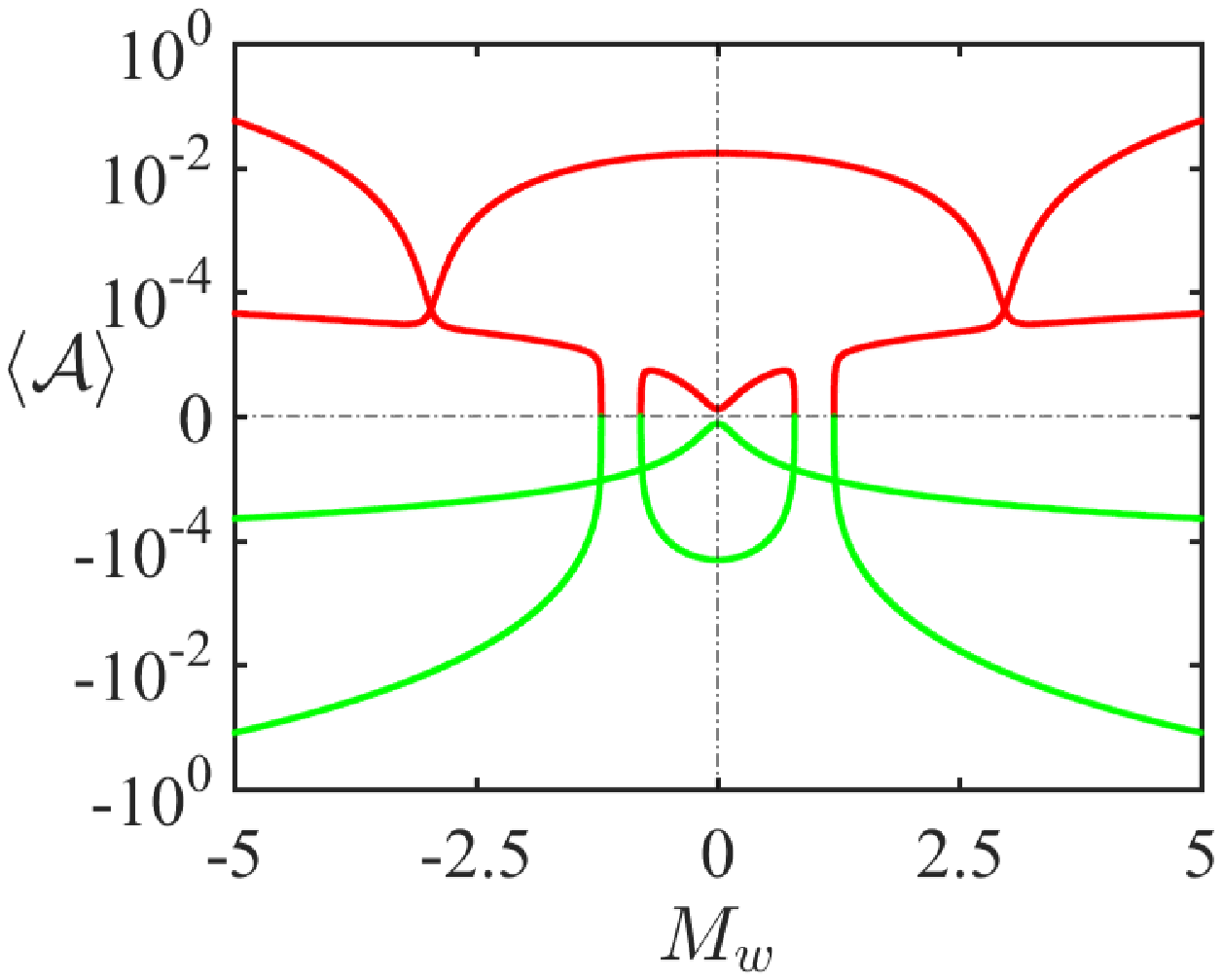}
    \caption{} \label{fig:10d}
  \end{subfigure}%
  \begin{subfigure}{0.33\textwidth}
    \includegraphics[width=\textwidth]{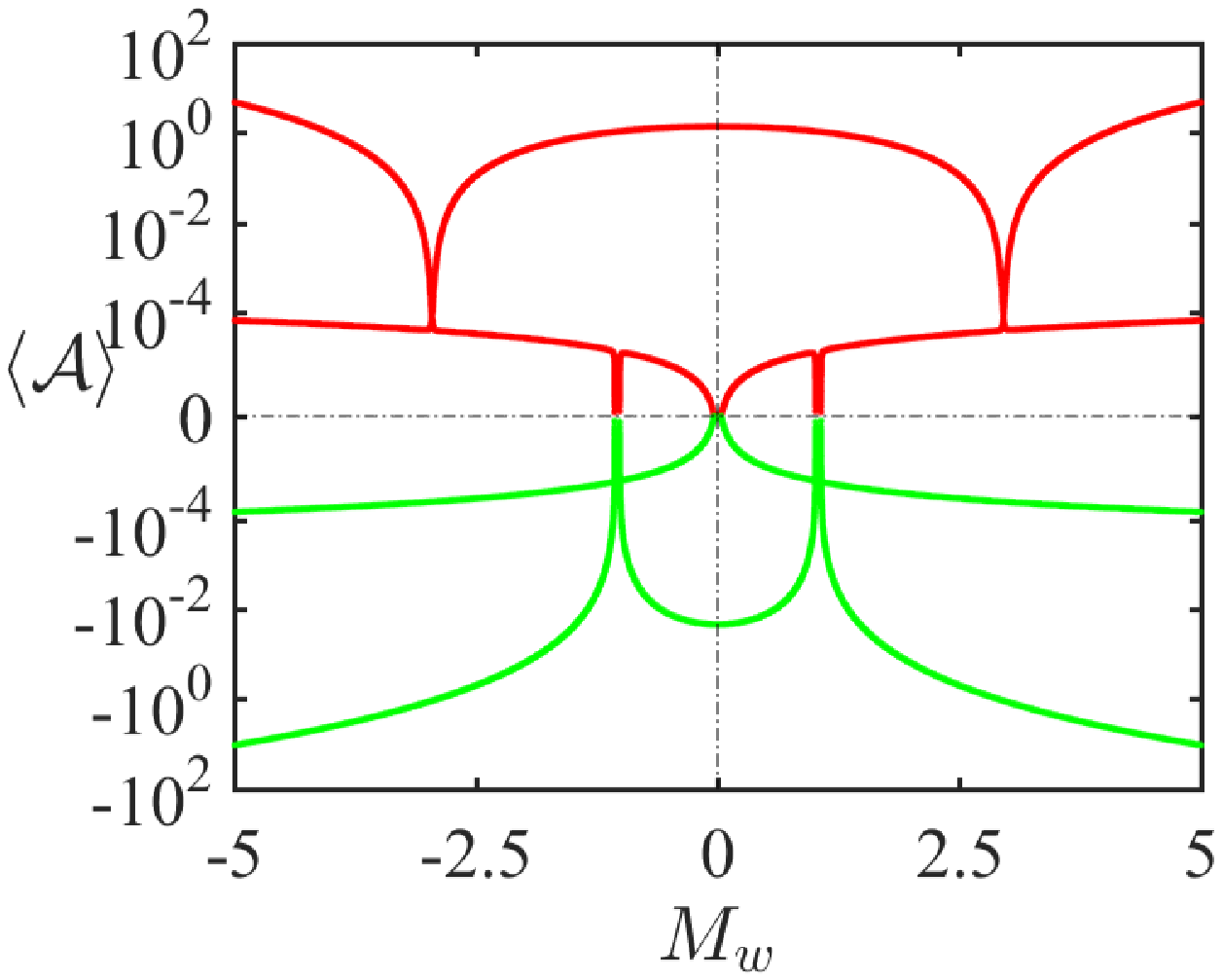}
    \caption{} \label{fig:10e}
  \end{subfigure}%
  \begin{subfigure}{0.33\textwidth}
    \includegraphics[width=\textwidth]{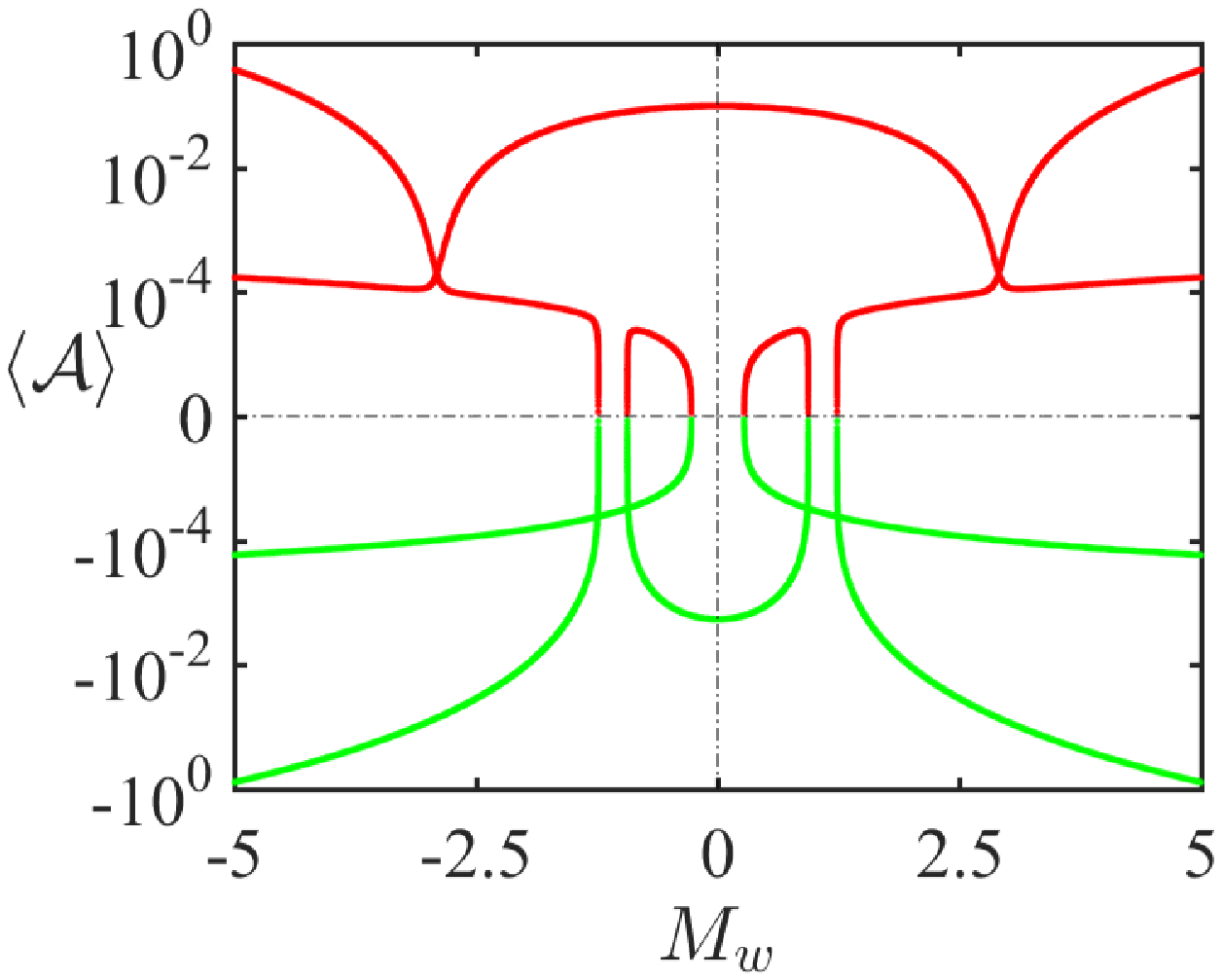}
    \caption{} \label{fig:10f}
  \end{subfigure}%

  \caption{The averaged wave energy (upper panels) $\left\langle\mathcal{E}\right\rangle$ given by the expression \rf{e_axi1} and the action (lower panels) $\left\langle\mathcal{A}\right\rangle = \left\langle\mathcal{E}\right\rangle / \omega$ over the Mach number $M_w$ for $\hat \xi=0.01$, $M=M_0=2$, and (a, d) $\beta=0.05$ and $\kappa=\kappa_0-0.1$, (b, e)  $\beta=10^{-3}$ and $\kappa=\kappa_0\approx 0.5218134478$, (c, f) $\beta=0.05$ and $\kappa=\kappa_0+0.3$. \label{energy_Mw}}
\end{figure}

The representation \rf{e_cairns} can be found, e.g. in \cite{C1979}, and can be derived in the frame of the general Lagrangian variational approach \cite{W1999,ORT1986}, see also the recent work by \cite{F2014} for historical notes and application to stability of vortices. Notice that according to \rf{e_cairns} the energy vanishes at the points where $\omega=0$ or ${\partial\mathcal{D}}/{\partial\omega}=0$, the latter condition corresponding to the existence of multiple roots of the dispersion relation.
Correspondingly, the ratio $\left\langle \mathcal{E} \right\rangle / \omega$, which is the averaged wave action $\left\langle \mathcal{A} \right\rangle $ \citep{W1999,HQ2016}, vanishes only at the locations of the multiple eigenvalues, cf. Fig.~\ref{fig:sigma_k_b} and Fig~\ref{fig:energy}.
In the latter figure as well as in Fig.~\ref{energy_Mw} we show several computations of the averaged wave energy  and wave action over the fluid Mach number $M$, and, respectively, the membrane Mach number $M_w$, where $\omega$ is calculated with the use of the dispersion relation \rf{domba}.


\begin{figure}
  \begin{subfigure}{0.33\textwidth}
    \includegraphics[width=\textwidth]{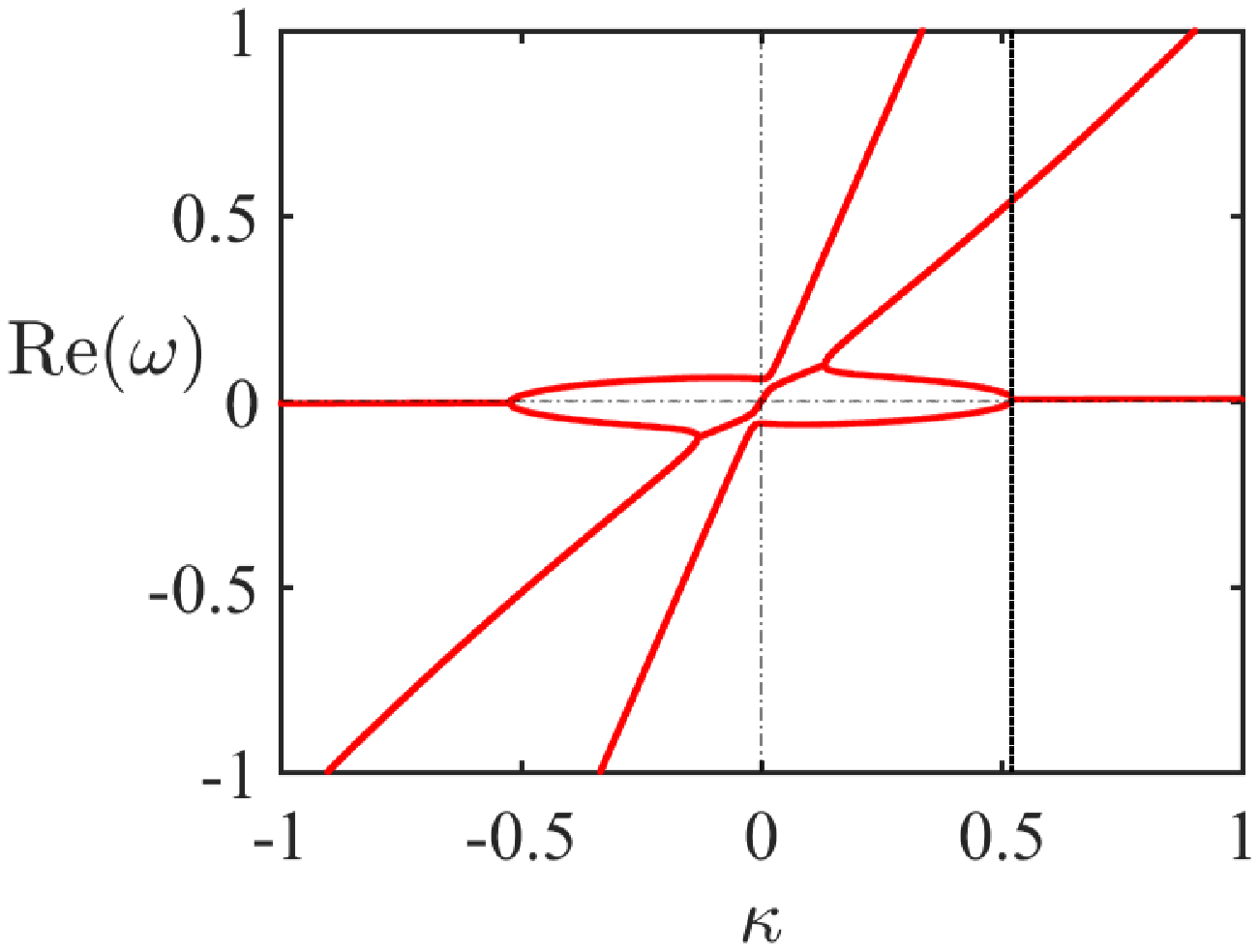}
    \caption{} \label{fig:11a}
  \end{subfigure}%
  \begin{subfigure}{0.33\textwidth}
    \includegraphics[width=\textwidth]{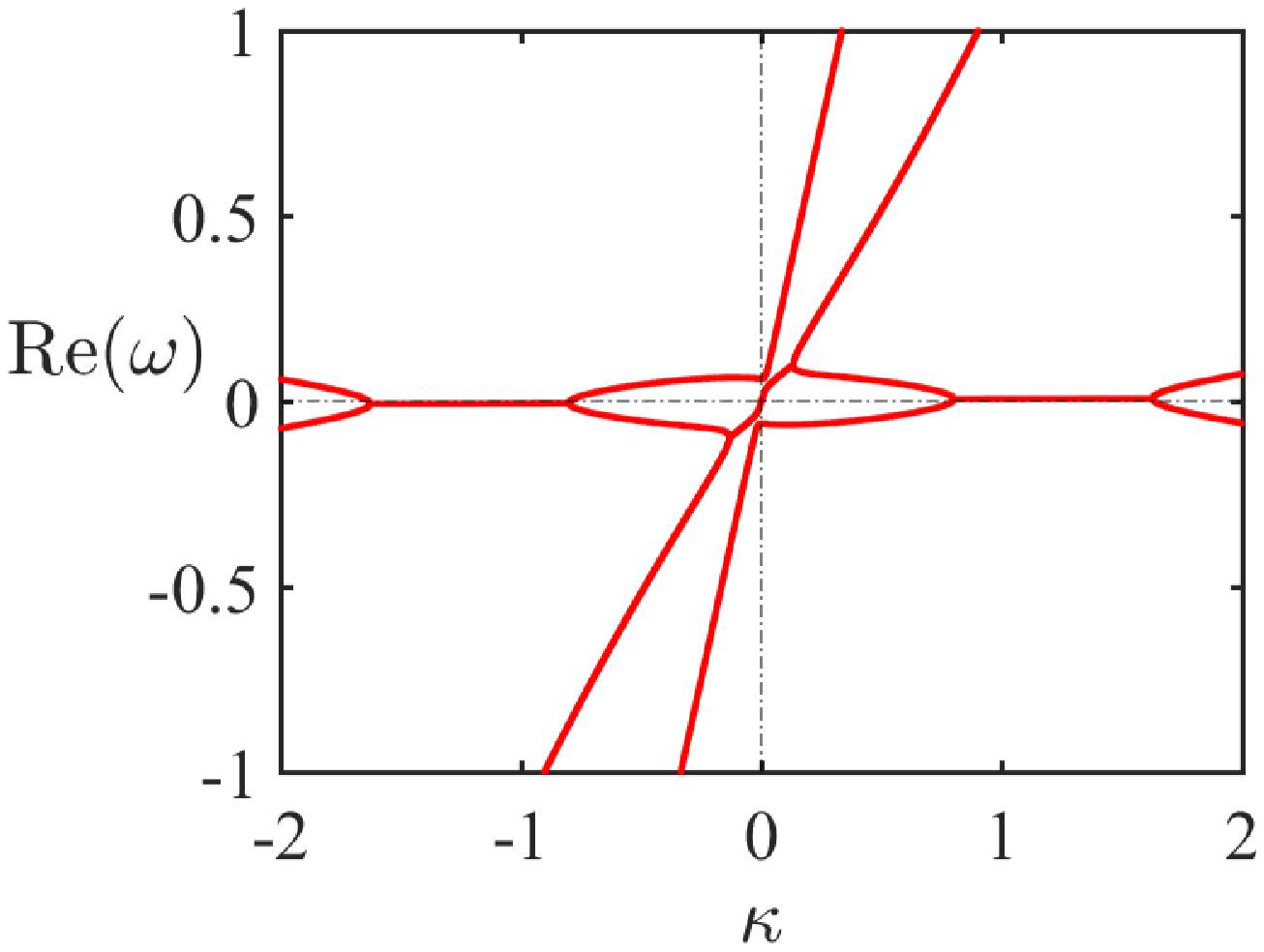}
    \caption{} \label{fig:11b}
  \end{subfigure}%
  \begin{subfigure}{0.33\textwidth}
    \includegraphics[width=\textwidth]{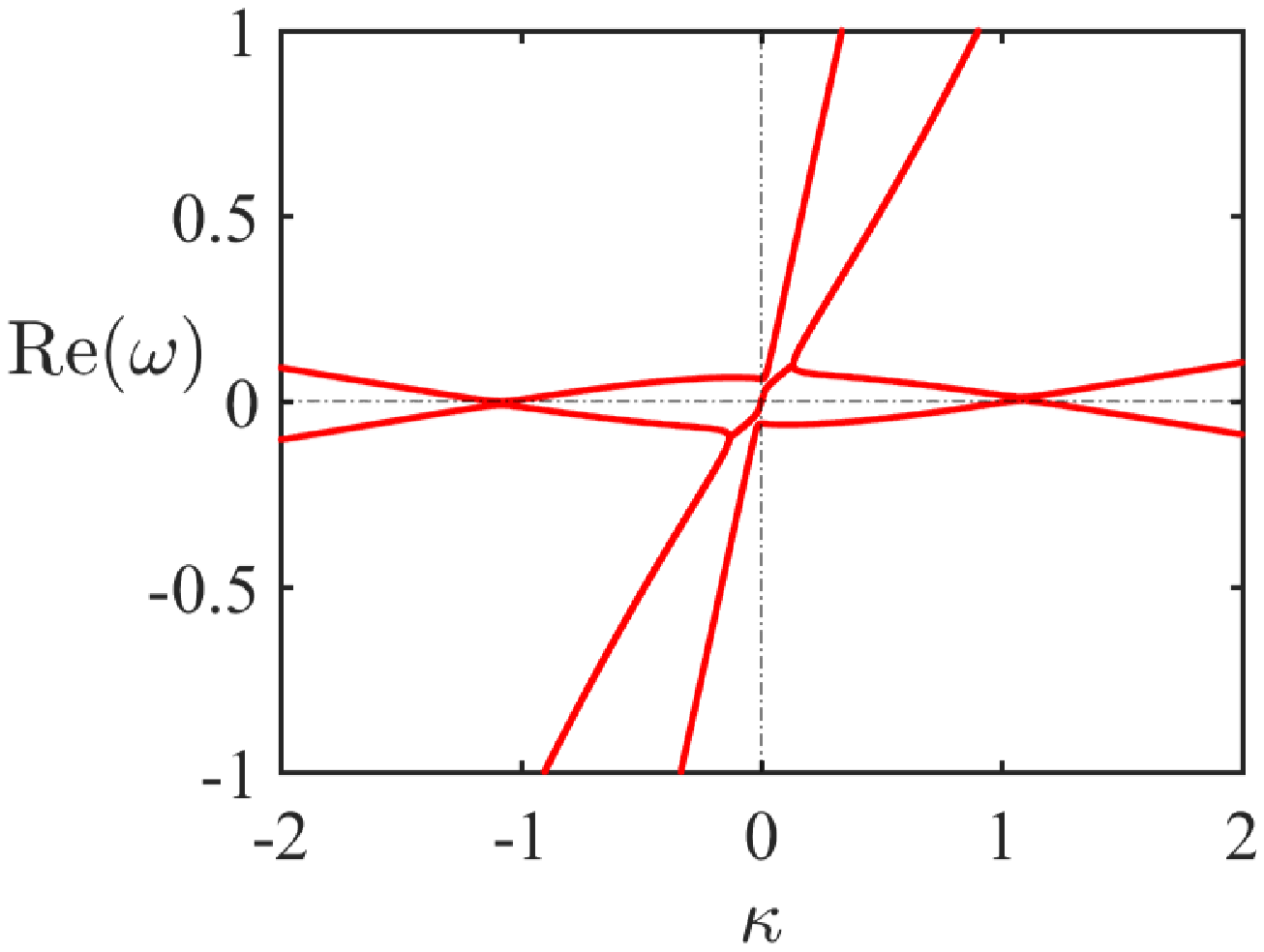}
    \caption{} \label{fig:11c}
  \end{subfigure}\\
  \begin{subfigure}{0.33\textwidth}
    \includegraphics[width=\textwidth]{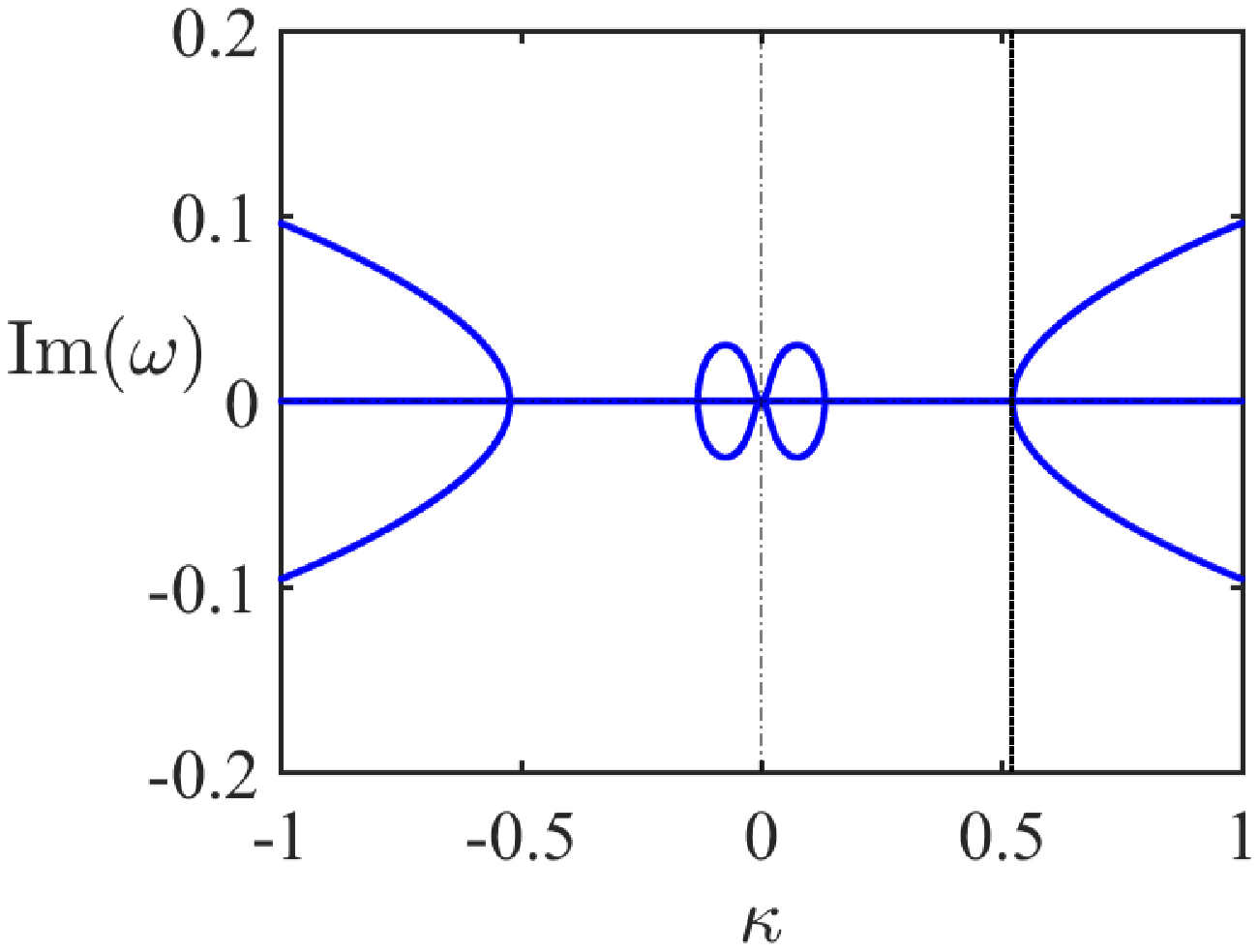}
    \caption{} \label{fig:11d}
  \end{subfigure}%
  \begin{subfigure}{0.33\textwidth}
    \includegraphics[width=\textwidth]{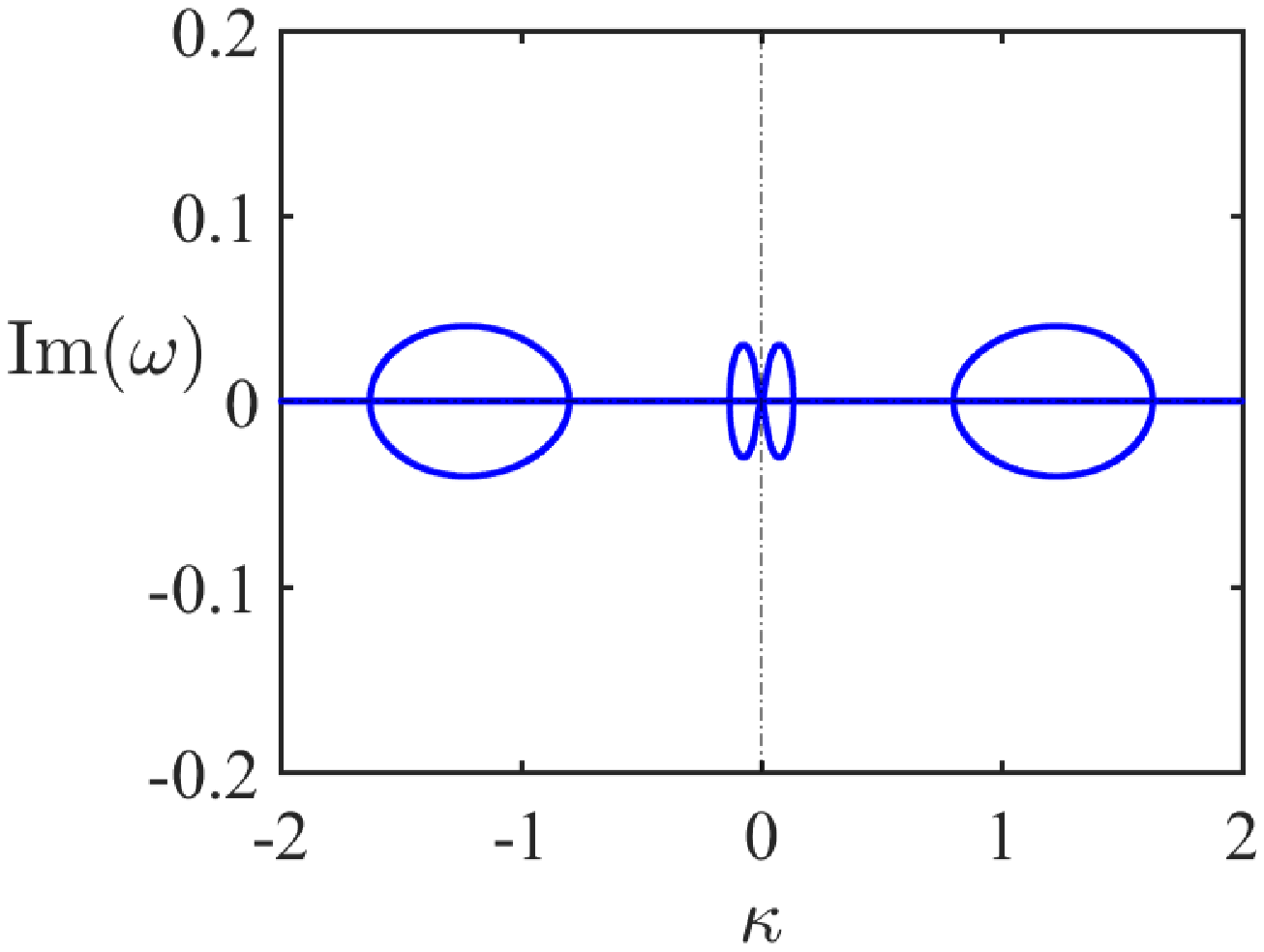}
    \caption{} \label{fig:11e}
  \end{subfigure}%
  \begin{subfigure}{0.33\textwidth}
    \includegraphics[width=\textwidth]{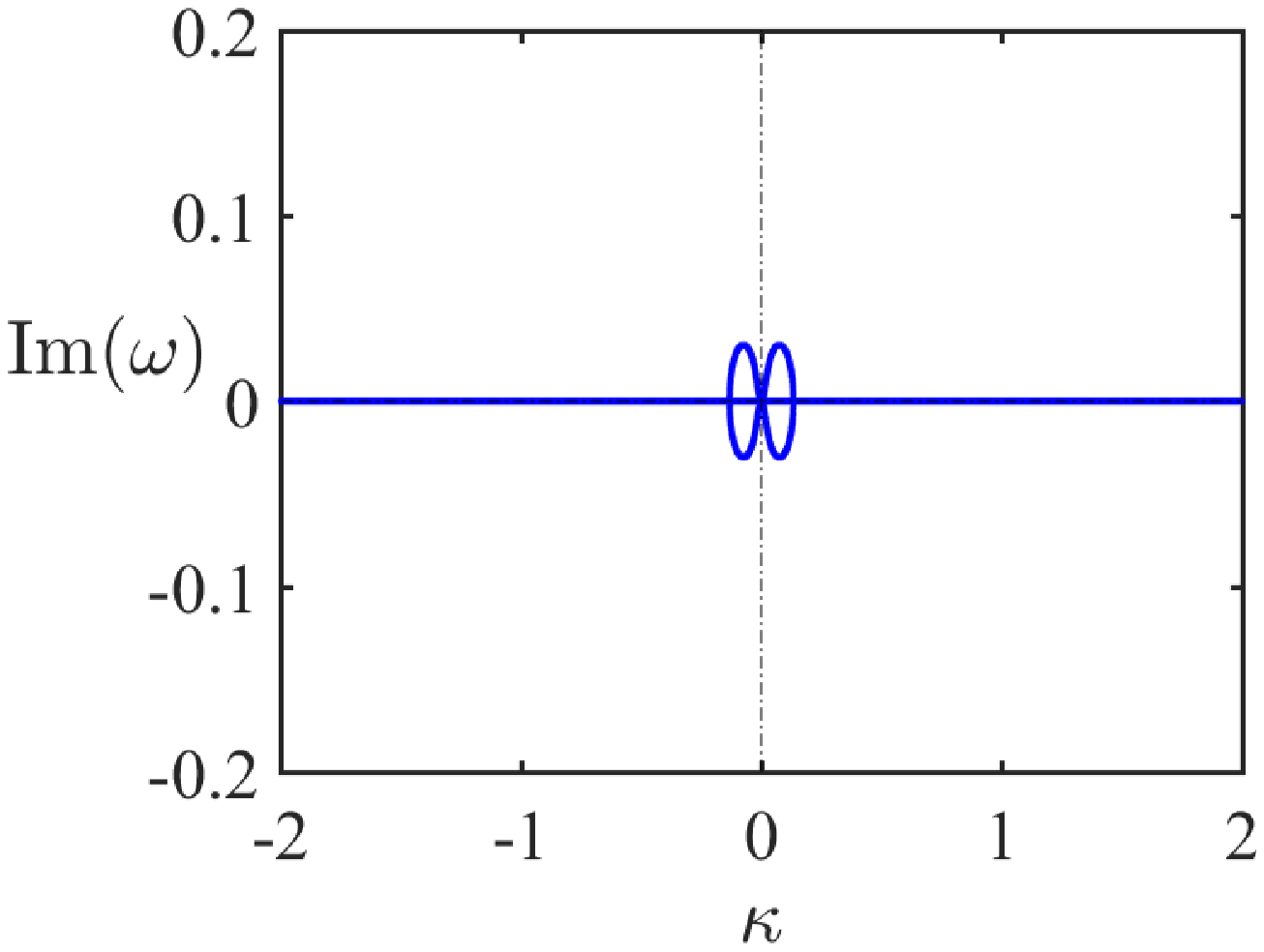}
    \caption{} \label{fig:11f}
  \end{subfigure}\\
  \begin{subfigure}{0.33\textwidth}
    \includegraphics[width=\textwidth]{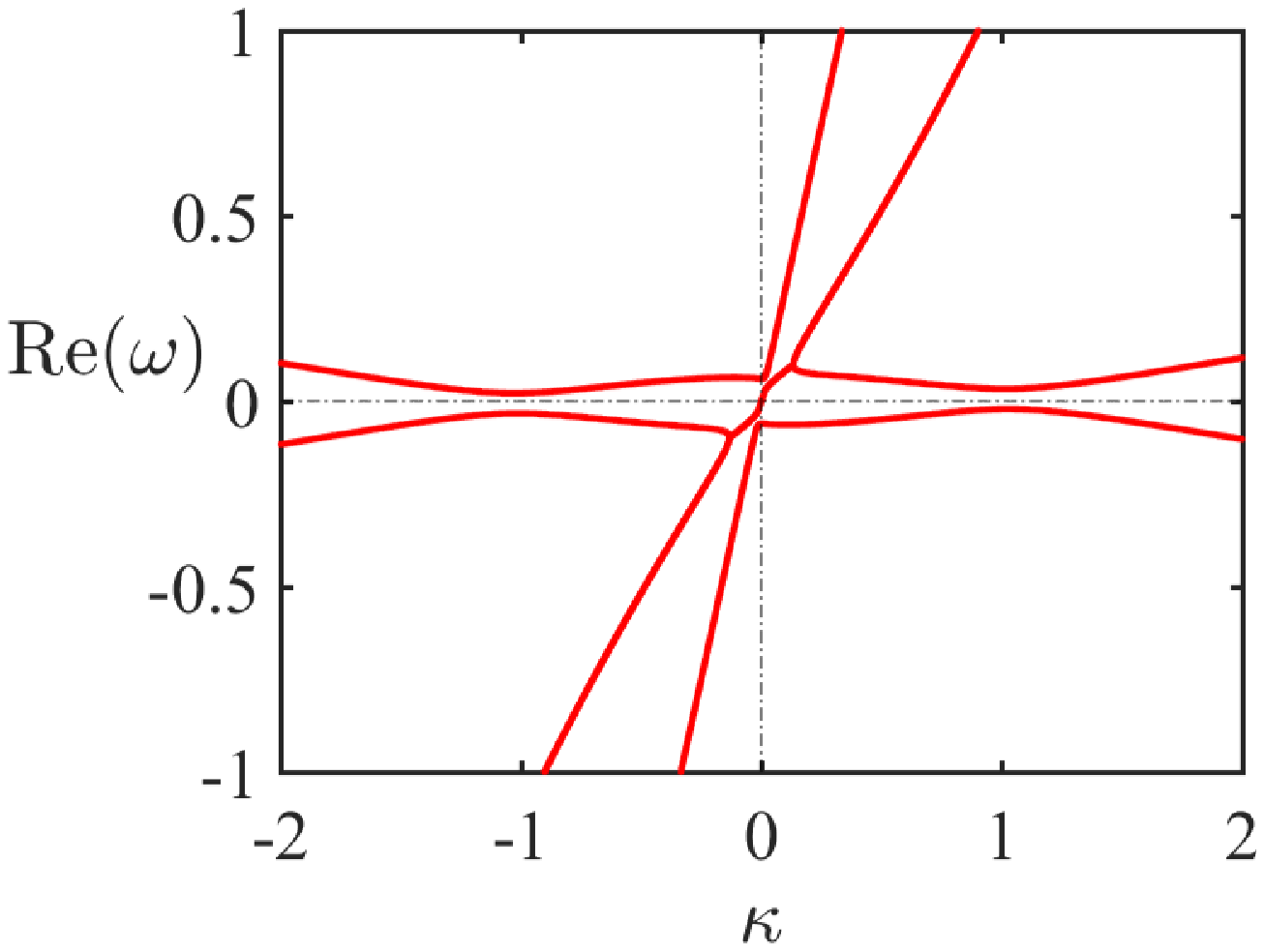}
    \caption{} \label{fig:11g}
  \end{subfigure}%
  \begin{subfigure}{0.33\textwidth}
    \includegraphics[width=\textwidth]{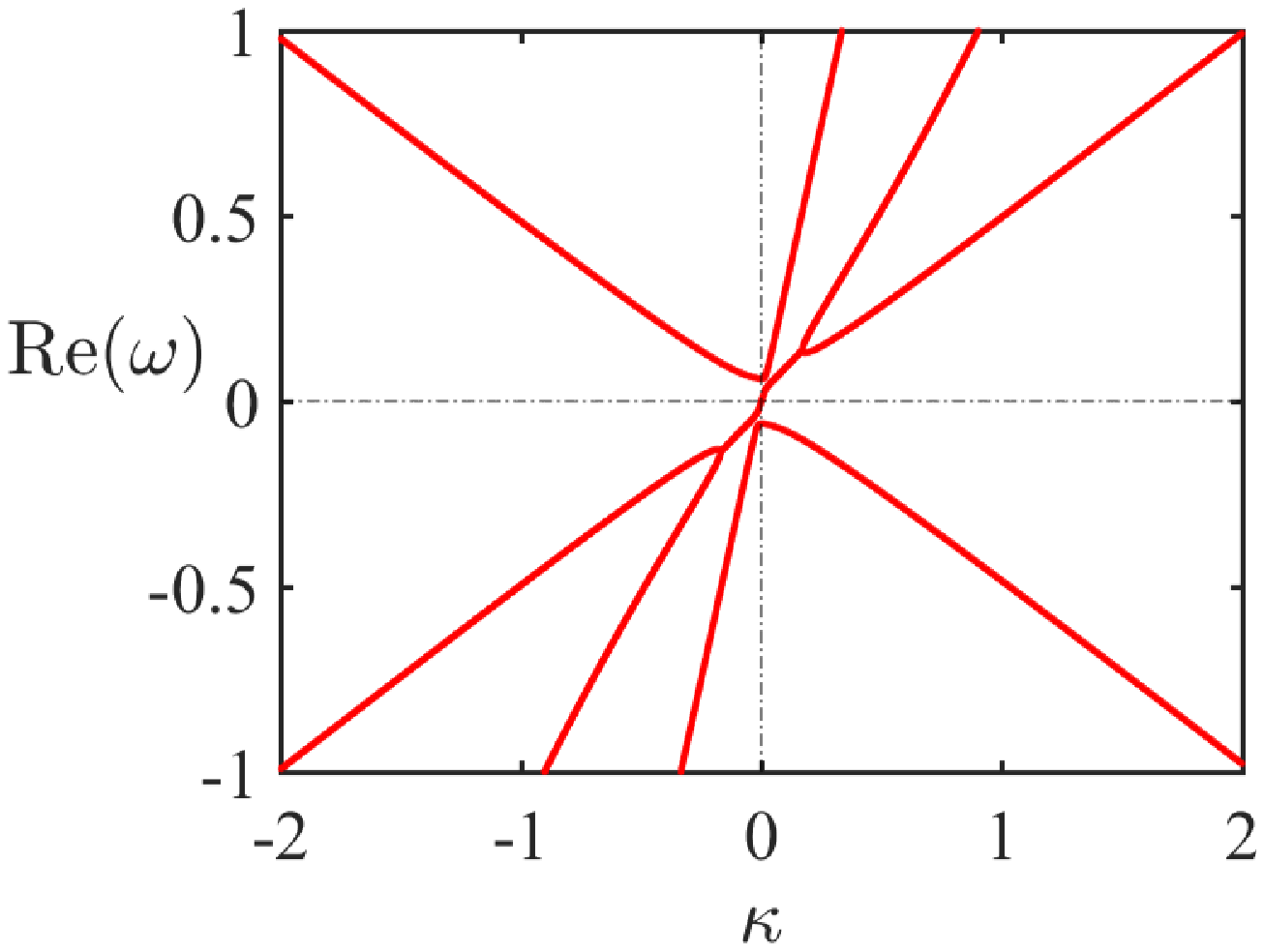}
    \caption{} \label{fig:11h}
  \end{subfigure}%
  \begin{subfigure}{0.33\textwidth}
    \includegraphics[width=\textwidth]{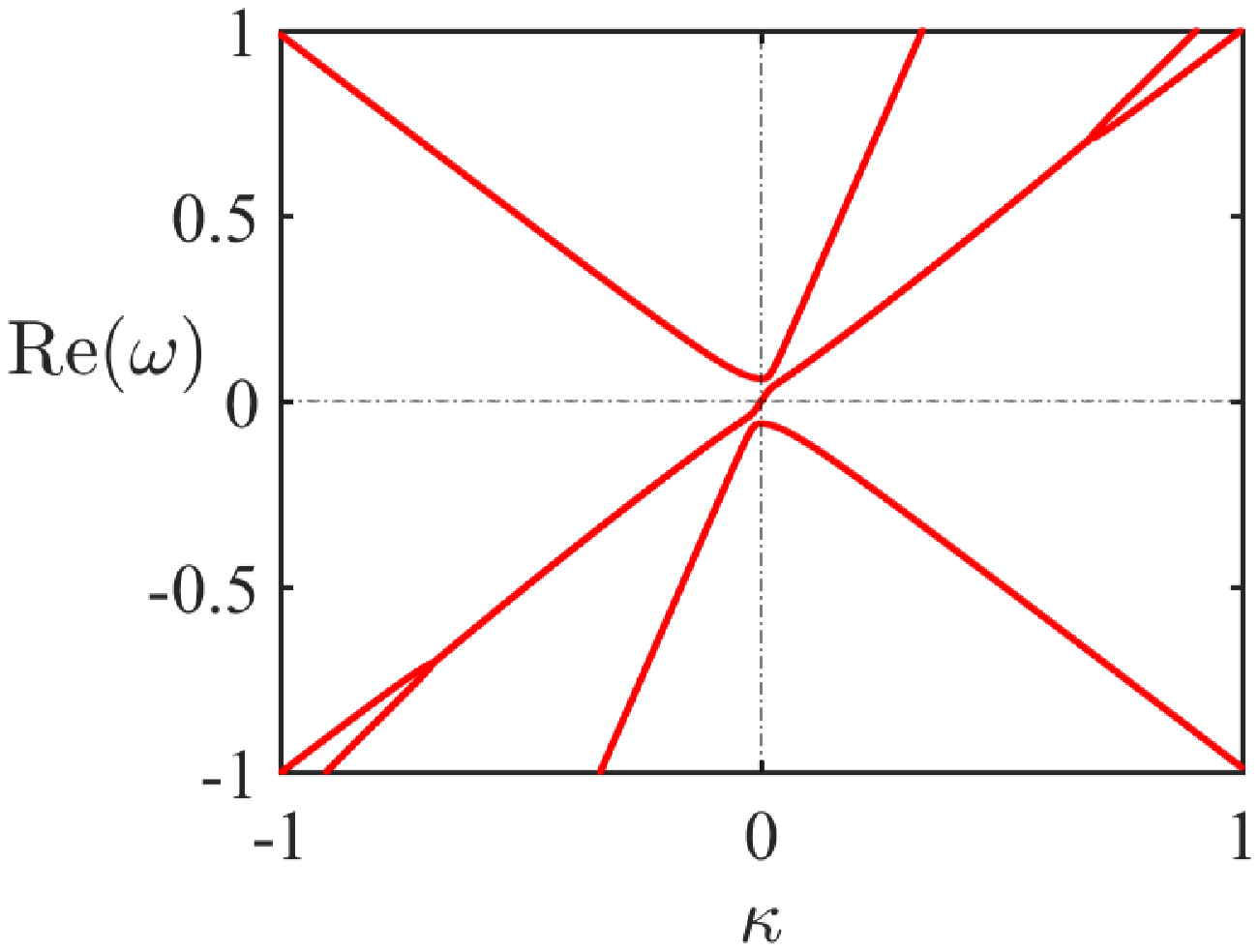}
    \caption{} \label{fig:11i}
  \end{subfigure}\\
  \begin{subfigure}{0.33\textwidth}
    \includegraphics[width=\textwidth]{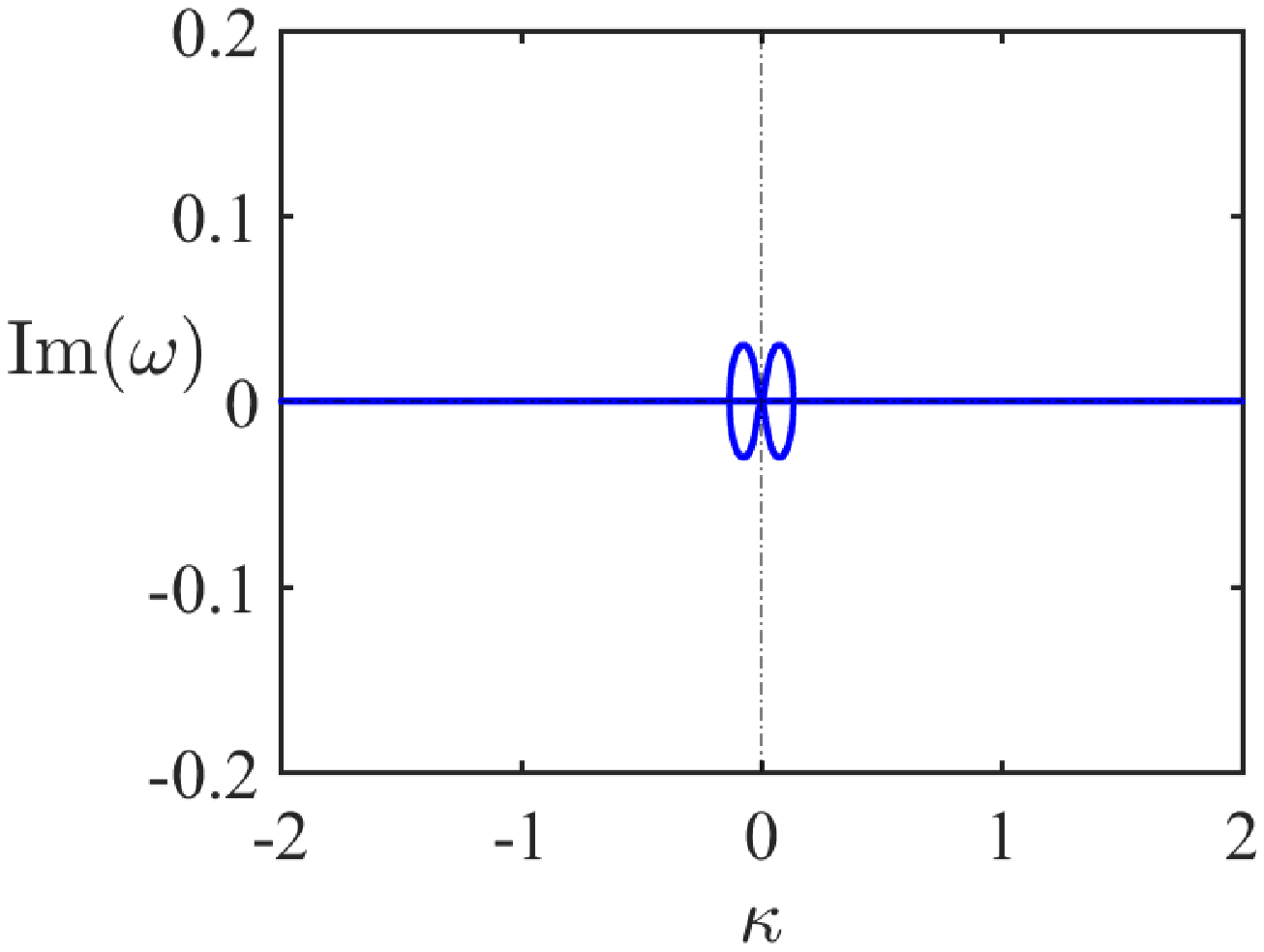}
    \caption{} \label{fig:11j}
  \end{subfigure}%
  \begin{subfigure}{0.33\textwidth}
    \includegraphics[width=\textwidth]{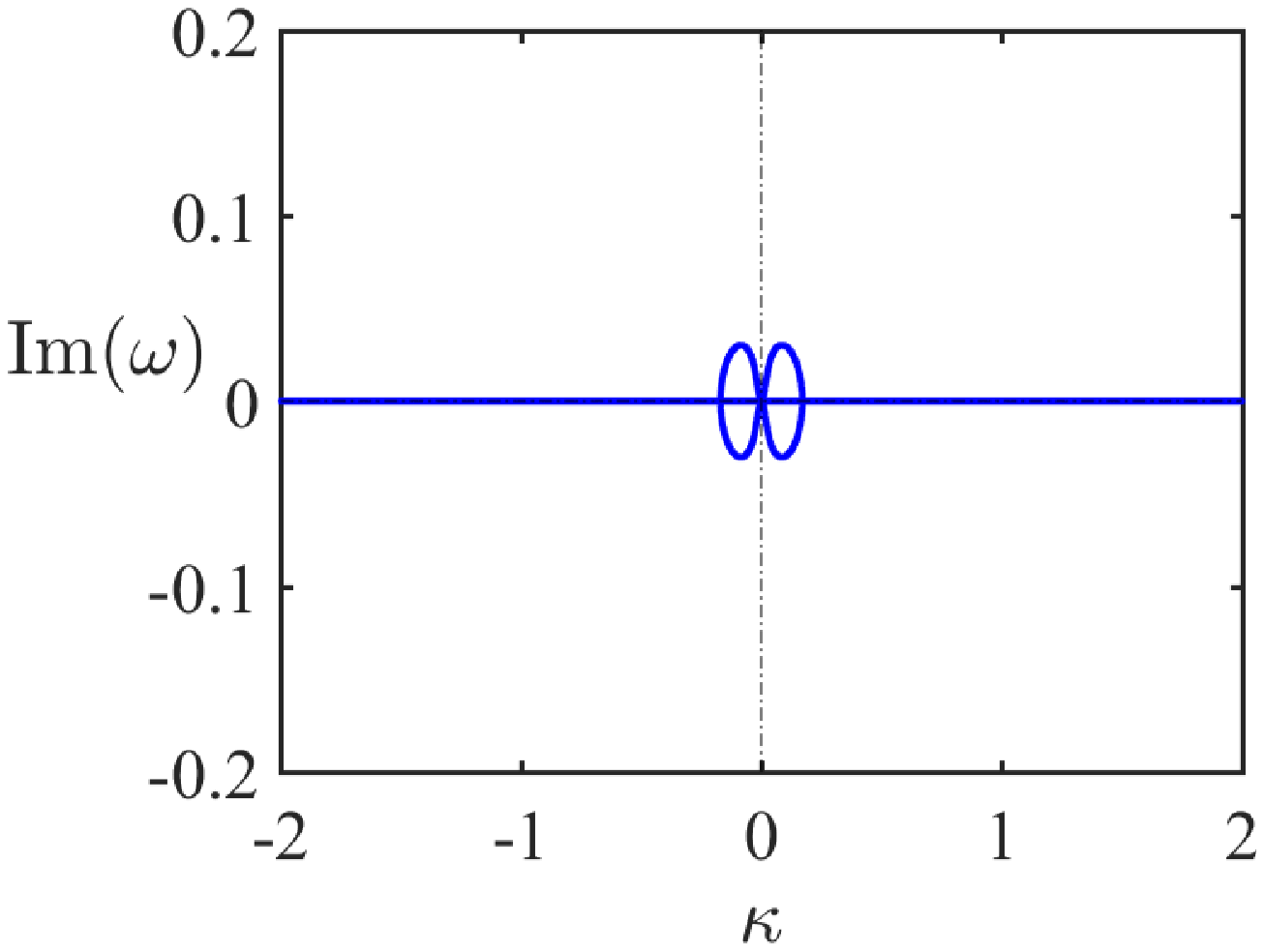}
    \caption{} \label{fig:11k}
  \end{subfigure}%
  \begin{subfigure}{0.33\textwidth}
    \includegraphics[width=\textwidth]{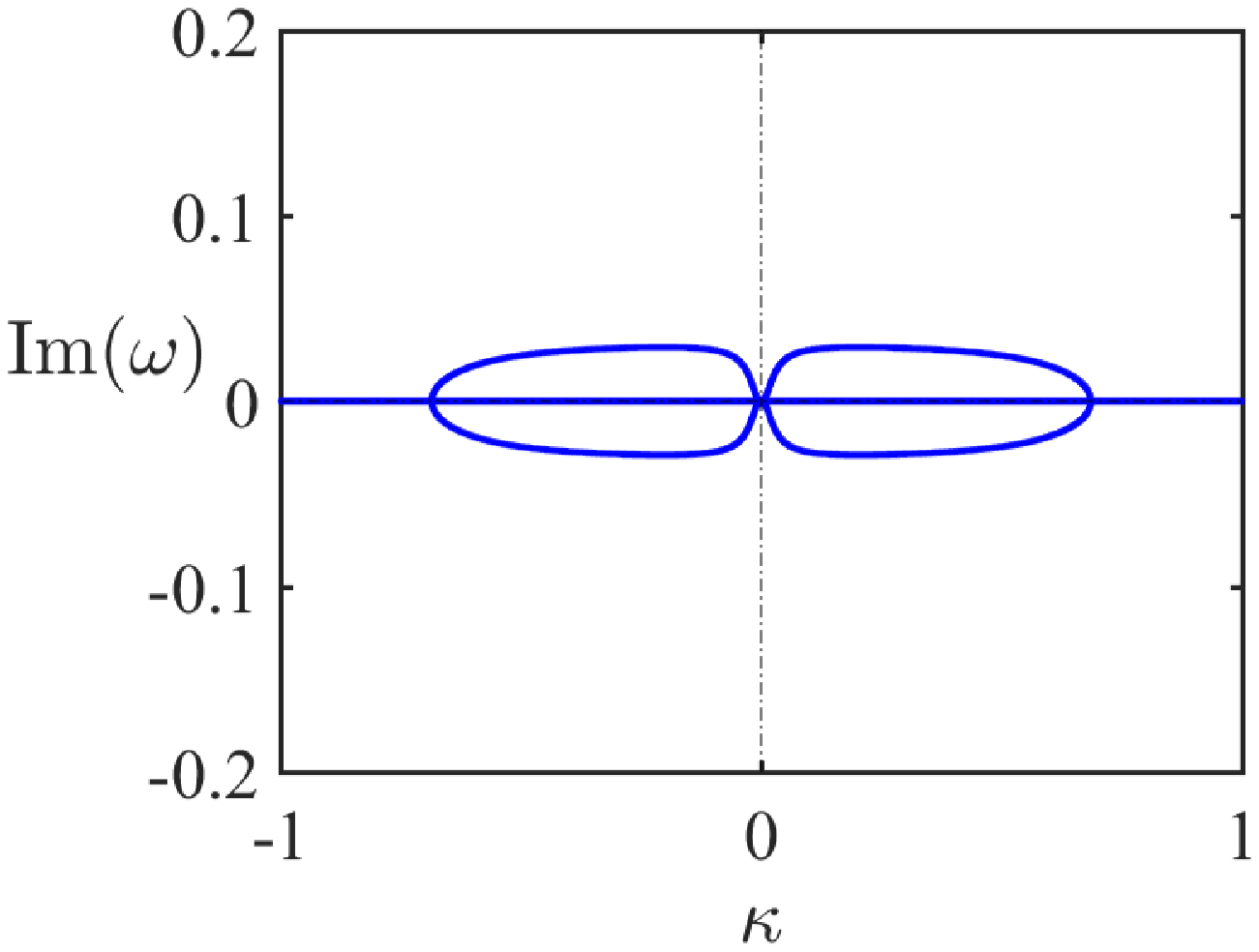}
    \caption{} \label{fig:11l}
  \end{subfigure}%

  \caption{Dispersion curves ((red) real and (blue) imaginary parts of the roots $\omega$ of the dispersion relation \rf{domba}) for $M=2$, $\alpha\approx 0.0036725648$, and (a, d) $M_w=0$, (b, e) $M_w=0.09$, (c, f) $M_w=0.0967$, (g, j) $M_w=0.1$, (h, k) $M_w=0.5$, (i, l) $M_w=1$. Vertical dashed lines in the panels (a) and (d) correspond to $\kappa=\kappa_0\approx 0.5218134478$ and mark the onset of instability corresponding to the central instability zone in Fig.~\ref{fig:cone1}(a) and the conical instability zone in Fig.~\ref{fig:cone1}(b). \label{fig11}}
\end{figure}

\section{Discussion}

Comparing the eigenvalue plots of Fig.~\ref{fig:sigma_k_b} and Fig.~\ref{fig6} with the corresponding to each branch averaged wave energy and wave action that are shown in Fig.~\ref{fig:energy} and Fig.~\ref{energy_Mw}, respectively, we notice that flutter instability is necessarily accompanied with the interaction of waves of opposite sign of energy/action. In contrast to the action, the energy changes sign also at the points where the phase velocity $\sigma$ changes sign, quite in accordance with \rf{e_cairns}.

Looking now at the roots \rf{rb0} of the \textit{decoupled} dispersion equation \rf{b0}, we conclude that the elastic waves $\sigma_1^{\pm}=\pm M_w$ propagating in the membrane always have positive energy
whereas among the surface gravity waves $\sigma_2^{\pm}=M\pm \sqrt{(\tanh{\kappa})/{\kappa}}$ it is the energy of the slow wave $\sigma_2^-$ that becomes negative for $M>0$ as soon as $M>\sqrt{(\tanh{\kappa})/{\kappa}}$. Therefore at the crossing \rf{phasy} corresponding to $M_0^{+} = M_w+\sqrt{(\tanh \kappa)/{\kappa}}$ the positive-energy/action elastic wave meets the slow surface gravity wave that carries negative-energy/action \citep{N1986}.

With $\beta$ increasing from zero, the crossing unfolds causing the eigenvalue branches to merge on the interval bounded by the points where $\partial_{\omega}D=0$. At these exceptional points \citep{K2013dg} both the energy and the action change sign, see Fig.~\ref{fig:energy} and Fig.~\ref{energy_Mw}. On the interval the roots are complex and form the bubble of instability, see Fig.~\ref{fig:sigma_k_b} and Fig.~\ref{fig6}.

Since the fast surface gravity wave carries positive energy, one needs to add energy to the flow in order to excite this wave on the flow. In contrast, in order for the slow surface gravity wave carrying negative energy to build up on the flow, the energy must be extracted from the flow \citep{Nezlin76} via some mechanism for dissipation of its energy. In the Nemtsov problem, such a mechanism is the energy transfer from the slow surface gravity wave to an elastic wave associated with the membrane, which is a stationary medium and therefore has positive energy \citep{Nezlin76}. One can say that this transferred energy yields flutter of the membrane due to emission of the slow surface gravity wave carrying negative energy.

In Fig.~\ref{fig:sigma_k_b} and Fig.~\ref{fig6} as well as in Fig.~\ref{fig:energy} and Fig.~\ref{energy_Mw}
we observe that the flutter instability of the membrane occurs only if the velocity of the flow is higher than the phase velocity of the oscillations on the surface of the flow, $\sigma<M$, i.e. the flow moves faster than the waves it can excite \citep{NE1987,Nezlin76}.
The condition $\omega=M\kappa$ or $\sigma=M$ is known as the Cerenkov condition for emission of radiation by a moving source \citep{Nezlin76,GF1947,G1996,BS1998,CR2013}. Being substituted into a dispersion relation, the Cerenkov condition transforms the latter into the expression defining a surface in the space of wave numbers that determines the wake pattern behind the source \citep{CR2013,S1987}. For the supercritical velocities $M>\sigma$ the surface in the space of wave numbers develops a conical singularity known as the Cerenkov cone \citep{CR2013,N1986} with the angular aperture $$2\arccos\left(\frac{\sigma}{M}\right).$$ Anomalous Doppler effect (ADE) is the change in the sign of the field frequency radiated into the Cerenkov cone as compared with the field radiated outside this cone  \citep{Nezlin76,N1986,GG1983,AMN1986,CR2013}. It is exactly the slow surface gravity wave that satisfies this condition
$$\sigma_2^- - M= -\sqrt{\frac{\tanh \kappa}{\kappa}}<0.$$

Hence, for the one-dimensional or, more precisely, plane two-dimensional waves, both the negative energy waves and the ADE correspond simply to waves with phase velocity lower than the flow velocity and wave-vector pointing in the same direction as the flow \citep{ORT1986,N1986}. In our case, the radiated slow gravity wave increases the energy of oscillations of the membrane at the expense of the energy of the flow that supports this wave.

Finally, we plot the dispersion curves $\omega(\kappa)$ in Fig.~\ref{fig11}, which show that the slow surface gravity wave branch and the membrane branch interact under the line ${\rm Re}(\omega)=\kappa M$, if $\kappa>0$. Substituting the Cerenkov condition in the dispersion relation \rf{domba} we reduce it to
$
(M_w^2-M^2)\tanh \kappa=0
$,
thus providing a rationale for the absence of instabilities for $M_w^2>M^2$ that is evident in all our stability diagrams.

\section{Conclusion}

Through the revival of a classical hydrodynamical model performed in this work, we have been able to extend the stability analysis and to enhance knowledge of the underlying physics by making connections with the fundamental concepts such as negative energy waves and the anomalous Doppler effect, supported by advanced mathematical tools.

Our continuation of Nemtsov's investigation of the radiation-induced flutter of a membrane in a uniform flow with the new derivation of the dispersion relation for the fluid layer of arbitrary depth and membrane of infinite chord length has led to a significant improvement in the computation of stability diagrams without any limitation on the range of the system parameters.

An exhaustive stability analysis has been performed using the original perturbation theory of multiple roots of the dispersion relation to compare with the exact stability domains, and both computations are proven to be in excellent agreement. More precisely, the crossings and avoided crossings of the dispersion curves are very well approximated by the simplified expressions for the phase speed of the membrane and fluid modes derived with the perturbation approach.

After computing the discriminant of the full dispersion relation, we have identified a new instability domain arising from a conical singularity in the parameter space that was not present in the early study of Nemtsov. This new domain is associated with a low-frequency flutter for short wavelengths and corresponds to the case when the velocity of propagation of elastic waves in the membrane is much smaller than the velocity of the flow.

Moreover, following the procedures used in previous studies on simplified hydrodynamical systems to calculate the averaged wave energy and after developing the method further to take into account the coupling between the free surface of the flow and the elastic membrane on the bottom, we have obtained an elegant and applicable expression for the total averaged energy. We have verified that, in the absence of the background flow, the system respects the equipartition of energy in accordance with the virial theorem, thus confirming that the existence of the negative energy waves can only be possible when the fluid is in motion.

We have shown that the formula for the total averaged energy recovered in our work by means of the direct integration of different physical fields is expressed via the derivative of the dispersion relation with respect to the frequency of oscillations and reduces exactly to the form described by \cite{C1979}.

The anomalous Doppler effect (ADE) is a direct consequence of the relative motion of an oscillator in a medium and more precisely, it occurs when the internal energy of the system increases due to the emission of negative energy waves (NEW). In our context, while the system is composed of a fluid layer and a membrane, such a phenomenon has been proved by Nemtsov to exist only when the conditions of phase synchronism and NEW emission are satisfied. The criterion for the phase synchronism in the system is easily identified in the computations of the dispersion curves as the crossings of the different branches that lead to the onset of positive growth rate and therefore to temporal instability. The latter phenomenon is a natural consequence of the highly excited state of energy that the system transits to due to the dominance of NEW over the waves carrying positive energy. Indeed, NEW emission is known as a process that increases the total energy of an oscillatory system while radiating energy away from the oscillator and, only when this gain in internal energy exceeds the losses from the contribution of positive energy modes, the total energy of the system starts growing in amplitude. Hence, it requires to have waves carrying energy of opposite signs that interact for the instability to develop.

Our expressions for the action and energy of the Nemtsov system demonstrated as expected the collision of waves carrying positive and negative energy as the onset for the radiative instability and the flutter of the membrane. Such a phenomenon is well-known in the community of dynamical systems, but in this context, it is associated with the emission of NEW in the region of ADE. Hence, in addition to improving the stability analysis of the Nemtsov system and computing the averaged wave energy, our study provides a further, more detailed, examination of the ADE in hydrodynamics. Despite our problem being restricted to the study of planar waves, with the latter being emitted only in the horizontal direction, it is still sufficient for exploring the connection between the ADE and flutter theory.

An extension of this work to the case of a membrane with a chord of finite size, as described by the system of equations \rf{eq:eom}, is a promising necessary next step requiring asymptotic methods for the global stability analysis and numerical computations that we leave for future work.

\begin{acknowledgements}
We thank an anonymous referee for bringing important early works to our attention. We are grateful to Prof. T. J. Bridges for helpful discussions.
J.~L. was supported by a Ph.D. Scholarship from Northumbria University. 
\end{acknowledgements}

\appendix

\section{Sensitivity analysis of dispersion equations}
\label{sadr}
In contrast to other works on frequency coalescence, e.g. \cite{TT1991}, we adapt a more systematic multiparameter sensitivity analysis that can be found, e.g.,  in \cite{KS2002, KS2004, K2007a, K2007b, K2009, KGS2009, K2010, K2013dg}.

Let us consider the dispersion equation
\be{dispe}
D(\omega,p,q)=0,
\ee
where $D$ is a smooth function of scalar arguments $\omega$, $p$, and $q$. Assume that $D(\omega)$ is a polynomial of degree $n$ in $\omega$.

\subsection{Sensitivity of simple roots}
Let at $p=p_0$ and $q=q_0$ \rf{dispe} have a simple root $\omega_0$ such that
$$
D_0:=D(\omega_0,p_0,q_0)=0,
$$
where we use the symbol $:=$ to indicate a definition.

Following \cite{KS2002, KS2004, K2007a, K2007b, K2010, K2013dg}, we assume that $p=p(\varepsilon)$ and $q=q(\varepsilon)$.
For $|\varepsilon|$ being sufficiently small we can represent these functions as Taylor series
\ba{pq}
p(\varepsilon)&=&p_0+\varepsilon \frac{d p}{d \varepsilon}+\frac{\varepsilon^2}{2} \frac{d^2 p}{d \varepsilon^2}+o(\varepsilon^2),\nn\\
q(\varepsilon)&=&q_0+\varepsilon \frac{d q}{d \varepsilon}+\frac{\varepsilon^2}{2} \frac{d^2 q}{d \varepsilon^2}+o(\varepsilon^2),
\ea
with the derivatives evaluated at $\varepsilon=0$, and $p_0:=p(0)$ and $q_0:=q(0)$.
Then, $\omega=\omega(\varepsilon)$ is also a root of \rf{dispe}, i.e., it satisfies the equation
\be{Deps}
D_{\varepsilon}:=D(\omega(\varepsilon),p(\varepsilon),q(\varepsilon))=0.
\ee

Differentiating \rf{Deps}, we find
$$
\frac{d}{d \varepsilon}D_{\varepsilon}=\partial_{\omega}D \frac{d \omega}{d \varepsilon}+\partial_{p}D \frac{d p}{d \varepsilon}+\partial_{q}D \frac{d q}{d \varepsilon}=0,
$$
where the partial derivatives are evaluated at $\omega=\omega_0$, $q=q_0$, $p=p_0$.

Denoting $\Delta \omega=\varepsilon \frac{d \omega}{d \varepsilon}\approx \omega-\omega_0$, $\Delta q=\varepsilon \frac{d q}{d \varepsilon}\approx q-q_0$, and $\Delta p=\varepsilon \frac{d p}{d \varepsilon}\approx p-p_0$, we find the expression for the increment of the simple root $\omega_0$ of \rf{dispe} when the parameters depart from the values $q_0$ and $p_0$:
\be{dofo}
\Delta \omega=- \frac{\partial_p D}{\partial_{\omega} D}\Delta p- \frac{\partial_q D}{\partial_{\omega} D}\Delta q+
o(|\Delta p|, |\Delta q|).
\ee

\subsection{Double root of the dispersion relation: Generic case}

Let at $p=p_0$ and $q=q_0$ the dispersion equation \rf{dispe} have a double root $\omega_0$, which implies
\ba{dr}
D_0&=&0,\nn\\
\partial_{\omega}D_0:=\partial_{\omega}D(\omega_0,p_0,q_0)&=&0.
\ea
Assume that the perturbation of the parameters \rf{pq} causes splitting of the double root $\omega_0$ which generically is described by the Newton-Puiseux series \citep{KS2002, KS2004, K2007a, K2007b, K2010, K2013dg}
\ba{pert}
\omega(\varepsilon)=\omega_0+\omega_1\varepsilon^{1/2}+\omega_2\varepsilon+\omega_3\varepsilon^{3/2}+\omega_4\varepsilon^{2}+o(\varepsilon^2).
\ea

Expanding $D_{\varepsilon}$ as
\be{De}
D_{\varepsilon}=\sum_{s=0}^n\frac{\left(\omega(\varepsilon)-\omega_0\right)^s}{s!}\left(\partial_{\omega}^s D+\varepsilon \partial_{\omega}^s D_1+\varepsilon^2\partial_{\omega}^s D_2 +o(\varepsilon^2) \right),
\ee
where
\ba{der}
D_1&:=&\partial_pD \frac{d p}{d \varepsilon}+\partial_qD \frac{d q}{d \varepsilon},\nn\\
D_2&:=&\frac{1}{2}\partial_p D \frac{d^2 p}{d \varepsilon^2}+\frac{1}{2}\partial_q D \frac{d^2 q}{d \varepsilon^2}+\frac{1}{2}\left( \partial_p^2 D \frac{d^2 p}{d \varepsilon^2}+2\partial^2_{pq} D \frac{d p}{d \varepsilon}\frac{d q}{d \varepsilon}+ \partial_q^2 D \frac{d^2 q}{d \varepsilon^2}\right),
\ea
substituting expansion \rf{pert} into \rf{De}, and collecting the coefficients at
the same powers of $\varepsilon$, we find
\ba{nonde}
D_0&=&0,\nn\\
\omega_1 \partial_{\omega}D_0&=&0,\nn\\
D_1+\omega_1^2\frac{1}{2}\partial^2_{\omega}D+\omega_2 \partial_{\omega}D_0&=&0.
\ea
Looking for the coefficient $\omega_1\ne 0$, we see that the first two equations of \rf{nonde} are satisfied in view that $\omega_0$ is a double root of the dispersion equation \rf{dispe}. Taking this into account, the last of the equations \rf{nonde} yields the expression for the coefficient $\omega_1$ in the expansion \rf{pert}:
\be{om1}
\omega_1^2=-D_1 \left(\frac{1}{2}\partial_{\omega}^2 D \right)^{-1}
\ee
where all the partial derivatives are calculated at $p = p_0$, $q = q_0$, $\omega = \omega_0$.

Therefore, if $D_1\ne 0$, the double root $\omega_0$ splits under variation of parameters \rf{pq} according to the formula
\be{split}
\omega=\omega_0 \pm \sqrt{-\varepsilon D_1\left(\frac{1}{2}\partial_{\omega}^2D\right)^{-1}}+o(|\varepsilon|^{1/2}).
\ee
In terms of the increments of the parameters, we can re-write \rf{split} as
\be{spliti}
\Delta\omega= \pm \sqrt{-(\partial_pD \Delta p+\partial_qD \Delta q)\left(\frac{1}{2}\partial_{\omega}^2D\right)^{-1}}+
o(|\Delta p|^{1/2}, |\Delta q|^{1/2}).
\ee

\subsection{Double root of the dispersion relation: Degenerate case}

The case $D_1 = 0$ is degenerate, because the leading term in \rf{split} of order $\varepsilon^{1/2}$ vanishes and
the expansion \rf{pert} is no longer valid, see e.g. \cite{KS2004}. Substituting expansion \rf{split} with $\omega_1=0$ into \rf{De} and collecting coefficients of the same powers of $\varepsilon$, we obtain
\ba{dc12}
D_1+\omega_2 \partial_{\omega} D_0&=&0,\nn\\
\omega_3 \partial_{\omega} D_0&=&0,\nn\\
D_2+\omega_2^2\frac{1}{2}\partial_{\omega}^2 D+\omega_2\partial_{\omega} D_1 +\omega_4\partial_{\omega} D_0&=&0.
\ea
Taking into account that $\partial_{\omega}D_0=0$ since $\omega_0$ is the double root and that $D_1 = 0$ due to our assumption,
we conclude that the first two of equations \rf{dc12} hold automatically. The third one simplifies as follows:
\be{deq}
\omega_2^2\frac{1}{2}\partial_{\omega}^2 D+\omega_2\partial_{\omega}D_1+D_2=0,
\ee
where all the derivatives are calculated at $\omega = \omega_0$, $p=p_0$, and $q = q_0$.

Therefore, the degeneracy, $D_1 = 0$, implies that the double root $\omega_0$ splits according to the formula
\be{desplit}
\omega = \omega_0+\omega_2\varepsilon+o(\varepsilon),
\ee
where the coefficient $\omega_2$ is a root of the polynomial \rf{deq}.

In combination with \rf{der} and \rf{desplit} the polynomial \rf{deq} transforms into
\ba{qede}
&(\Delta\omega)^2\frac{1}{2}\partial_{\omega}^2 D+\Delta\omega(\partial^2_{\omega p}D \Delta p+\partial^2_{\omega q}D \Delta q)+\frac{1}{2}\left[ \partial_p^2 D (\Delta p)^2+2\partial^2_{pq} D \Delta p \Delta q+ \partial_q^2 D (\Delta q)^2\right]\nn \\
&+ \partial_p D \Delta p+\partial_q D \Delta q=0.
\ea

Extension to the case of more than two parameters is straightforward, see e.g. \cite{KS2002, KS2004, K2007a, K2007b, K2009, KGS2009, K2010, K2013dg}.

\end{document}